\documentclass[aps,preprint,amsmath,amssymb,eqsecnum,showpacs]{revtex4-1}
\usepackage[colorlinks=true,citecolor=blue,linkcolor=blue]{hyperref}
\usepackage{graphics}
\usepackage{graphicx}
\usepackage{epsfig}
\usepackage{setspace}
\usepackage{amssymb,amsmath}
\usepackage{color}
\def \be {\begin{equation}}
\def \ee {\end{equation}}
\def \ba {\begin{eqnarray}}
\def \ea {\end{eqnarray}}
\def \bm {\begin{displaymath}}
\def \em {\end{displaymath}}
\def \br {{\bf r}}

\newcommand{\Rnum}[1]{\uppercase\expandafter{\romannumeral #1\relax}}
\begin{document}
\title{ Intrusion of liquids into liquid infused surfaces \\           
        with nanoscale roughness 
         }
\author{Swarn Lata Singh{$^{1,2,3}$}, Lothar Schimmele{$^{1,2}$}, and S. Dietrich{$^{1,2}$}}
 \affiliation{$^{1}$ Max-Planck-Institut f\"{u}r
 Intelligente Systeme, D-70569 Stuttgart, Heisenbergstr. 3, Germany, \\$^{2}$ IV. Institut f\"{u}r
 Theoretische  Physik, Universit\"{a}t Stuttgart, Pfaffenwaldring 57,
 D-70569 Stuttgart, Germany
 \\ $^{3}$ Department of Physics, Mahila Mahavidyalaya (MMV), Banaras Hindu University, 
           Varanasi, UP, 221005, India
 }
\email{swarn@bhu.ac.in}
\begin{spacing}{2.5}
\begin{abstract}
We present a theoretical study of the intrusion of an ambient liquid into the pores of 
a nano-corrugated wall $w$. The pores are prefilled with a liquid lubricant which adheres to
the walls of the pores more strongly than the ambient liquid does. The two liquids are
modeled as a binary liquid mixture of two species of particles, $A$ and $B$.
The mixture can decompose into a liquid rich in $A$ particles, representing the ambient liquid, and
another one rich in $B$ particles, representing the liquid lubricant. The wall is taken to 
attract the $B$ particles more strongly than the $A$ particles.  
The ratio $w$--$A$/$w$--$B$ of these interaction strengths is changed in order to tune
the contact angle $\theta_{AB}$ formed by the $A$-rich/$B$-rich liquid interface between 
the two fluids and a planar wall, composed of the same material as the one forming the pores.
We use classical density functional theory, in order to capture the effects of microscopic details on 
the intrusion transition, which occurs as the concentration of the minority component 
or the pressure in the bulk of the ambient liquid is varied, moving away from
bulk liquid--liquid coexistence within the single-phase domain of the $A$-rich bulk ambient liquid. 
These liquid structures have been studied as a function of the contact angle $\theta_{AB}$ and 
for various widths and depths of the pores.
We also studied the reverse process in which a pore initially filled
with the ambient liquid is refilled with the liquid lubricant. 
The location of the intrusion transition, with respect to its dependence on the contact angle $\theta_{AB}$ 
and the width of the pore, qualitatively follows the corresponding
shift of the capillary-coexistence line away from the bulk liquid--liquid coexistence line, 
as predicted by a macroscopic capillarity model.  
Quantitatively, the
transition found in the microscopic approach occurs somewhat closer
to the bulk liquid--liquid coexistence line than predicted by the macroscopic capillarity model.
The quantitative discrepancies become larger for narrower cavities.
In cases in which the wall is completely wetted by the lubricant
($\theta_{AB} = 0$) and for small contact angles the reverse transition follows the same path 
as for intrusion; there is no hysteresis.
For larger contact angles hysteresis is observed. The width of the hysteresis increases with
increasing contact angle. A reverse transition is not found inside the 
domain within which the ambient liquid forms a single phase in the bulk, once $\theta_{AB}$
exceeds a geometry dependent threshold value. According to the macroscopic capillarity theory, for 
the considered geometry, this is the case for $\theta_{AB} > 54.7^\circ$.
Our computations show, however, that nanoscale effects shift this threshold value to much higher values.
This shift increases strongly if the widths of the pores become smaller
(below about ten times the diameter of the $A$ and $B$ particles). 
\end{abstract}

\maketitle
\section{Introduction}	
Super-liquid-repellent surfaces are of much interest for designing surfaces with 
anti-fouling, -fogging, -icing,
or self-cleaning properties, for microfluidic applications, 
and in order to reduce drag in flow along these interfaces 
\cite{Ref1, Ref2, Ref3, Ref4, Ref5, Ref6, Ref7, Ref8, Ref9, Ref10}, 
to provide a few examples. 
Most of the synthetic liquid-repellent surfaces are inspired by the lotus
effect \cite{Ref11}, which is based on forming micro/nano structures on the
surfaces. The liquid-repellency of these surfaces is due to a combination of surface
chemistry, which facilitates to create a lyophobic surface, and surface roughness. This combination
allows air pockets to stay within the surface roughness, which inhibit the effective contact
between the liquid and the solid surface, enabling the liquid to roll
off easily \cite{Ref12, Ref13, Ref14, Ref15, Ref16}. 
A potential problem concerning the applicability of such surfaces arises, because the
trapped air pockets cannot sustain pressure and the surface looses
liquid-repellency under increased pressure, e.g., as a drop of liquid impacts on the surface. 
Furthermore, for surfaces with micronscale roughness the
liquid-repellency cannot be restored easily after a pressure induced collapse of the air pockets. 
By adding nanoscale roughness this problem can be avoided in principle
so that robust liquid-repellent surfaces can be fabricated \cite{Alberto1, Alberto2}. 
Other problems might arise from the occurrence of surface defects and irreparable damages and 
contaminations which accumulated over time
\cite{Ref2, Ref17, Ref18, Ref19, Ref20, Ref21, Ref22, Ref23}.
The challenge of preparing robust and durable liquid-repellent
surfaces increases further for liquids with low surface tension
such as organic liquids \cite{Ref24, Ref25, Ref26}.

A different route towards the development of stable liquid repellent surfaces
utilizes low surface tension liquids (lubricant) infused into the surface protrusions.
The lubricating liquid is chosen such that it is immiscible
with the ambient liquid and has a higher affinity to
the solid walls in comparison to that of the ambient liquid.
Such arrangements are known as slippary liquid infused porous surfaces
(SLIPS) \cite{Ref2, Ref27, Ref28}. 
It is expected that the 'incompressibility' of the lubricating liquid provides SLIPS the
robustness which air pockets, being highly compressible, cannot. 
In addition SLIPS may exhibit other useful properties such as the ability of the lubricant 
to flow into damaged
sites on the underlying surface and to stabilize them \cite{Ref2, Ref16}.
Numerous studies have been devoted to the application of SLIPS
as anti-fouling \cite{Ref29, Ref30, Ref31} and anti-icing \cite{Ref34, Ref35, Ref36} surfaces,
and to surfaces which facilitate condensation and heat transport
\cite{Ref37, Ref38, Doris1}.
In addition to generally demonstrating the potential of SLIPS,
most further efforts focus on novel methods of fabricating such surfaces
\cite{Ref32, Ref33, Ref39, Ref39b, Ref40, Ref41, Ref42, Ref43, Ref44, Ref45}.  
However, detailed investigations of the wetting behavior of lubricant infused surfaces, 
as a function of various system parameters, remain sparse \cite{Ref9, Ref46, Ref47, Doris2}.

The purpose of the present theoretical study is to provide an understanding 
of the stable and metastable thermodynamic states of lubricant infused surfaces,
which are in contact with an ambient liquid.
These states are investigated as functions of the interactions of the ambient liquid 
and of the lubricant liquid, respectively, with the solid surface.
The dependence on the linear dimensions of the surface corrugations is discussed
as well. 
In this context, as the simplest case we consider a binary liquid mixture composed of, say, $A$ and $B$ 
particles (molecules), which decomposes into a liquid rich in $A$
particles, representing the ambient liquid, and an other liquid rich in
$B$ particles, representing the lubricant. The lubricant adheres to the pore walls 
more strongly than the ambient fluid, i.e., in the present model the $B$ particles are attracted
by the walls more strongly than the $A$ particles. The mutual solubility might be
very low, but is nonzero. The concentration of the minority component
in the ambient liquid (i.e., the concentration of the $B$ particles) is varied 
within the single-phase domain of the A-rich bulk ambient liquid in order
to find out how this variation affects the stability of the lubricant infused configuration.

We are especially interested in nanoscale effects, which become relevant if the
corrugations become very narrow with linear dimensions of nanometer size.
For this reason we use classical density functional theory (DFT), which
captures microscopic details of the system, such as the extended range of
the interactions between the liquid particles and between the liquid particles 
and the particles forming the solid surfaces, packing effects due to the
strong repulsion between particles at short distances, and the widths of the
various interfaces as well as their molecular structure. 
We are particularly interested in the intrusion of the ambient liquid into
the lubricant infused nano-pores, triggered by variations of the composition
of the ambient liquid and of its pressure. Furthermore, we study the
reformation of the lubricant infused configuration from a configuration
in which the pores are filled with the ambient liquid.
These processes are studied as a function of the lubricant--wall interaction strength
and of the width and depth of the pores. 

Our study is also relevant for designing channels and pores with 
sharp gating abilities.
These channels, if filled with a 'gating liquid' (i.e., a lubricant) strongly wetting 
the pore walls, prevent passage of an ambient liquid. Upon changing conditions
in the ambient liquid, the 'gating liquid' gives way and the ambient liquid 
can pass the channel \cite{Ref50, Ref51}.
In this context selectivity and reversibility of this mechanism are important issues.

Examples of such pores are actually found in nature, such as the nuclear pore complex in
eukaryotic cells and 
Stomata and  Xylem in plants, which use liquids to mechanically reconfigure the
pores \cite{Ref48, Ref49}.
Another example related to our studies consists of porous surfaces the physical properties 
of which can be changed by switching between a state, in which the pores are filled 
with a lubricant, and one in which the pores are filled with an ambient liquid.

It should be noted, that in our studies transitions between thermodynamically stable or 
metastable states are considered. In some of the experimental studies and applications,
chemical equilibrium might not be reached within the typical duration of the experiment
or of the considered process.
Nevertheless, also in these cases, knowledge of the equilibrium configurations
is needed for a proper interpretation of the experimental results and this knowledge
should also be useful in designing systems for specific applications.

In the next section (Sec. \Rnum{2}), details of our model system are described,
as well as the computational method. Furthermore, we discuss liquid--liquid
coexistence in the bulk and we describe the modifications to be expected,
based on macroscopic capillarity theory, if these liquids coexist within a capillary.
Based on this discussion, we can make a macroscopic prediction concerning the intrusion 
of the ambient liquid into a lubricant infused pore.
In the following sections we present and discuss the results of the microscopic
DFT computations.
   
\section{Model system}
In order to model a liquid infused porous surface, we consider a fluid which
is composed of $A$ and $B$ particles. The parameters for the fluid--fluid
interactions and the thermodynamic conditions are chosen such that in the bulk the
fluid is in a liquid state which phase separates into an $A$-rich and into 
a $B$-rich liquid. In the following the $A$-rich liquid is denoted as $L_{A}$, 
the $B$-rich liquid as $L_{B}$. For a suitable choice of parameters, the $A$-rich liquid 
may contain up to a few percent of $B$ particles, and vice versa the $B$-rich liquid 
contains a few percent of $A$ particles.  
The porous surface is modeled by engraving square nanopits into an otherwise 
flat solid surface, also called a wall (see Fig. \ref{schem}). The pits are periodically 
repeated with the periodicity length chosen to be sufficiently large such that the periodicity 
is irrelevant for the present investigations. The interaction strength between
the $B$ particles and the wall is taken to be stronger than the one between the 
$A$ particles and the wall.
Thus the $B$-rich liquid $L_{B}$ adheres strongly to the walls of the pits. 
The $B$-rich liquid $L_{B}$ represents the lubricant whereas the $A$-rich liquid $L_{A}$
represents an ambient liquid, which is always present above the wall
and may or may not intrude into a pit initially filled with the lubricant $L_{B}$.
We consider stable or metastable equilibria in a grand canonical ensemble. 
This implies that starting from an initial configuration the $A$ and $B$ particles 
are allowed to 'diffuse' between the ambient liquid 
above the wall and the liquid inside the pit until chemical equilibrium is reached. 
Since on the nanoscale interdiffusion is very fast, this scenario is of practical
relevance. Even on larger length scales the configurations, which are stable
on a longer time scale, correspond to the ones considered here.
This phenomenon exhibits obvious similarities with the Cassie--Wenzel transition, which
refers to the intrusion of a liquid into vapor-filled surface asperities  
(see, e.g., Refs. \cite{Ref12, Ref13, Ref15, Ref16, Ref17, Alberto1, Alberto2, mypre2}).
In the following we shall use a corresponding terminology. 
The wetting state, in which the pits are completely
filled with the lubricant ($L_{B}$) and the ambient liquid ($L_{A}$)
remains suspended above the wall, is called the Cassie state, whereas the Wenzel state 
corresponds to a state in which the ambient liquid intrudes the pit 
and replaces the lubricant. The Cassie--Wenzel transition 
(i.e., the intrusion of ambient liquid $L_{A}$ into a pit filled with lubricant $L_{B}$) 
and the reverse transition (removing the ambient liquid $L_{A}$ from the pit and
restoring the filling with lubricant $L_{B}$), called Wenzel--Cassie transition,
is studied as a function of various system parameters.
We also want to point out, that there are similarities to the filling transitions 
which are observed if a wall with surface asperities is exposed to vapor and
is partially filled by the coexisting liquid phase 
(see, e.g., Refs. \cite{RDN, Alex2, Alex3, Alex}). The correspondence is less
obvious in this case, due to different conditions and dependences studied
in these investigations.   
In the following subsections the various pieces of the system, the numerical method, 
which is based on classical density functional theory (DFT), and that part of 
the bulk phase diagram, which is relevant for the present investigations, 
are explained in more detail.
\subsection{Binary liquid mixture}
The particles $A$ and $B$ forming the binary liquid mixture are considered to 
interact via Lennard-Jones (LJ) pair potentials. In the spirit of DFT as used here 
and discussed in more detail below, each pair potential is replaced by the sum of
hard sphere interactions
\begin{equation} \label{eq:pre3}
U^{hs}_{ij}(r) =   \left\{
\begin{array}{r l}
\infty,  &  r\leq\sigma_{ij} = R_{i} + R_{j} , \\
 0,      &  r > \sigma_{ij}
\end{array}
\right.
\end{equation}
and of a soft attractive part \cite{Weeks}
\begin{eqnarray}\label{eq:pre4}
U^{att}_{ij}(r)=-\epsilon_{ij} \Theta({2}^{1/6}\sigma_{ij}-r)
+\Phi^{LJ}_{ij}(r)\Theta(r-{2}^{1/6}\sigma_{ij})
\end{eqnarray}
with the Heaviside function $\Theta$ and
\begin{equation}\label{eq:pre5}
\Phi^{LJ}_{ij}=4\epsilon_{ij}\left[{\left(\frac{\sigma_{ij}}{r}\right)}^{12}-{\left(\frac{\sigma_{ij}}{r}\right)}^{6}\right].
\end{equation}
Here, $i$ and $j$ represent the two species $A$ and $B$ of
the binary liquid mixture, $R_{i}$ is the radius of the fluid particles of species $i$,
$-\epsilon_{ij}$ is the potential depth for the $ij$-pair
potential at $r=2^{1/6}\sigma_{ij}$, $r$ is the center-to-center interparticle separation,
and $\sigma_{ij}$ is the distance of contact between the centers of two hard sphere
particles. For simplicity, the binary liquid mixture is considered to be symmetric
with $R_{A} = R_{B} = R$ so that $\sigma_{ij}=\sigma=2R$ and $\epsilon_{AA} = \epsilon_{BB}$,
where $\epsilon_{ij}$ is the strength of the $ij$-pair potential.
The fluid--fluid interaction is rendered effectively
short-ranged by introducing a cut-off at $r=R_{cut}$.
In the following we adopt $R_{cut} = 5R$, which is implemented by a cut-off function.
The interaction parameters $\epsilon_{ij}$ for the cut-off
potential are rescaled such that the integrated interaction is equal
to the original one resulting from the potential in Eq. (\ref{eq:pre5}) with $R_{cut} \to \infty$.
For further details see Refs. \cite{mypre1, mypre2}.
\subsection{Fluid--wall interaction and wall topography}
\begin{figure}   
\hspace*{-2.0cm}\includegraphics[scale = 0.55]{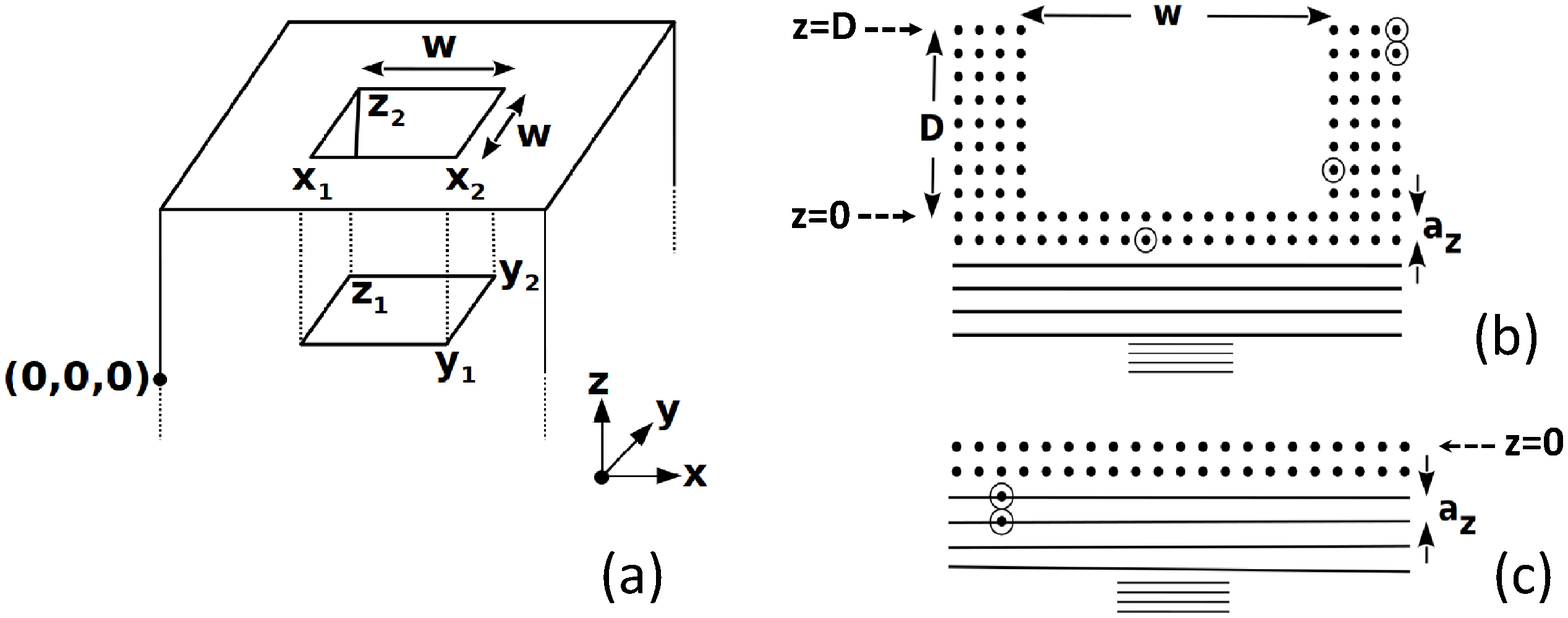}
	\caption{\label{schem}   
 $(a)$ Schematic representation of the structured wall studied
 here. The structure is modeled by square pits $w \times w \times D$
 of width $w$ and depth $D$. The pits are carved out from a planar
 wall. In $(a)$ and $(b)$ a single pit is shown. The solid substrate
 is modeled as a simple cubic lattice occupied densely by particles of type $C$.
 The lattice spacing is $a_{z}$ which equals twice the radius $R_{C}$ of the
 $C$ particles. The circles indicate the size of the $C$ particles.
 In $(b)$, the vertical cross section of the pit is shown.
 For lattice planes sufficiently far from the liquid volume, discrete
 sums (dots) are replaced by integrals (full horizontal lines), 
 in order to speed up the calculations.
 In $(c)$ the reference configuration of the planar substrate is shown,
 for which the surface tensions $\sigma_{sA}$ and $\sigma_{sB}$ 
 (see Eq. (\ref{eq:Young})) between the solid substrate and the
 $A$-($B$-) rich liquid are determined.
 The lengths $w$ and $D$ are measured between the respective
 loci of the nuclei of the $C$ particles. These definitions
 imply that they encompass the corresponding depletion zone
		for the density profiles of the fluid particles.    
  }
\end{figure}
The wall is modeled as a block of solid particles $C$ occupying a simple cubic lattice.
Certain lattice sites are left unoccupied, such that a wall with the desired corrugation
is formed (see Fig. \ref{schem}). 
The solid particles $C$ interact with the liquid particles via a Lennard-Jones
potential. Summing over all interactions one obtains the wall potential acting on 
a fluid particle of sort $i$ ($i=$ $A$ or $B$) as 
\begin{equation}\label{eq:pre7}
V_{i,ext}({\bf r}) = -4\epsilon_{i}\sum_{l}{\left[{\left(\frac{\sigma_{i}}{|{\bf r}-{{\bf r}_{l}}|}\right)}^{12}
- {\left(\frac{\sigma_{i}}{|{\bf r}-{{\bf r}_{l}}|}\right)}^{6}\right]} .
\end{equation}
The parameter $\epsilon_{i}$ defines the strength of the interaction between a
solid particle $C$ at lattice site ${\bf r}_{l}$ and a fluid particle of sort $i$
at ${\bf r}$,
$\sigma_{i} = R_{i} + R_{C} = R + R_{C}$, and $R_{C}$ is the radius
of the solid particles.
In Eq. (\ref{eq:pre7}) the sum over $l$ amounts to a sum over all
lattice sites occupied by $C$ particles.
In order to speed up the computations of $V_{i,ext}$ some modifications
are introduced, as described in Ref. \cite{mypre2}. However, these
have negligible effects on $V_{i,ext}({\bf r})$.
%
\subsection{Classical density functional theory (DFT)}
The grand canonical potential $\Omega$ of a classical system
of an $N$-component mixture follows from the variational functional
\begin{equation}\label{eq:9}
\Omega[\{\rho_{i}\}]=F[\{\rho_{i}\}]+\sum_{i=1}^{N}\int d^{3}r \rho_{i}({\bf r})(V_{i, ext}({\bf r})-\mu_{i})
\end{equation}
of the one-particle number densities $\rho_{i}({\bf r}), i = 1,...., N$.
$F$ is the Helmholtz free energy functional, $V_{i, ext}({\bf r})$
is the external potential, and $\mu_{i}$ is the chemical potential of
species $i=A,B$, respectively. The equilibrium number densities $\rho_{i, 0}
({\bf r})$ minimize $\Omega$:
\begin{equation}\label{eq:10}
\left. \frac{\delta \Omega[\rho_{i}]}{\delta \rho_{i}({\bf r})}\right\vert_{\rho_{i}(\br)=\rho_{i, 0}(\br)}=0 .
\end{equation}
$\Omega[\{\rho_{i, 0}\}]$ is the equilibrium grand canonical potential of the system \cite{Evans2, gurug}.
The free energy functional can be divided into two parts:
\begin{equation}\label{eq:11}
F[\{\rho_{i}\}]=F_{id}[\{\rho_{i}\}]+F_{ex}[\{\rho_{i}\}],
\end{equation}
where $F_{id}$ is the ideal gas part
\begin{equation}\label{eq:12}
F_{id}[{\rho_{i}}]=k_{B} T \sum_{i=1}^{N} \int d^{3}r \rho_{i}(\br)\left[\ln\left(\rho_{i}(\br)\Lambda_{i}\right)-1\right],
\end{equation}
$\Lambda_{i}={(\frac{h^{2}}{2\pi m_{i} k_{B}T})}^{3/2}$ is
the cube of the thermal wavelength associated with a particle
of species $i$ and mass $m_{i}$, $h$ is Planck's constant, and
$k_{B}$ is the Boltzmann constant.
The excess part $F_{ex}$ arises due to the interparticle
interactions. We approximate the excess part as the
sum of two distinct contributions: one
arising due to the hard core repulsion $(F_{hs})$, and the
other due the attractive part of the interaction
$(F_{att})$:
\begin{equation}\label{eq:13}
F_{ex}=F_{hs}+F_{att}.
\end{equation}
$F_{hs}$ is treated within the framework of fundamental measure
theory (FMT), as described in, e.g., Refs. \cite{Roth3, Tarazona, Yasha1, mypre1, mypre2};
we have chosen the variant proposed by Rosenfeld {\it et al.} \cite{Yasha2}. 
$F_{att}$ is approximated within a simple random phase approximation.
In order to approximate $F_{att}$, we have used
the following truncation of the corresponding functional perturbation expansion:
\begin{equation}
F_{att}=\frac{1}{2}\sum_{i,j = 1}^{N}\int d^{3} r \int d^{3}r^{'}\rho_{i}({\bf r})\rho_{j}({\bf r^{'}})U^{att}_{ij}({\bf r}-{\bf r^{'}}) \nonumber
\end{equation}
with $U^{att}_{ij}$ defined via Eqs. (\ref{eq:pre4}) and (\ref{eq:pre5}). Here, the
minimization of  $\Omega[\{\rho_{i}\}]$
has to be carried out numerically. The number density
is discretized on a simple cubic grid, and a Picard
iteration scheme is used in order to minimize $\Omega$ and to
determine the equilibrium number densities. Details about the
computational techniques can be found in Refs. \cite{mypre1, mypre2, Roth4}.
%
\subsection{Bulk liquid--liquid coexistence}
\begin{figure}   
\hspace{-0.0cm}\includegraphics[scale = 0.33]{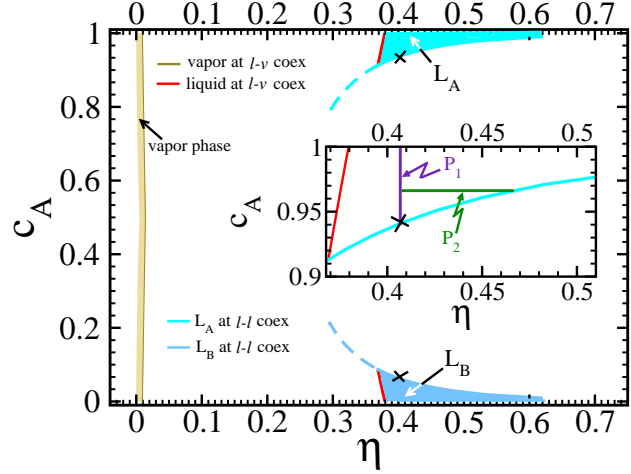}
{\vspace*{-0.0cm}
\caption{\label{pd} 
Cut through the bulk phase diagram of the symmetric binary liquid, which is composed
of $A$ and $B$ particles ($\epsilon_{AA} = \epsilon_{BB}$, $\epsilon_{AB} = 0.77\epsilon_{AA}$,
$R_{A} = R_{B} = R$), at fixed temperature $T = 0.899\epsilon_{AA}/k_{B}$;
$c_{A}$ is the concentration of $A$ particles and $\eta = \eta_A + \eta_B$ is the total packing fraction. 
The right edge of the brownish stripe, at low $\eta$, represents the $c_{A}$ and $\eta$ values of 
the vapor coexisting with 
an $A$-rich liquid ($L_{A}$) or $B$-rich liquid ($L_{B}$); the red lines represent the respective 
$c_{A}$ and $\eta$ values of the coexisting liquids. 
The cyan line defines the $c_A$ and $\eta$ values of the liquid $L_{A}$ at coexistence
with the liquid $L_{B}$; the $c_A$ and $\eta$ values for $L_{B}$ at liquid--liquid coexistence
is represented by the blue line. In the studies of the Cassie--Wenzel transition, $c_A$
and $\eta$ in the ambient liquid $L_{A}$ are controlled. $L_{A}$ ($L_{B}$) 
is the stable liquid for concentrations $c_{A}$ above (below) the cyan (blue) line.  
The dashed lines extend the liquid--liquid coexistence lines into the range within which
liquid is only metastable. The crosses on the liquid--liquid coexistence lines mark the pair of 
reference points used in the linearization described in Sect. II E.
In the inset $P_1$ and $P_2$ indicate two thermodynamic paths (see the main text). 
                             }}
\end{figure}

In what follows we study, as already mentioned, a symmetric binary liquid mixture, 
i.e., the radii of the $A$ and the $B$ particles are equal ($R_{A} = R_{B} =R$) 
and the strengths of the $A$--$A$ and $B$--$B$ interactions are equal, too, i.e., 
$\epsilon_{AA} = \epsilon_{BB}$. We further fix the temperature
such that $T = 0.899\epsilon_{AA}/k_{B}$. The strength of the $A$--$B$ 
interaction relative to the $A$--$A$ ($B$--$B$) interaction is fixed to the value
$\epsilon_{AB}/\epsilon_{AA} = 0.77$. The parameters are chosen such
that thermodynamically the fluid is away from criticality and also from solidification. 
Moreover the parameters are selected such that the liquid is demixed. Otherwise, the choice of the 
precise numbers is rather arbitrary.   
In addition to the number densities $\rho_{A}$ and $\rho_{B}$
of the $A$ and $B$ particles, respectively, we also introduce the
concentrations $c_A = \rho_{A}/(\rho_{A} + \rho_{B})$ and
$c_B = \rho_{B}/(\rho_{A} + \rho_{B})$ (with $c_B = 1 - c_A$), 
as well as the total fluid packing fraction
$\eta = \eta_{A} + \eta_{B}$ with $\eta_{A} = (4\pi/3)R^3_A\rho_A$ and
$\eta_{B} = (4\pi/3)R^3_B\rho_B$; $\eta_{A}$ and $\eta_{B}$
are the fluid volume fractions blocked by the
hard cores of the $A$ and $B$ fluid particles, respectively.
A cut through the bulk phase diagram of this liquid at the temperature 
$T = 0.899\epsilon_{AA}/k_{B}$, as computed by using the DFT version described above,
is shown in Fig. \ref{pd} in
a plane spanned by the variables $c_A$ and $\eta$. 
The cyan line gives the concentration $c_A$ and the total packing fraction $\eta$
of the $A$-rich liquid $L_{A}$ at coexistence with the $B$-rich liquid $L_{B}$,
the $c_A$ and $\eta$ values of which are defined by the blue line;
the two coexisting liquids always have the same $\eta$ values. 
For completeness we also show as red lines the $c_A$ and $\eta$ values for the liquids
$L_{A}$ and $L_{B}$, respectively, at coexistence with vapor, the $c_A$ and $\eta$ 
values of which are defined by the right edge of the brownish stripe at small total packing fraction $\eta$.
The liquids, in general, have concentrations $c_A$, which are very different from
the ones of the coexisting vapor. When the red lines meet the cyan and blue lines, one has three-phase
coexistence between the two liquids $L_{A}$ and $L_{B}$, and the vapor with
concentration $c_A = 0.5$. Upon varying temperature the three-phase coexistence
points form the triple line of binary liquid mixtures.

In the following studies of the Cassie--Wenzel or Wenzel--Cassie transitions the
concentration $c_A$ and the total packing fraction $\eta$ in the ambient liquid 
$L_{A}$ are controlled. The changes of parameters create the vertical path $P_1$
(variation of $c_A$ at fixed $\eta$) or the horizontal path $P_2$ (variation of $\eta$ at fixed $c_A$),
as indicated in the inset of Fig. \ref{pd}. 
%
%
%
\subsection{Macroscopic theory of liquid--liquid coexistence in a capillary }
\begin{figure}
\includegraphics[scale = 0.35]{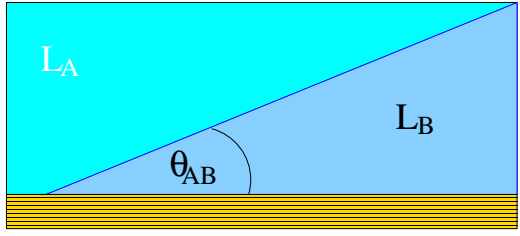}
\caption{\label{tAB} The $L_{A}$--$L_{B}$ interface and the surface of the solid wall
		     form the contact angle $\theta_{AB}$ (see Eq. (\ref{eq:Young})). 
                                                    } 
\end{figure}

If the liquid is confined between macroscopicly extended walls, the liquid--liquid coexistence lines 
are shifted.
The size and the direction of this shift depend also on the fluid--wall interactions.
In order to study these shifts we first consider predictions based 
on the macroscopic capillarity theory. 
In order to facilitate comparisons with the full DFT calculations, the corresponding
macroscopic parameters are computed by using DFT.
In this context, the most important macroscopic parameter is the contact angle $\theta_{AB}$,
at which the $L_{A}$--$L_{B}$ interface meets the surface of a solid wall
(see Fig. \ref{tAB}).  
The contact angle $\theta_{AB}$ follows from Young's law 
\be
\label{eq:Young}
\cos\theta_{AB} = \frac{\sigma_{sA} - \sigma_{sB}}{\sigma_{AB}} \, ,
\ee
where $\sigma_{sA}$, $\sigma_{sB}$, and $\sigma_{AB}$ are the interfacial
tensions of the solid--$L_{A}$, the solid--$L_{B}$, and the $L_{A}$--$L_{B}$
interface, respectively; $\sigma_{AB}$ and thus $\theta_{AB}$ are properly defined only at
bulk $L_{A}$--$L_{B}$ coexistence, but not for complete wetting \cite{D01}.  

At bulk $L_{A}$--$L_{B}$ coexistence the following three conditions hold:
\be
\label{eq:Coex_bulk}
\mu^{L_{B}}_{A}(c^{L_{B},co}_{A},\eta^{L_{B},co}) = \mu^{L_{A}}_{A}(c^{L_{A},co}_{A},\eta^{L_{A},co});
\hspace{0.2cm} 
\mu^{L_{B}}_{B}(c^{L_{B},co}_{A},\eta^{L_{B},co}) = \mu^{L_{A}}_{B}(c^{L_{A},co}_{A},\eta^{L_{A},co});
\hspace{0.2cm} 
 p^{L_{B}} = p^{L_{A}} \, .
\ee
Here, $\mu^{L_{B}}_{i}$ and $\mu^{L_{A}}_{i}$ are the chemical potentials of species 
$i$ in the liquid $L_{B}$ and $L_{A}$, respectively.
The concentrations of $A$ particles in the coexisting liquids $L_{B}$ and $L_{A}$, 
respectively, are denoted as $c^{L_{B},co}_{A}$ and $c^{L_{A},co}_{A}$
(with $c_B = 1 - c_A$); the corresponding total packing fractions are 
$\eta^{L_{B},co}$ and $\eta^{L_{A},co}$, respectively.  
The pressures $p$ in the two liquids are equal.
However, in confinement the pressures of the two coexisting liquids are no longer equal. 
Within the macroscopic capillary theory, the grand potential can be expressed in terms of volume 
and interfacial contributions. Equating the grand potentials for a large but finite capillary 
with square cross section (see Fig. \ref{schem}), which is filled with either liquid $L_{A}$  
or liquid $L_{B}$, one obtains
\be
\label{eq:Coex_cap}
 p^{L_{A}} - p^{L_{B}} = \frac{4(\sigma_{sA} - \sigma_{sB})}{w} \,  ,
\ee
which replaces the last equation in Eq. (\ref{eq:Coex_bulk}).
According to Eq. (\ref{eq:Young})  and with 
$(\sigma_{sA} - \sigma_{sB})/\sigma_{AB} \in [1,-1]$, 
Eq. (\ref{eq:Coex_cap}) can be expressed as
\be
\label{eq:Coex_cap_B}
 p^{L_{A}} - p^{L_{B}} = \frac{4\sigma_{AB} \cos\theta_{AB}}{w} \,  .
\ee  
The change of the pressure conditions, from equal pressure in the two liquids at bulk
coexistence to a pressure difference as given by Eq. (\ref{eq:Coex_cap}) in the 
two coexisting liquids confined by a capillary, leads to a shift of the 
coexistence lines for capillary coexistence away from the bulk coexistence lines.
This shift can be approximately determined by linearizing the pressure 
and the chemical potentials for both species, as a function of concentration
and total packing fraction, about a pair of reference points on the 
bulk coexistence lines. Such a pair of coexisting liquids $L_{A}$ and $L_{B}$
is indicated by the two crosses on the coexistence lines in Fig. \ref{pd}.
This linear expansion is given by
%
\begin{eqnarray}
\label{eq:expansion}
&&p^{L_{A}} = p^{ref} + \frac{\partial p}{\partial c_{A}}|_{ref,A}
            \Delta c^{L_{A}}_{A}+ 
             \frac{\partial p}{\partial \eta}|_{ref,A} \Delta \eta^{L_{A}} \, , \\ \nonumber 
&&p^{L_{B}} = p^{ref} + \frac{\partial p}{\partial c_{A}}|_{ref,B}
            \Delta c^{L_{B}}_{A}+ 
             \frac{\partial p}{\partial \eta}|_{ref,B} \Delta \eta^{L_{B}} \, , \\ \nonumber 
&& \mathrm{ and \, \, similarly \, for \, } \mu^{L_{A}}_{A} \, , \mu^{L_{B}}_{A} \, ,
                              \mu^{L_{A}}_{B}, \,  \mathrm{and} \, \mu^{L_{B}}_{B} \, . 
\end{eqnarray}
%
In Eq. (\ref{eq:expansion}) $p^{ref}$ is the coexistence pressure of the liquids $L_{A}$ and $L_{B}$
at the chosen reference points; ($\Delta c^{L_{A}}_{A}$, $\Delta \eta^{L_{A}}$) and
($\Delta c^{L_{B}}_{A}$, $\Delta \eta^{L_{B}}$) are the deviations of the concentration and the
packing fraction from their reference values for the $A$-rich and the $B$-rich liquid,
respectively. The derivatives are taken at the reference points on the $A$-rich and the 
$B$-rich side, indicated as $|_{ref,A}$ and $|_{ref,B}$, respectively. 
Inserting these linear expansions (Eq. (\ref{eq:expansion})) into the
conditions for capillary coexistence, i.e., the equality of the chemical potentials in the
$A$-rich and the $B$-rich liquid of both species (the first two of Eqs. (\ref{eq:Coex_bulk})) 
and the pressure condition (Eq. (\ref{eq:Coex_cap})),
leads to three linear equations. We are mainly interested in the shift $\Delta c^{L_{A}}_{A}$
of the capillary coexistence line of the $A$-rich liquid at a given packing fraction;
in the following, the concentration and total packing fraction in the $A$-rich liquid 
are prescribed and fix the chemical potentials of both species.
Solving the system of linear equations for $\Delta \eta^{L_{A}} = 0$ 
(the total packing fraction in the $A$-rich liquid is fixed and equal to the reference
value) we find
\begin{equation}
\label{eq:shift}
\Delta c^{L_{A}}_{A} = \frac{\sigma_{AB}\cos\theta_{AB}}{w} f(T, \eta_{ref}) \quad \mathrm{or} \quad
\Delta c^{L_{A}}_{A} = \frac{\sigma_{sA} - \sigma_{sB}}{w} f(T, \eta_{ref})
 \, ;
\end{equation}
the function $f(T, \eta_{ref})$ depends only on bulk fluid properties, not on the wall properties,
and not on the geometry.
(Capillary coexistence of two liquids is discussed also in Ref. \cite{Evans3}.)

For $\theta_{AB} = 0^o$ (complete wetting), the macroscopic estimate (Eq. (\ref{eq:shift}))
for capillary coexistence is no longer based on solid ground. In this case the equilibrium
structure of a wall--$L_A$ interface involves a macroscopic film of $L_{B}$ between the 
liquid $L_{A}$ and the wall. The equilibrium surface tension of this composite interface 
is thus the sum of the wall--$L_B$ surface tension and of the $L_B$--$L_A$ surface tension
($\sigma_{sA} = \sigma_{sB} + \sigma_{AB}$, which implies 
$(\sigma_{sA} - \sigma_{sB})/\sigma_{AB} = 1$; see below).
However, in deriving Eq. (\ref{eq:shift}) two configurations are compared, one of which is a
capillary filled with the $L_{A}$ liquid. A surface tension involving a
macroscopic $L_{B}$ film is not useful for computing the surface contribution to the 
free energy of such a configuration, because having a thick $L_{B}$ film does not
leave space for a $L_{A}$ liquid in a nanoscopic capillary.
In this context one should also note that capillary coexistence, for the interaction strengths considered
here, is shifted into the
$L_{A}$ single-phase domain. For this thermodynamic condition the wall--$L_{A}$
interface structure contains only a microscopicly thin '$L_{B}$ film'.
In this case the surface contribution to the free energy is obtained by replacing
the equilibrium surface tension $\sigma_{sA}$ by the 'surface tension' 
of a configuration without a macroscopicly thick wetting film.  At bulk 
liquid--liquid coexistence this non-equilibrium 'surface tension' might
be inferred from a constrained minimization of the variational free energy functional.
With this non-equilibrium 'surface tension' $'\sigma_{sA}'$ we can define the ratio 
$('\sigma_{sA}' - \sigma_{sB})/\sigma_{AB}$, which now becomes larger than 1, 
because $'\sigma_{sA}'$ must be larger than the equilibrium value. 
The macroscopic estimate for capillary coexistence, in the case of complete wetting,
should be based on Eq. (\ref{eq:shift}), but replacing $\cos\theta_{AB} = 1$ by
$('\sigma_{sA}' - \sigma_{sB})/\sigma_{AB} > 1$.
We have refrained from carrying out the required constrained minimizations. Instead, in the case of complete wetting,
we only give a lower bound to the macroscopic estimate for the concentration $c^{L_{A}}_{A}$ 
for capillary coexistence, which follows from Eq. (\ref{eq:shift}) with $\cos\theta_{AB} = 1$.  

\begin{figure}[h]           
\vspace*{0.0cm}
\hspace{0.0cm}\includegraphics[scale = 0.3]{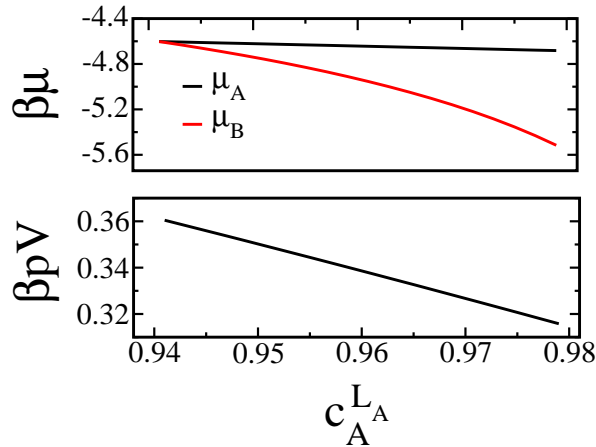}
\vspace*{-0.2cm}
\caption{ \label{pcp}Variation of the chemical potentials $\mu_{A}$ and $\mu_{B}$ of the 
             two species and of the pressure in the $A$-rich liquid $L_{A}$ along 
             path $P_1$ (see Fig. \ref{pd}); $V = R^3$ and $\beta = 1/(k_\mathrm{B}T)$. 
	     At liquid--liquid coexistence $\mu_{A}$ and $\mu_{B}$ are 
	     equal due to the $A$--$B$ symmetry }
\end{figure}

For $\theta_{AB} < 90^o$, the capillary coexistence line shifts towards higher concentrations
$c^{L_{A}}_{A}$ of $A$ particles in the $A$-rich liquid. We now consider this case and study
the force balance at the interface between the liquid $L_{A}$, which fills, say, 
the upper part of the capillary shown in Figs. \ref{schem} (a) and (b), 
and the liquid $L_{B}$, which fills the lower part. The capillary forces, 
due to the interfacial tensions, tend to drive the $A$-rich liquid out of the capillary. 
These forces can be balanced by the pressure force, provided the pressure in the 
$A$-rich liquid $L_{A}$ is higher than
it is in the $B$-rich liquid $L_{B}$ (see the pressure condition in Eq. (\ref{eq:Coex_cap})).  
Now we monitor the system if one moves in the phase diagram along path 
$P_1$ (see Fig. \ref{pd}),
starting at the bulk coexistence line and then enhancing the concentration $c^{L_{A}}_{A}$
towards capillary coexistence at $c^{L_{A}}_{A} + \Delta c^{L_{A}}_{A}$ (not shown in Fig. \ref{pd}).  
Figure \ref{pcp} shows that the pressure in the $L_{A}$ liquid decreases as one moves
towards higher $c^{L_{A}}_{A}$. However, in order to maintain chemical equilibrium
the concentration and the total packing fraction in the $L_{B}$ liquid move away from
their values at bulk coexistence such that the pressure in the $B$-rich liquid drops
even more than it does in the $A$-rich liquid. At capillary coexistence, which is
reached at the concentration $c^{L_{A},co}_{A} + \Delta c^{L_{A}}_{A}$ (see Eq. (\ref{eq:shift})), 
the pressure difference reaches the value required for a balance of pressure and
capillary forces. At higher values of $c^{L_{A}}_{A}$ the pressure in the liquid $L_{B}$
becomes too low, the position of the interface becomes unstable, and moves downwards
(see Fig. \ref{schem}), expelling the liquid $L_{B}$
from the capillary. In the case that $c^{L_{A}}_{A}$ is below its 
capillary-coexistence value $c^{L_{A},co}_{A} + \Delta c^{L_{A}}_{A}$, the position of the
interface is unstable, too, 
but now the liquid $L_{A}$ is expelled from the capillary.
  
Based on this picture we can describe the intrusion of the liquid $L_{A}$ into a pit
initially filled with the liquid $L_{B}$. 
Within a macroscopic description, the force balance at the $L_{A}$--$L_{B}$ interface
is the same for all positions of the interface between the upper opening of the
pit (Fig. \ref{schem}) and a position just above the pit floor. Therefore, on the level of a macroscopic
description it is predicted, that the liquid $L_{A}$ abruptly intrudes the
pit once $c^{L_{A}}_{A}$ is increased above the capillary-coexistence value 
$c^{L_{A},co}_{A} + \Delta c^{L_{A}}_{A}$. 
%
%
\subsection{Computational details}
In all computations discussed below, the relative strength
of the $A$--$B$ to the $A$--$A$ ($B$--$B$) interaction has been fixed to the value
$\epsilon_{AB}/\epsilon_{AA} = 0.77$.
Furthermore the temperature is fixed to $T = 0.899\epsilon_{AA}/k_{B}$.
These are the values used in Fig. \ref{pd}. 
The wall--$A$ interaction ($\epsilon_{A}$) is fixed as well, and it is
chosen such that a pure $A$ liquid, at liquid--vapor coexistence,
forms a contact angle of $\approx 157^{\circ}$ with the corresponding
planar wall surface. The wall--$B$ interaction $(\epsilon_{B})$ is varied, 
and it is quantified in terms of the contact angle $\theta_{AB}$ shown in 
Fig. \ref{tAB};
$\theta_{AB}$ is computed from Young's law (Eq. (\ref{eq:Young})).

In the following we present our results of the full DFT computations. In these studies 
we determine the conditions under which the $A$-rich ambient liquid $L_{A}$ intrudes
into the pits filled with the $B$-rich lubricant $L_{B}$. The reverse transitions
are studied as well.
The DFT computations require the numerical minimization of the 
grand canonical potential  $\Omega[\{\rho_{i}\}]$ with respect to the two number
densities. The number densities are discretized
on a simple cubic grid, and a Picard iteration scheme is used in order to 
reach the minimum of $\Omega$ and to determine the equilibrium number densities of
the two species.
The calculations have been carried out in a rectangular computational box.
A corrugated wall containing one pit with a square cross section is placed 
at the bottom of the computational box with the outward normal of the wall
pointing into the $z$-direction (see Fig. \ref{schem}). The radius of the 
wall particles has been chosen as $R_C=R_{A}/3$. This choice of the
size of the wall particles turns out to be a sound compromise
between minimizing the effect of wall roughness on interfacial
structures and the computational cost.
Periodic boundary conditions have been applied in the $x$- and $y$-
directions. At the upper end of the computational box, boundary
conditions are prescribed for the number densities
$\rho_{A}$ and $\rho_{B}$ corresponding to the specified bulk $A$-rich liquid $L_{A}$.
Due to the periodic boundary conditions in $x$- and $y$- directions,
effectively a wall with a periodic array of pits is analyzed.
However, the lateral dimensions of the box are chosen such that
the distance between the pits is sufficiently large and the walls 
separating them are sufficiently thick such that effectively
intrusion (extrusion) of the ambient liquid into (out of) 
isolated lubricant infused pits is studied. 
The size of the box in $z$-direction is also taken sufficiently large 
such that the bulk densities are attained near the upper end of the 
computational box.

In what follows, we always control the concentrations and the total packing
fraction in the $A$-rich ambient liquid $L_{A}$, moving along the paths
$P_1$ and $P_2$ indicated in Fig. \ref{pd}. The properties of the
$B$-rich liquid $L_{B}$ inside the pit adjust to the 
thermodynamic conditions prescribed by the ambient liquid.
The Cassie to Wenzel and Wenzel to Cassie transitions are studied as a
function of $\theta_{AB}$ and for various widths and depths of the pits.
In order to study hysteresis we carried out two sets of calculations
in each case. In the first set, the iterative determination of
the number densities is initialized with a distribution of number densities
similar to the Cassie state. The pit is initially filled with the homogeneous $B$-rich
lubricant $L_{B}$; the densities correspond to the bulk state,
$\eta_{L_{B}} = 0.4069093$, $c^{L_{B}}_{B} = 0.941$, and $c^{L_{B}}_{A} = 0.059$, 
marked by a cross on the blue coexistence line in Fig. \ref{pd}.
Above the pit the initial densities correspond to the ambient $A$-rich bulk liquid
$L_A$ with the proper concentrations and total packing fraction. 
In the following, calculations initialized in this way are denoted as I:C 
(initialized: Cassie). 
In the second set of calculations initial conditions are chosen which correspond
to the Wenzel state. In this case, the initial densities correspond to the 
ambient $A$-rich bulk liquid $L_A$ not only above the wall surface but also inside the pit. 
The corresponding calculations are denoted as I:W (initialized: Wenzel).
%
%
\section{Results and discussion}
In order to quantify the extent of the intrusion/extrusion transitions, 
we have calculated the average densities
of both species inside the pit volume $V_{p}$, defined as
\be
\label{eq:avd}
\overline{\rho}_{A/B}
  = \frac{1}{V_{p}}\int_{V_{p}}d^{3} r \,\rho_{A/B}{(\bf r)} \approx \frac{1}{N}\sum_{i}
             \rho_{A/B}(x_{i}, y_{i}, z_{i}),
\ee
where $i$ indicates a point on the discretization grid inside the pit 
($x_{1}\leq x_{i}\leq x_{2}$, $y_{1}\leq y_{i}\leq y_{2}$, $0\leq z_{i}\leq D$ (see Fig. $1$))
and $N$ is the total number of grid points inside the pit.
This definition takes into account also the vanishing density within the depletion zones adjacent
to the inner walls of the pits.

An alternative representation for the degree of intrusion is given by the position of 
the $L_{A}$--$L_{B}$ interface on the four-fold symmetry axis along the $z$ direction,
passing through the centers of the square cross sections of the pit. 
The location of the interface is defined as the position $z_{cr}$
where $\rho_{A}(z)$ and $\rho_{B}(z)$ {\it cr}oss, i.e., attain the same value. The distance $z$ is
measured from the bottom wall of the pit.
\subsection{Cassie--Wenzel/Wenzel--Cassie transition at fixed total packing fraction}
%
In the following the results of the full DFT computations along path $P_1$ (see Fig. \ref{pd})
are presented. There, the concentration $c^{L_{A}}_{A}$ is varied
at a fixed total packing fraction $\eta^{L_{A}} = 0.4069093$.
According to Eq. (\ref{eq:shift}) the shift of the concentration at capillary coexistence,
away from the bulk coexistence line,
is therefore computed based on the two reference points
($\eta^{L_{A}} = 0.4069093$, $c^{L_{A}}_{A} = 0.941$) [$c^{L_{A}}_{B} \, = 1 - c^{L_{A}}_{A} = 0.059$]
\hspace{0.05cm} and \hspace{0.05cm}
($\eta^{L_{B}} \,= \eta^{L_{A}} = 0.4069093$, $c^{L_{B}}_{A} = 0.059$) 
[$c^{L_{B}}_{B}$ = $1 - c^{L_{B}}_{A} = 0.941$].
These points are marked by the two crosses on the phase boundaries shown in Fig. \ref{pd}.
In order to compute the shift of the concentration at capillary coexistence
(based on Eq. (\ref{eq:shift})) via macroscopic capillary
theory, the macroscopic parameters entering into it are determined by using exactly
the same DFT as for the full DFT computations. 
The pertinent quantities follow from the equations of state for the two bulk liquids
(i.e., homogeneous densities). The interfacial tensions and 
Young's contact angle $\theta_{AB}$ are determined from DFT computations
for systems with planar interfaces.
The locations and the nature of the transitions, observed within the full DFT computations, 
are compared with the ones predicted by the macroscopic theory. The latter
predicts an abrupt intrusion at the shifted coexistence concentration.
%
\subsubsection{Dependence on the contact angle $\theta_{AB}$}
\begin{figure}[h]           
\includegraphics[scale = 0.36]{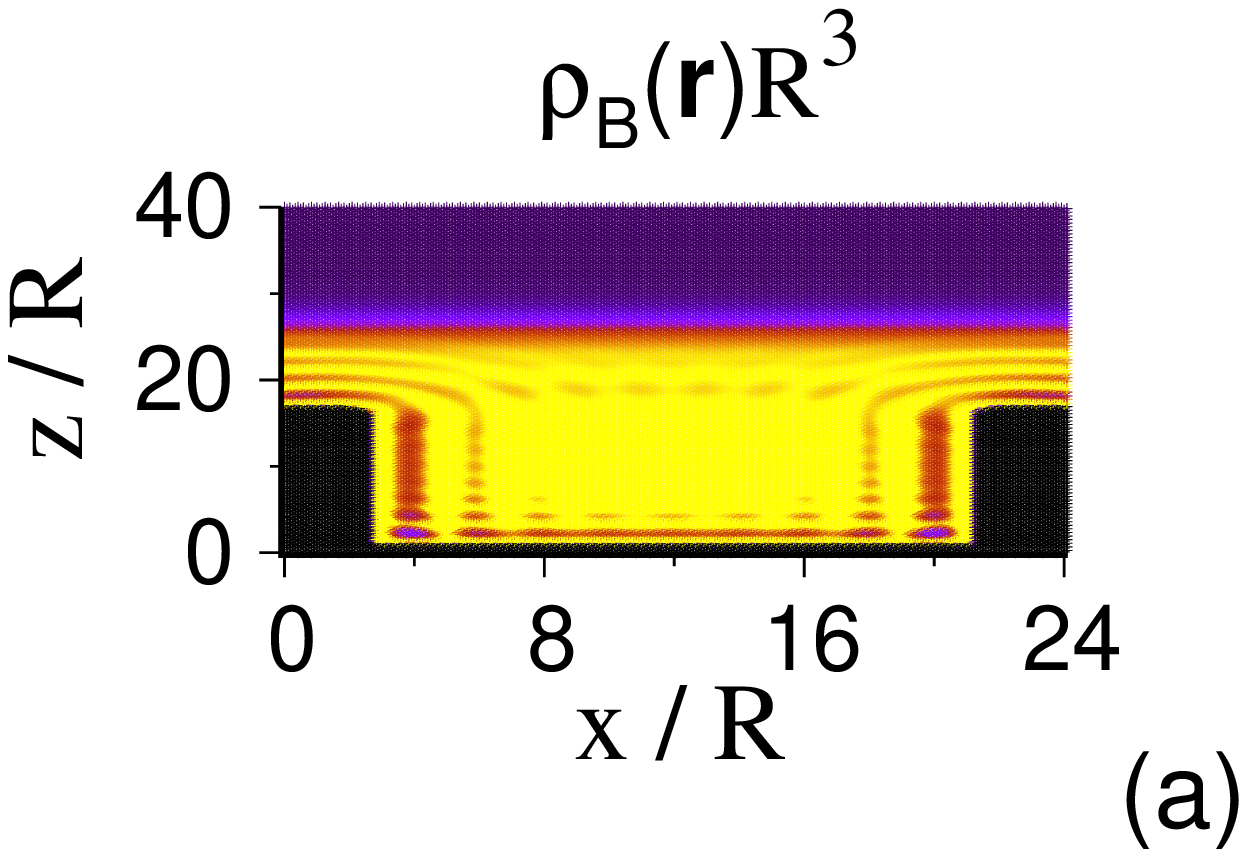}
\includegraphics[scale = 0.36]{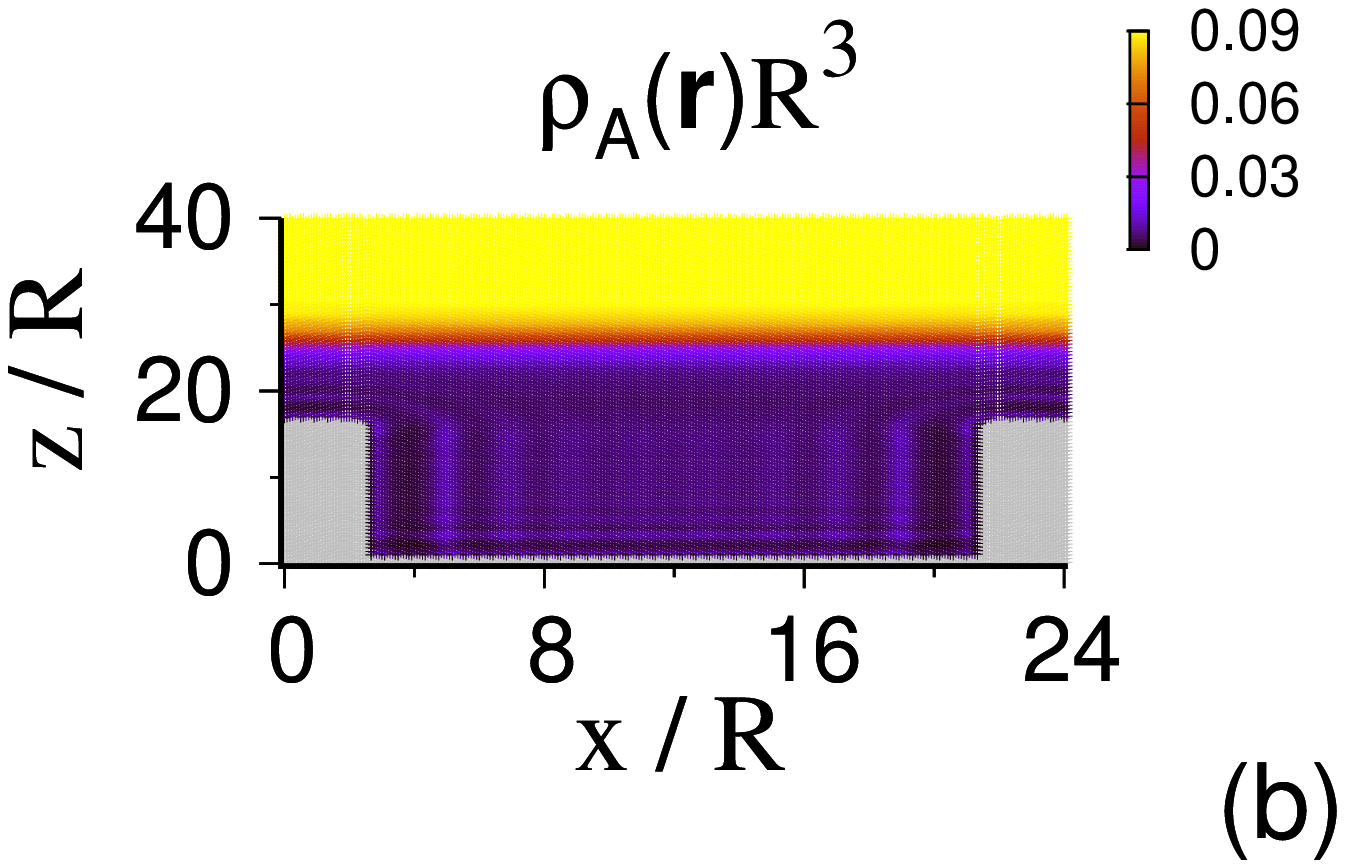}
\includegraphics[scale = 0.39]{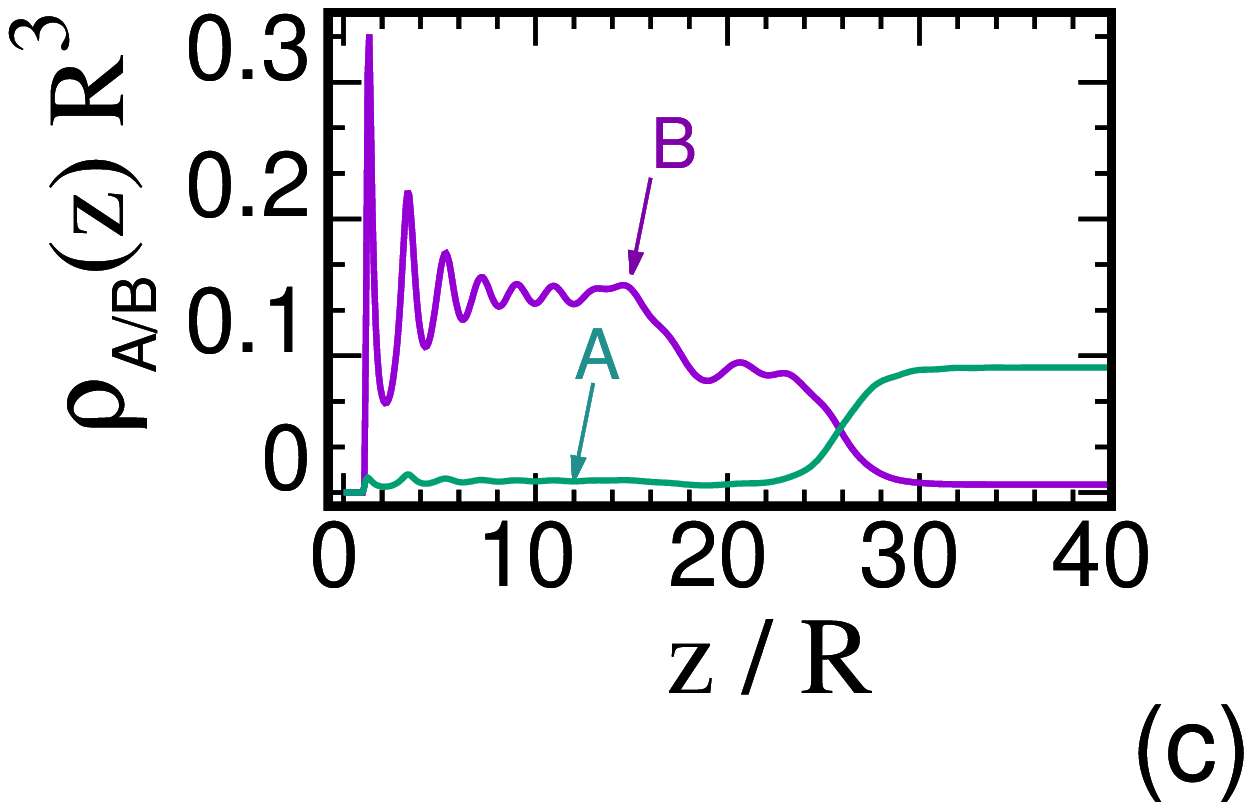}
{\vspace*{-0.0cm}
\caption{\label{fig3}  
Equilibrium number density distributions $\rho_{B}(\br)$ $(a)$ and $\rho_{A}(\br)$ $(b)$,
at liquid--liquid coexistence
($\eta^{L_{A}} = \eta^{L_{B}} = 0.4069093$, $c^{L_{A}}_{A} = c^{L_{B}}_{B} \approx 0.941$),
in the $xz$ plane passing through the middle of the pit. The color bar applies to
both (a) and (b). In (b) the regions of zero densities inside the wall and in the
depletion zones are shown in gray, in order to enhance the contrast.
The ratio of the wall--$A$ and the wall--$B$ interaction strengths is 
$\epsilon_{B}/\epsilon_{A} = 4$, which corresponds to complete wetting
of the wall by the $B$-rich liquid $L_{B}$. Above the wall the $A$-rich liquid
$L_{A}$ is found, as revealed by the high number density $\rho_{A}(\br)$ and the low
number density $\rho_{B}(\br)$ for large $z$. 
For the complete wetting scenario shown here, in reality the $L_{B}$ wetting layer
above the wall is macroscopically thick.
The depth of the pit is $16R = 8\sigma$, and its width is $20R = 10\sigma$.
Panel $(c)$ shows the densities along the symmetry axis parallel to $z$, 
through the centers of the square cross sections. The interface position is
chosen to be located at the crossing $z_{cr}$ of $\rho_{A}(z)$ and $\rho_{B}(z)$.
The remaining parameters in this and all following figures are 
$\epsilon_{AA}/(k_{B}T) = 1.112$ ($\epsilon_{BB} = \epsilon_{AA}$),
$\epsilon_{AB} = 0.77 \epsilon_{AA}$, and $\epsilon_{A}$ is chosen such that
for a pure $A$ type liquid (given the above value of $\epsilon_{AA}/(k_{B}T)$)
the contact angle formed by its liquid--vapor interface
is $\theta_{Y} \approx 157^{\circ}$. 
                                           }}
\end{figure}

\begin{figure}   
\hspace*{-0.4cm}\fbox{\includegraphics[scale = 0.28]{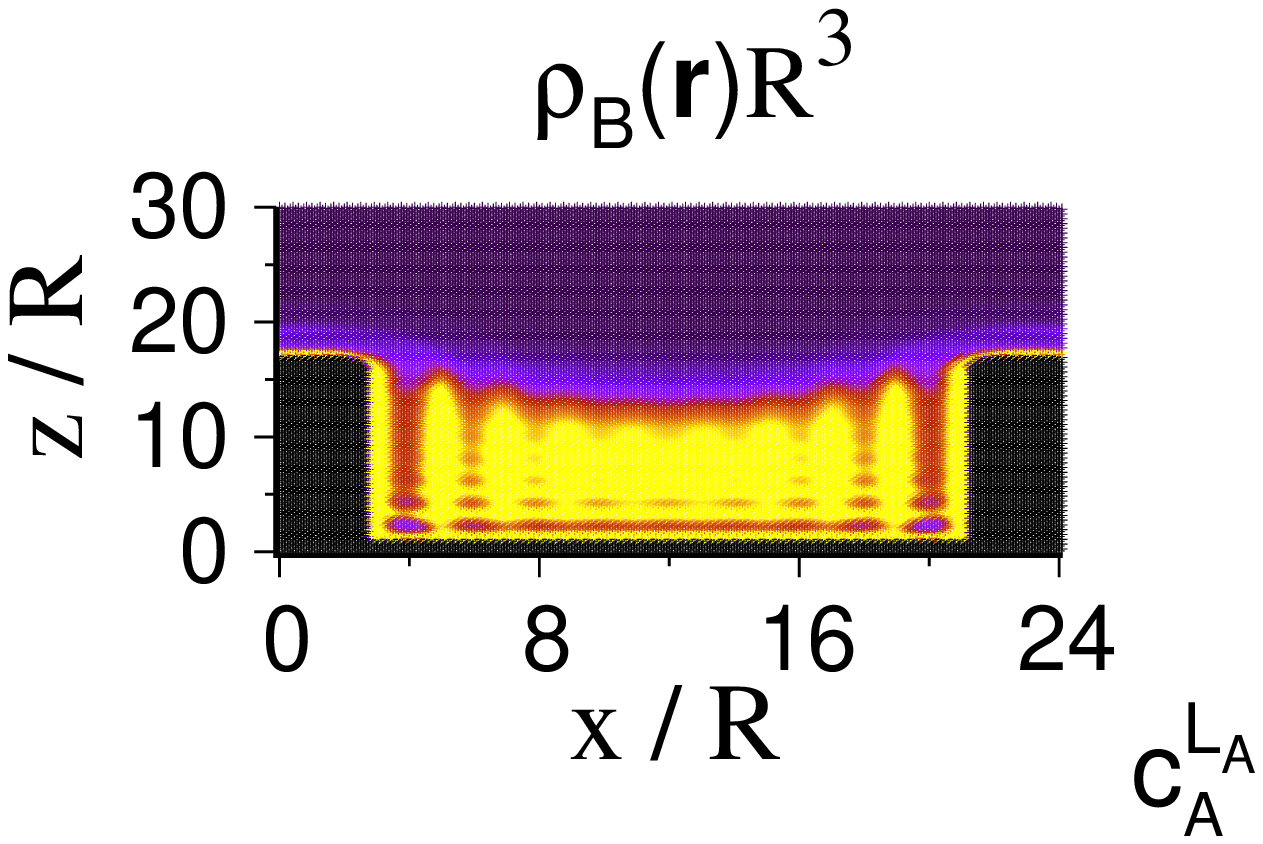}
\hspace*{ 0.40cm}\includegraphics[scale = 0.28]{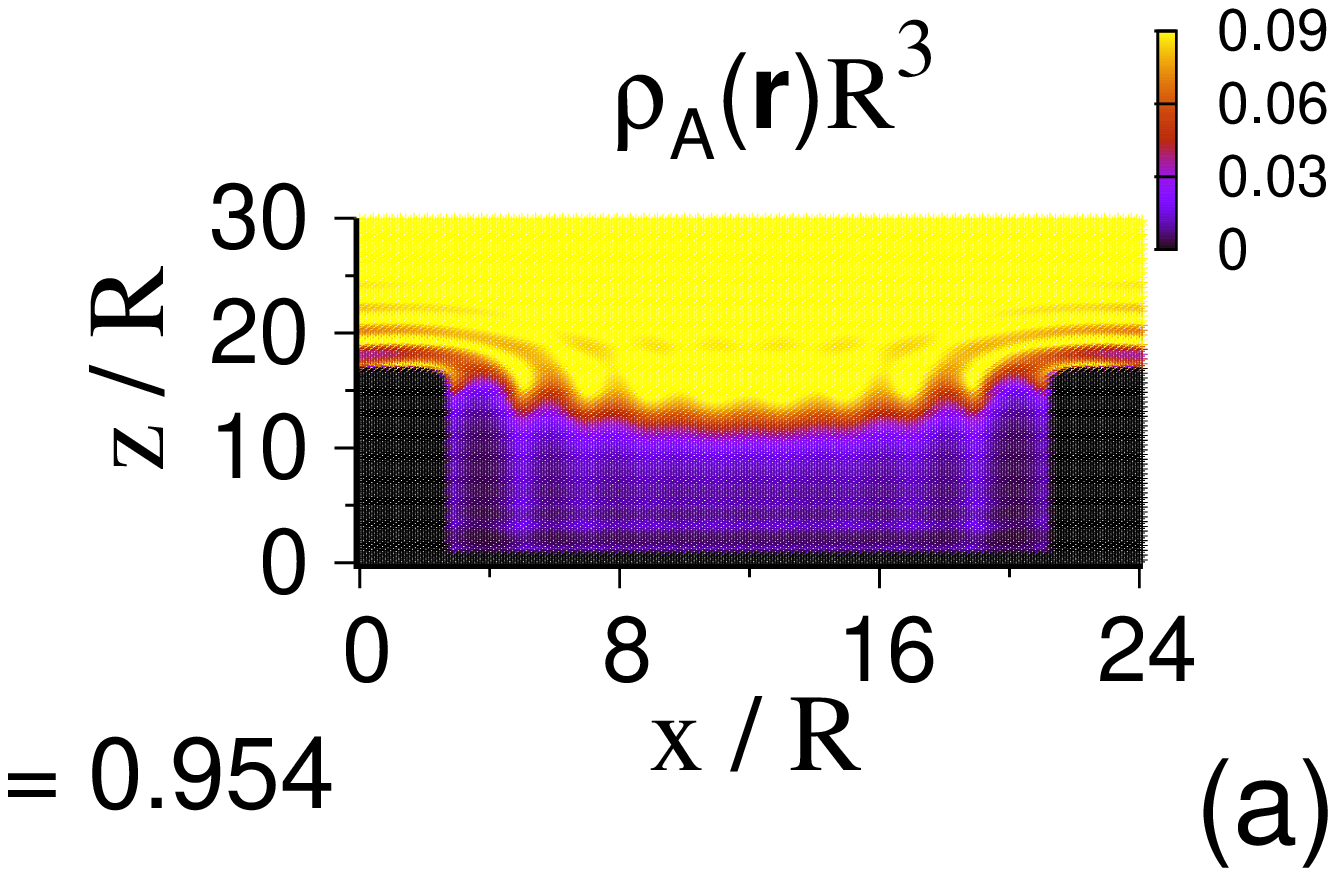}}
\hspace*{0.6cm}\fbox{\includegraphics[scale = 0.28]{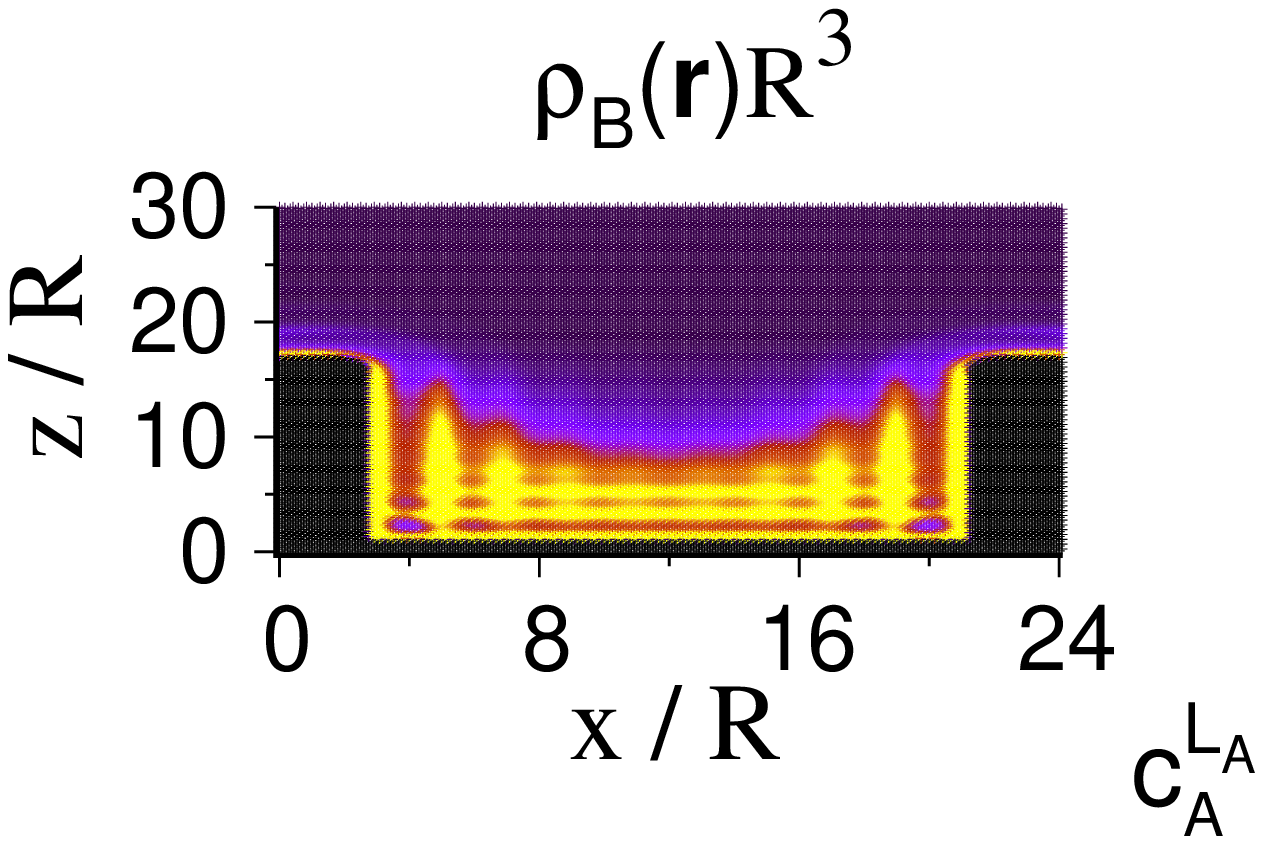}
\hspace*{ 0.40cm}\includegraphics[scale = 0.28]{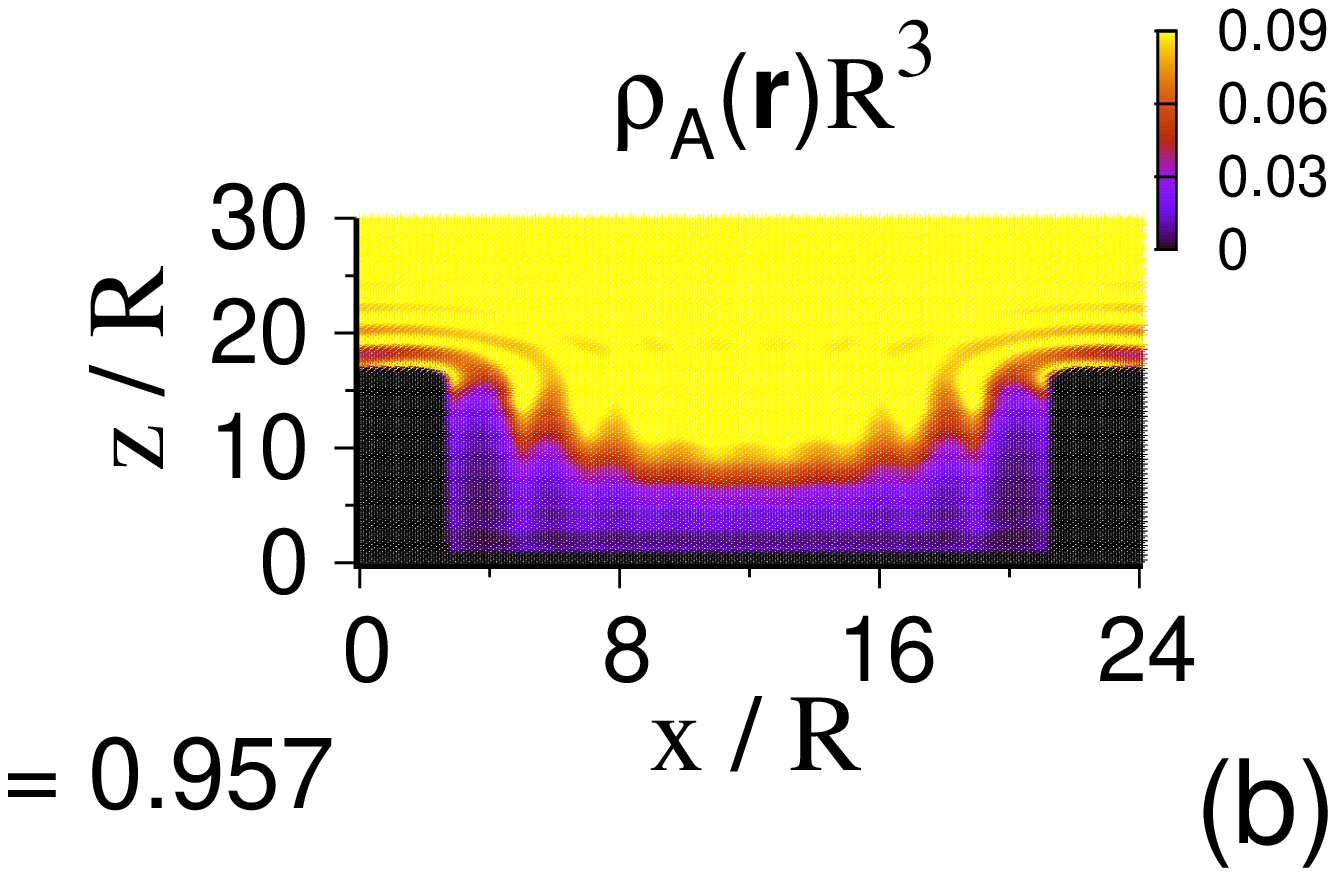}}

\vspace*{0.5cm}

\hspace*{-0.4cm}\fbox{\includegraphics[scale = 0.28]{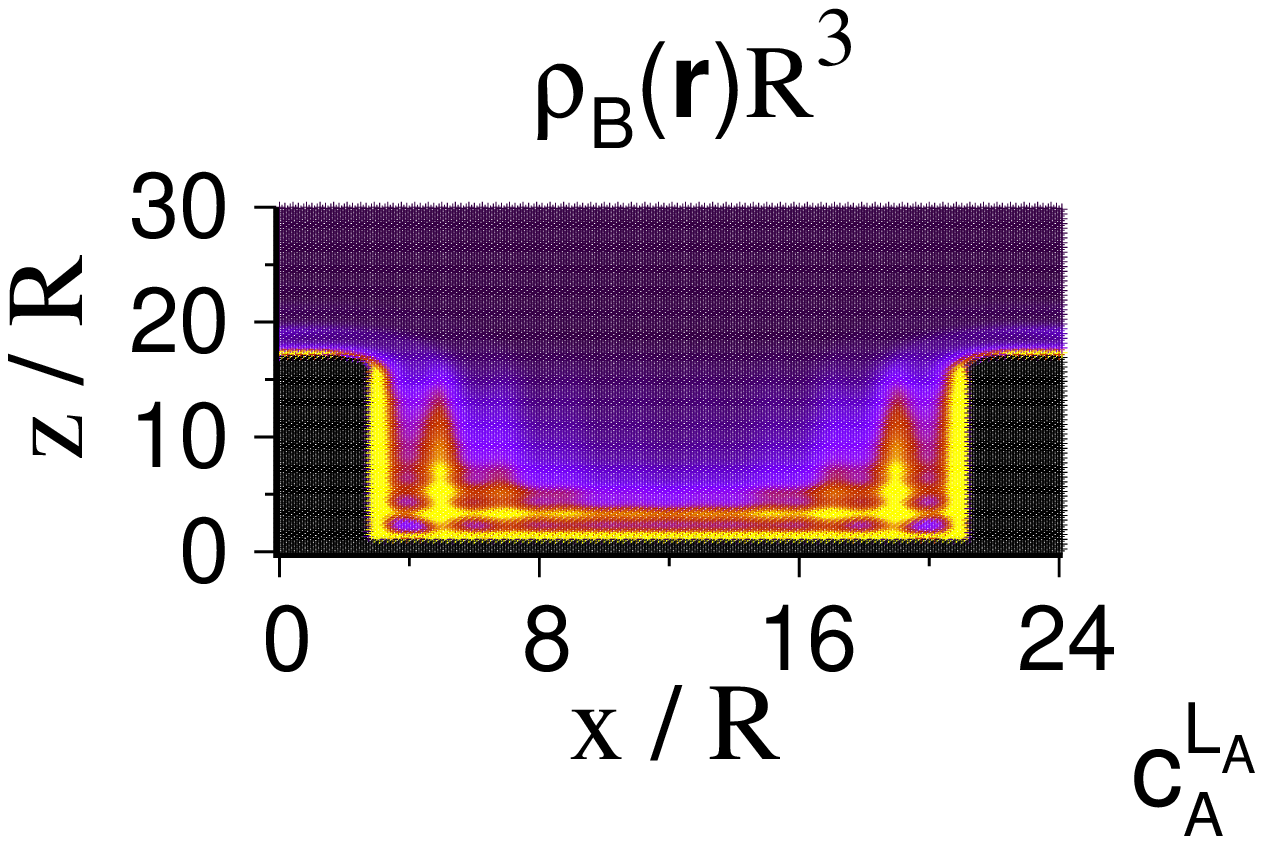}
\hspace*{ 0.40cm}\includegraphics[scale = 0.28]{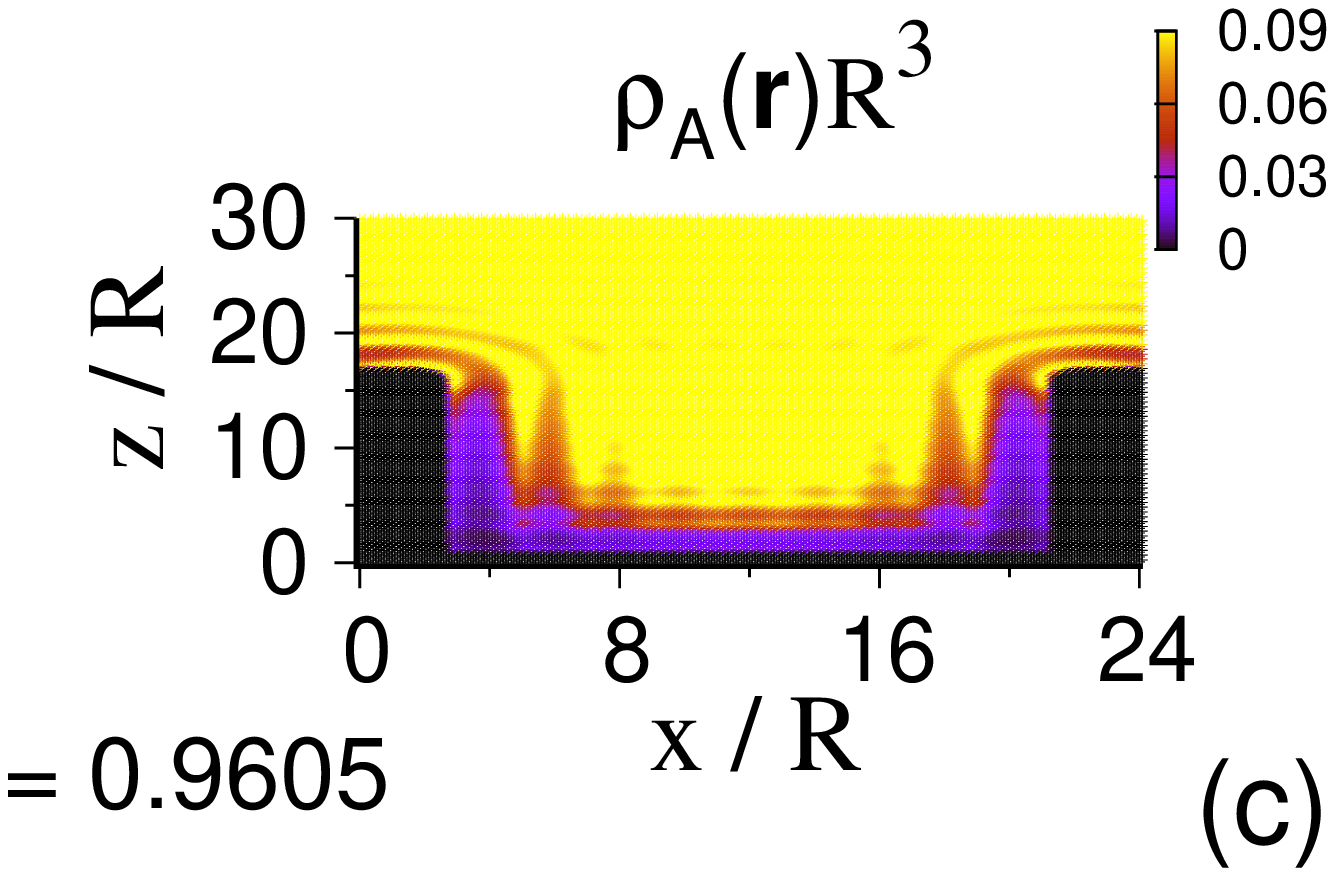}}
\hspace*{0.6cm}\fbox{\includegraphics[scale = 0.28]{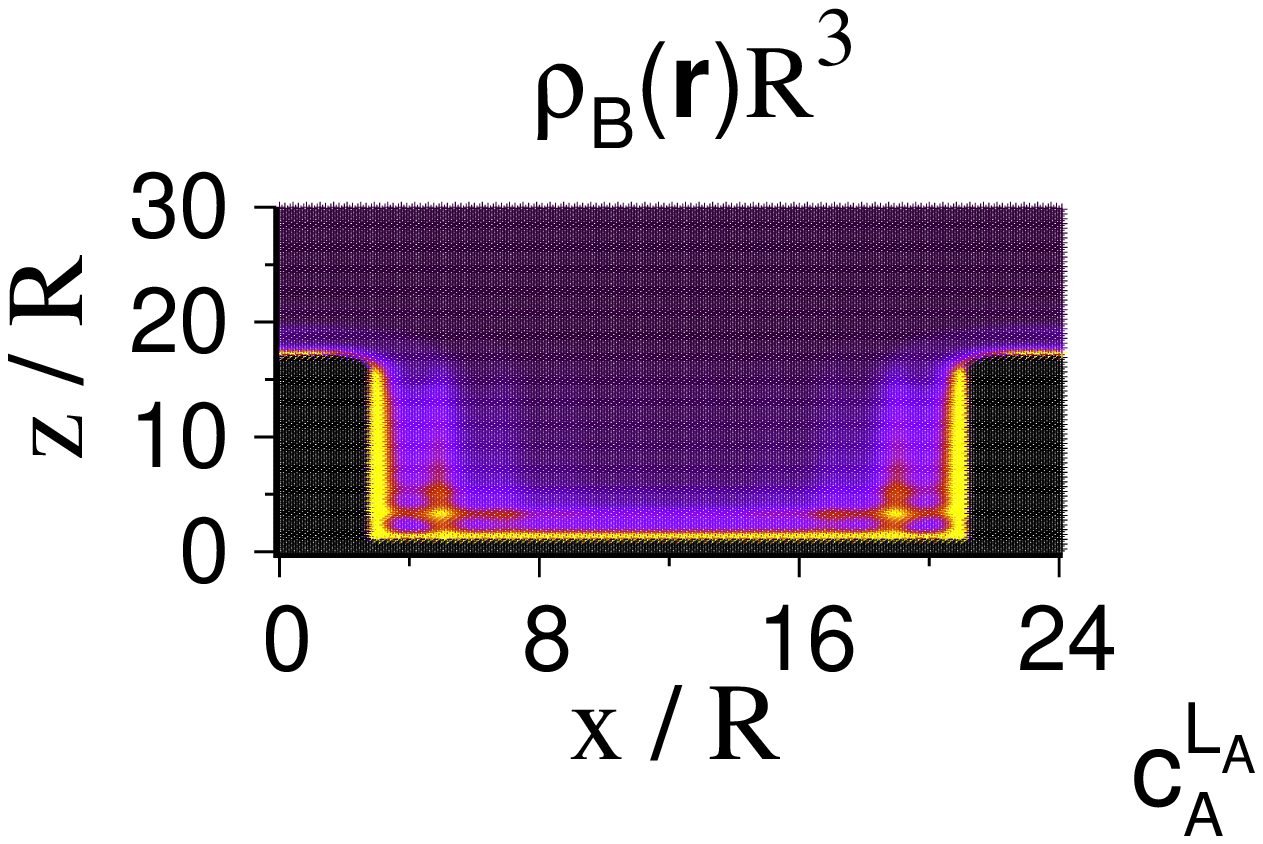}
\hspace*{ 0.40cm}\includegraphics[scale = 0.28]{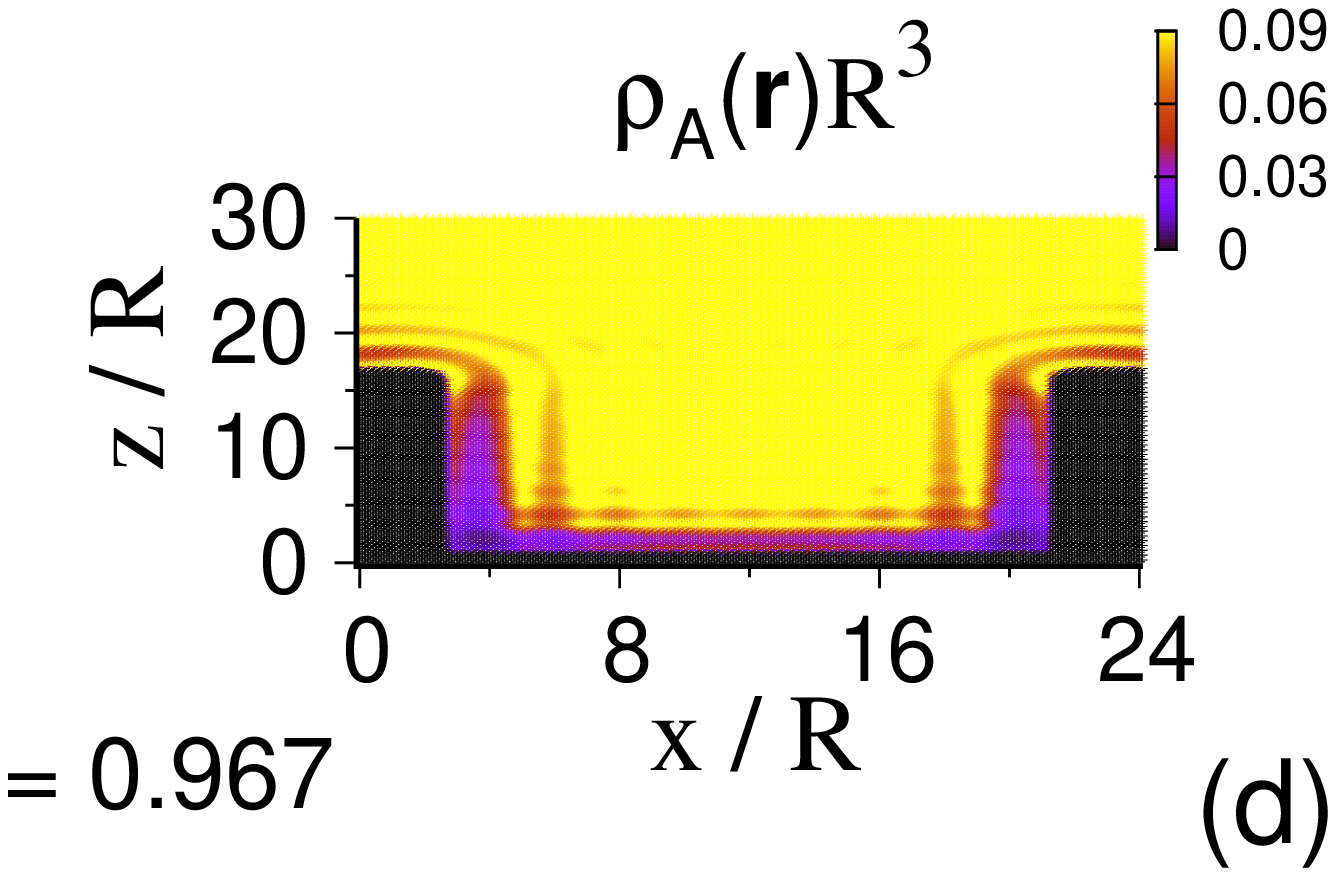}}
\caption{\label{fig4}  \baselineskip=2\baselineskip 
	Evolution of the system, which is shown in Fig. \ref{fig3} at liquid--liquid coexistence,
as the concentration of $A$ particles in the ambient liquid $L_{A}$ is increased at
constant total packing fraction along path $P_1$ in Fig. \ref{pd}.
Here, the equilibrium number densities are independent of whether
the numerical iterations are initialized in a Cassie type or in a Wenzel type
configuration, as anticipated.   
                            }
\end{figure}

\begin{figure}
\hspace*{-2.2cm}\includegraphics[scale = 0.42]{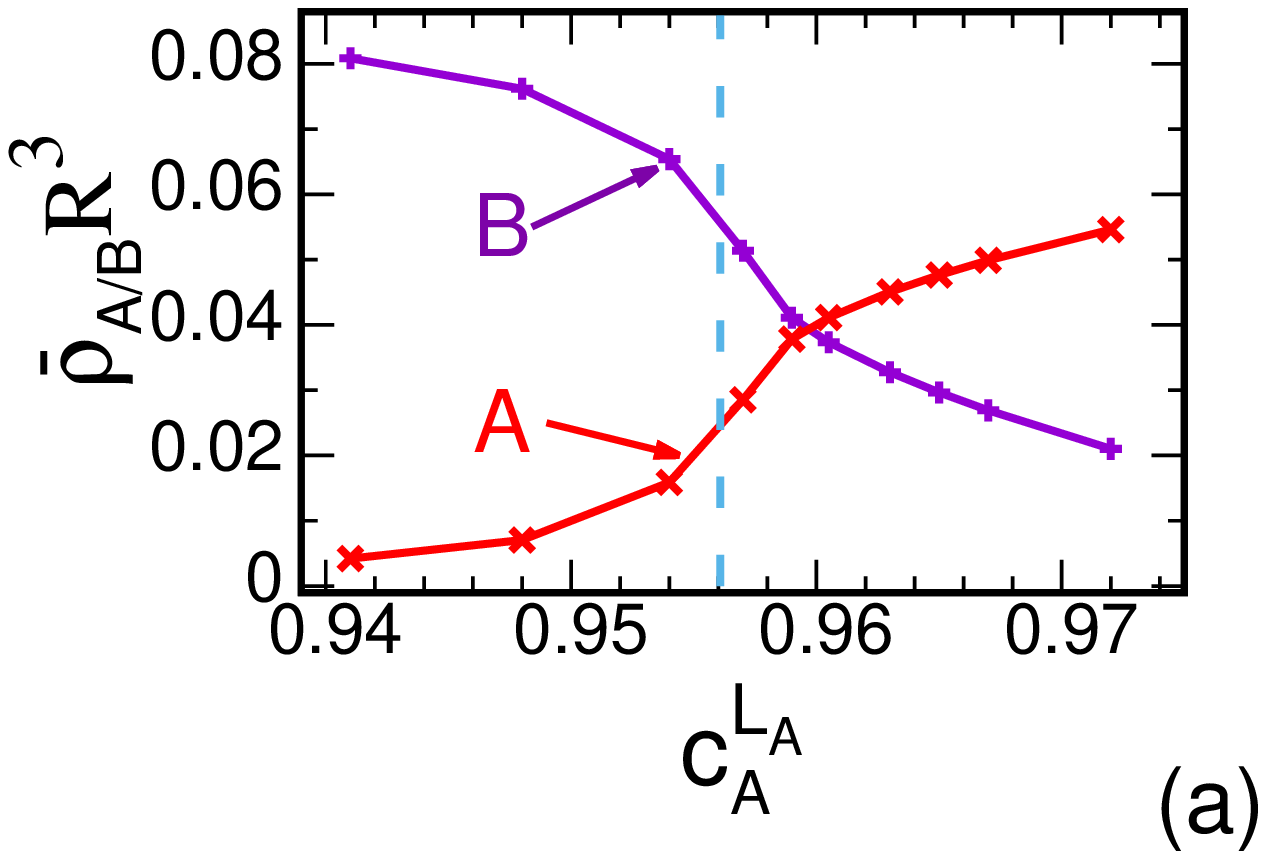}
{\hspace*{1.20cm}\includegraphics[scale = 0.42]{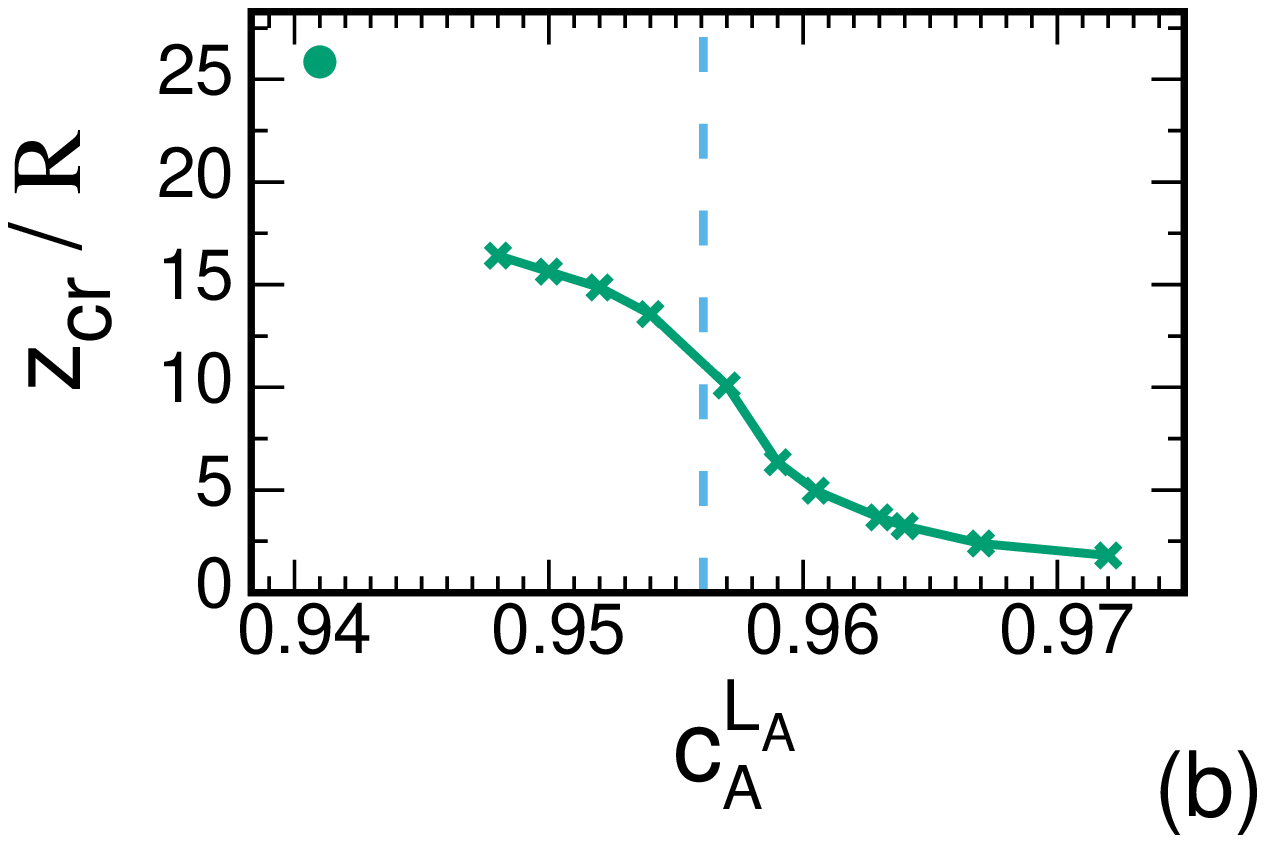}}
\caption{\label{fig5}  \baselineskip=2\baselineskip 
Averaged number densities $\bar\rho_{A/B}$ of the $A$ and the $B$ particles 
inside the pit (Eq. (\ref{eq:avd}), (a)) and the position $z_{cr}$ 
of the $L_{A}$--$L_{B}$ interface (panel (b)) as the concentration $c^{L_{A}}_{A}$
of $A$ particles in the ambient liquid is varied along path $P_1$ in Fig. \ref{pd}. The system parameters
are the same as the ones for Fig. \ref{fig3}. Due to complete wetting
at bulk liquid--liquid coexistence, one observes at the coexistence concentration 
$c^{L_{A}}_{A} \approx 0.941$ 
a large value of $z_{cr}$, representing an interface at a macroscopic distance above the wall
(marked by a dot in panel (b)).
At the next higher value of $c^{L_{A}}_{A}$ which has been investigated, $z_{cr}$ jumps to a much smaller value,
which corresponds to an interface position at the pit entrance. 
As $c^{L_{A}}_{A}$ is increased further,   
the interface gradually moves down towards the bottom of the pit (panel (b)).
There is a gradual transition of the averaged number densities $\bar\rho_{A/B}$
of the $A$ and the $B$ particles inside
the pit (panel (a)). No hysteresis has been observed. The number density profiles shown 
in Figs. \ref{fig3} and \ref{fig4} are in line with the transition shown here. 
The dashed vertical line indicates the lower bound to the macroscopic prediction for capillary coexistence 
of $L_{A}$ and $L_{B}$, for which an abrupt intrusion of $L_{A}$ into the pit occurs
(Eq. (\ref{eq:shift}) with $\cos\theta_{AB} = 1$).    
}
\end{figure}

We have fixed the pit dimensions at $w = 10\sigma$ and $D = 8\sigma$.
The ratio $\epsilon_{B}/\epsilon_{A}$ (Eq. (\ref{eq:pre7})) is varied in order to change the contact
angle $\theta_{AB}$, which is computed by using Young's law
$\cos\theta_{AB} = (\sigma_{sA} - \sigma_{sB})/\sigma_{AB}$. 
First, we choose $\epsilon_{B}/\epsilon_{A} = 4$, which
corresponds to a scenario in which, at $L_{A}$--$L_{B}$ coexistence in the bulk
(i.e., for $c^{L_{A}}_{A} = 0.941$), the $B$-rich liquid wets a planar wall completely.
In Fig. \ref{fig3} we present the
corresponding number densities of the $A$ and $B$ species. 
A thick wetting layer of the $B$-rich liquid $L_{B}$ is observed above the wall. 
Of course, the configuration shown cannot correspond to a fully converged iteration;
the equilibrium thickness of the wetting layer should be macroscopicly thick at coexistence in the bulk.
However, increasing the thickness of the wetting layer beyond the one shown, lowers
the free energy barely. 
Because of the very slow convergence of the iteration with respect to the one
particular, collective coordinate, which describes the increase of the wetting layer thickness,
the densities shown in the figure virtually are the
equilibrium densities for a constrained position of the $L_{B}$--$L_{A}$ interface
(i.e., a constrained thickness of the wetting layer). 
The pit is filled with the $B$-rich liquid $L_{B}$. Increasing the
concentration $c^{L_{A}}_{A}$ of $A$ particles in the ambient liquid $L_{A}$
to values away from coexistence (see the cyan line in Fig. \ref{pd}, 
moving along path $P_1$ in Fig. \ref{pd}),
a gradual intrusion of the ambient liquid $L_{A}$ into the pit is
observed. Typical equilibrium number density distributions are shown in
Fig. \ref{fig4}. One observes that once the concentration of $A$ particles
is increased to $c^{L_{A}}_{A} = 0.967$ (Fig. \ref{fig4} (d)) 
the lubricant $L_{B}$ is replaced inside the pit completely by the ambient liquid $L_{A}$. 
However, because the wall
attracts the $B$ particles much stronger than the $A$ particles, a
thin $B$ rich layer remains at the surfaces of the walls 
(see Fig. \ref{fig4}).
For this choice of parameters there is no hysteresis; the 
equilibrium number densities are found to be independent of whether 
the numerical iterations are initialized in a Cassie type or in a Wenzel type
configuration. The average number densities of the $A$ and the $B$ particles inside
the pit as well as the position $z_{cr}$ of the interface as a function
of $c^{L_{A}}_{A}$ are shown in Fig. \ref{fig5}. These curves
show a gradual decrease of the average number density of $B$ type particles
(and an increase of the average number density of $A$ type particles) inside the pit
and an interface gradually moving downwards; the strong increase of
$z_{cr}$ (panel (b) in Fig. \ref{fig5}) at liquid--liquid coexistence
is due to the formation of the wetting layer with $L_{B}$. 
For comparison, in Fig. \ref{fig5} the lower bound to the sharp transition, predicted
by the macroscopic theory at a concentration corresponding to capillary
coexistence (see Eq. (\ref{eq:shift}) with $\cos\theta_{AB} = 1$) is indicated as a dashed vertical
line in the figure. 

\vspace*{0.3cm}

We carried out the same type of computation as described above, by decreasing the ratio 
$\epsilon_{B}/\epsilon_{A}$ step by step.
In this way we change from a regime, in which the $B$-rich liquid wets the wall completely,
towards and into the regime of partial wetting, in the case of which an $L_{A}$--$L_{B}$ interface
meets the planar wall at a nonzero contact angle $\theta_{AB}$.  
Down to a ratio $\epsilon_{B}/\epsilon_{A} = 3.5$ 
we do not observe a pronounced jump in the averaged number densities $\bar\rho_{A/B}$ or
in the position $z_{cr}$ of the $L_{A}$--$L_{B}$ interface as the concentration $c^{L_{A}}_{A}$ is
varied in steps of 0.001. 
The transition
seems to be continuous although it becomes steeper and the interval, within which the steep 
changes occur, becomes 
narrower as the ratio $\epsilon_{B}/\epsilon_{A}$ becomes smaller. Down to 
$\epsilon_{B}/\epsilon_{A} = 3.2$, corresponding to $\theta_{AB} = 12^{\circ}$,
no hysteresis is observed, i.e., irrespective of whether the calculations are initialized
in a Cassie type configuration (I:C) or in a Wenzel type configuration (I:W),
the same equilibrium number densities are obtained. 
A small hysteresis emerges first, if $\epsilon_{B}/\epsilon_{A}$ is
reduced to the value 3.1. Starting in the Cassie configuration, the $L_{A}$
liquid intrudes the pit for $0.953 < c^{L_{A}}_{A} < 0.954$ 
(the interval is defined by the steps in which the concentration is varied),
whereas starting in the Wenzel configuration the Cassie configuration
is restored for $0.952 < c^{L_{A}}_{A} < 0.953$,
i.e., one has to move closer to the liquid--liquid coexistence
line by the amount $\Delta c^{L_{A}}_{A} = 0.001$ in order to recover
the Cassie state in which a pit is filled with the liquid $L_{B}$.
A number of computations have been carried out for several smaller ratios  
$\epsilon_{B}/\epsilon_{A}$; the smallest ratio considered has been
$\epsilon_{B}/\epsilon_{A} = 2.0$, corresponding to a contact angle
$\theta_{AB} = 59^{\circ}$.
In general, the hysteresis increases with decreasing
$\epsilon_{B}/\epsilon_{A}$, whereas the value of $c^{L_{A}}_{A}$, 
at which intrusion of the ambient liquid occurs, decreases, i.e., the transition from
the Cassie to the Wenzel state occurs closer to the
bulk liquid--liquid coexistence line. Qualitatively, the shift of the intrusion concentration
as observed within the full DFT computations agrees with the shift of the capillary coexistence
as predicted by the macroscopic theory. 
The numbers, characterizing quantitatively the findings from the full DFT computations,
are tabulated in Table I together with the macroscopic predictions
for the intrusion concentrations.
A more detailed discussion of the DFT results is given below for three selected
contact angles $\theta_{AB}$. 

\begin{figure}[h]           
\hspace{-0.4cm}\fbox{\includegraphics[scale = 0.28]{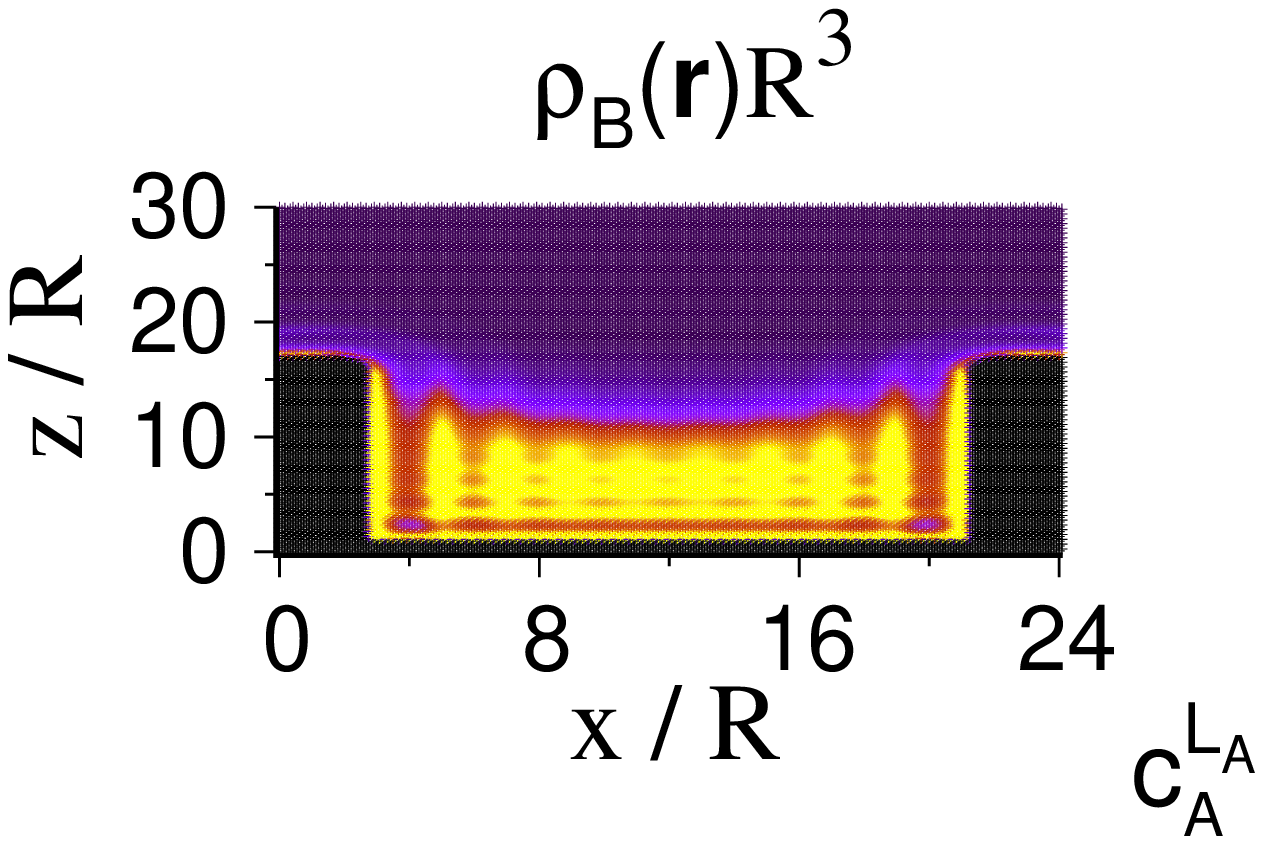}
\vspace*{0.00cm}\hspace*{0.40cm}\includegraphics[scale = 0.28]{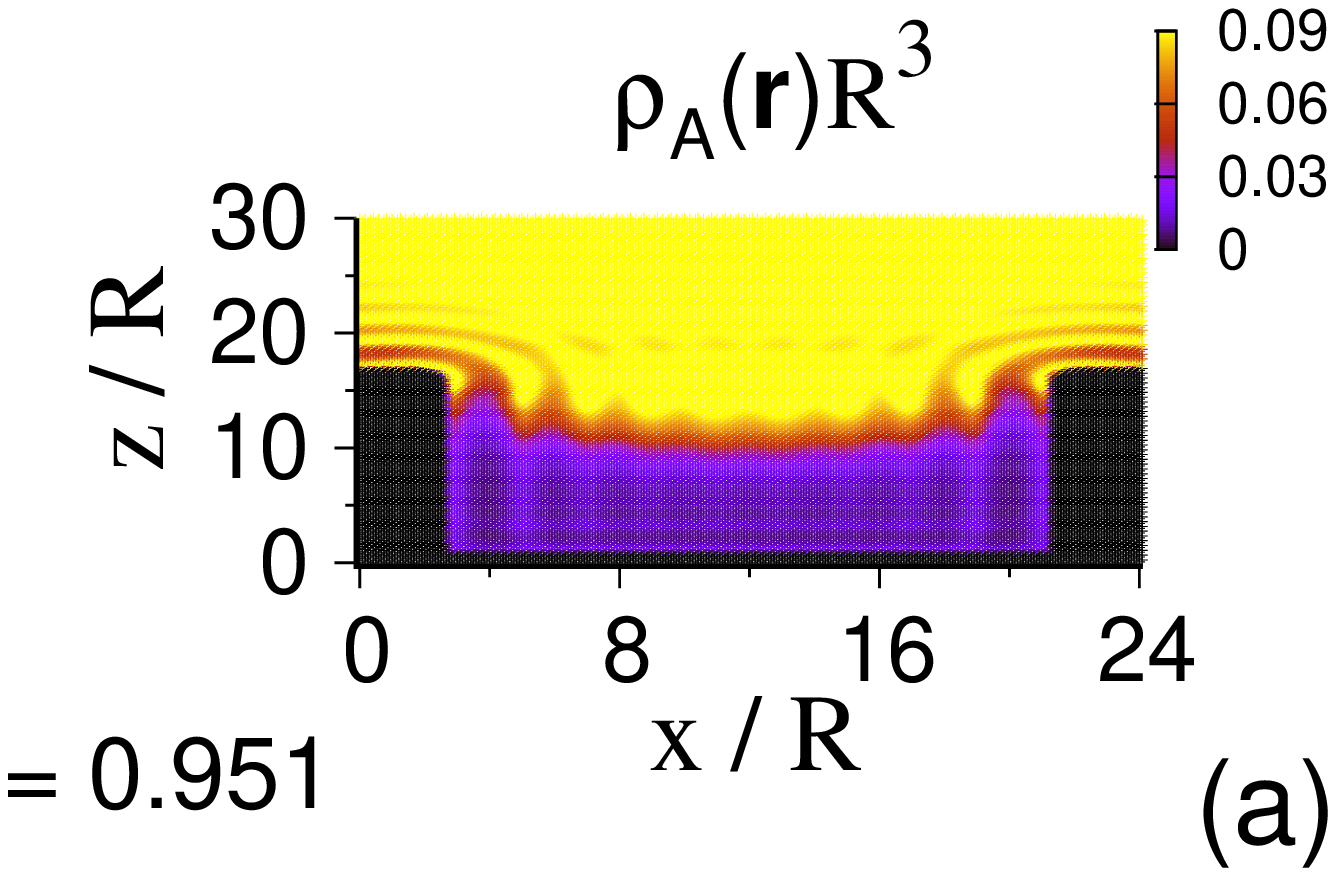}}
\hspace{0.60cm}\fbox{\includegraphics[scale = 0.28]{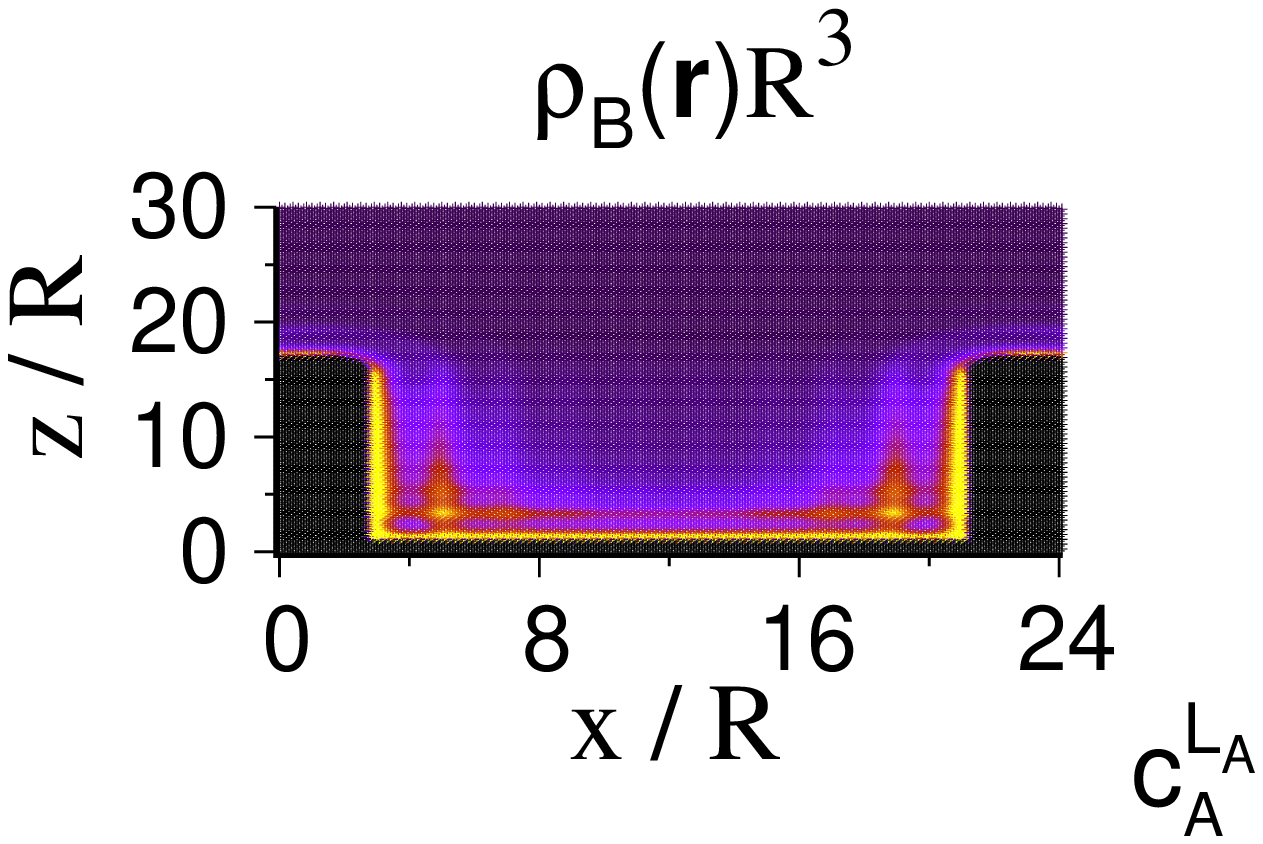}
\vspace*{0.00cm}\hspace*{0.40cm}\includegraphics[scale = 0.28]{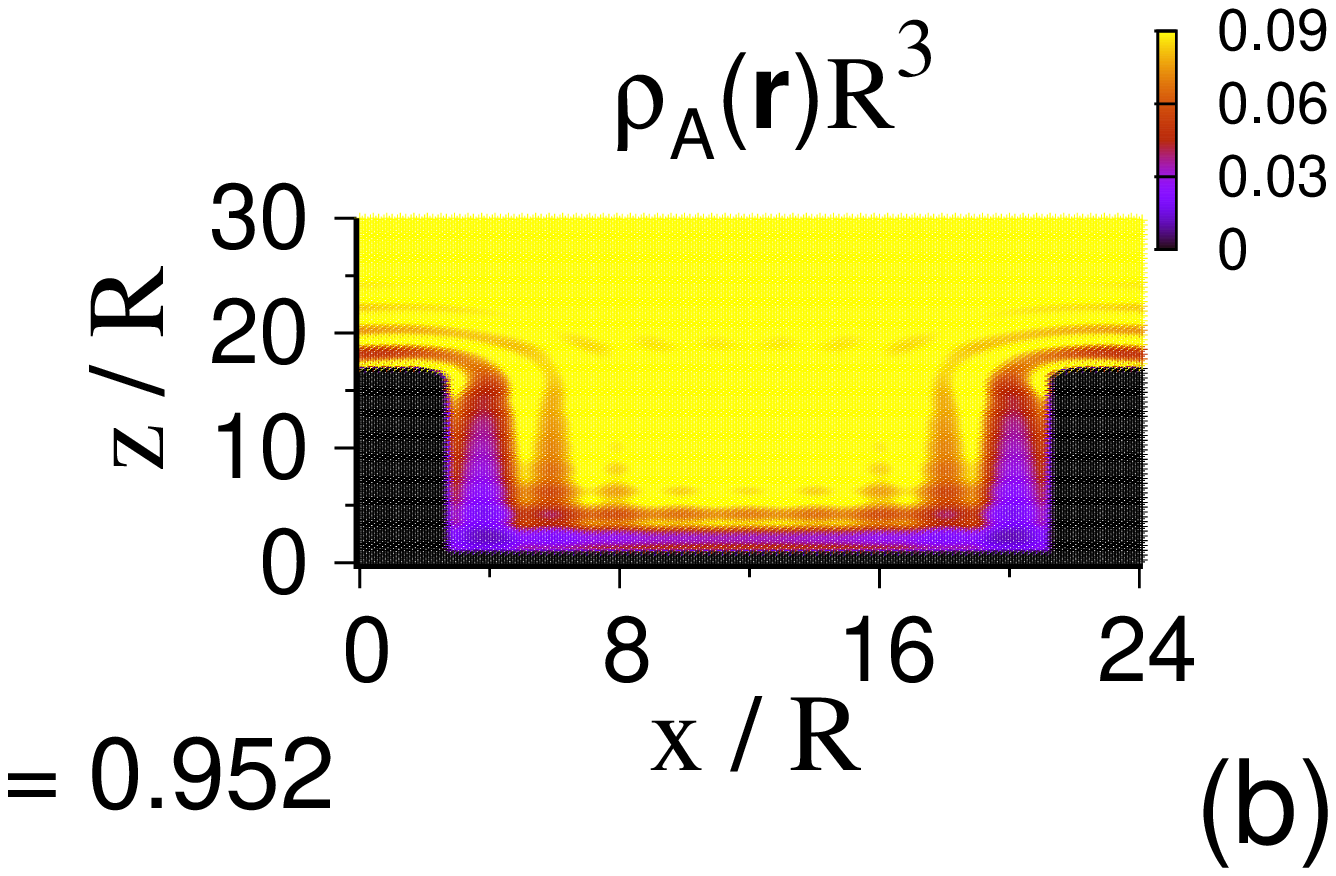}}
\caption{\label{i81030-1} \baselineskip=2\baselineskip 
Equilibrium number densities $\rho_{B}(\br)$  and $\rho_{A}(\br)$ 
in the $xz$ plane passing through the middle of the pit for a ratio of the 
wall--$A$ to the wall--$B$ interaction strengths which corresponds to a contact angle
$\theta_{AB} \approx 30^{\circ}$. 
Here, the numerical iterations are initialized in a Cassie state, i.e., the pits are initially filled
with the lubricant $L_{B}$.
The total packing fraction in the ambient liquid is fixed to $\eta^{L_{A}} = 0.4069093$,
i.e., its value along path $P_1$ in Fig. \ref{pd}.
In panel (a) the concentration of $A$ particles in the ambient liquid $L_{A}$ is
$c^{L_{A}}_{A} \approx 0.951 $, whereas in panel (b) one has $c^{L_{A}}_{A} \approx 0.952 $.
Between these two concentrations an abrupt intrusion by the ambient liquid $L_{A}$ is observed.
The width of the pit is $20R = 10\sigma$ and the depth is $16R = 8\sigma$.
The remaining parameters are the same as in all other figures.
                          }
\end{figure}

The first example which we discuss corresponds to 
$\epsilon_{B}/\epsilon_{A} = 2.8$ (i.e., $\theta_{AB} \approx 30^\circ$).
If the iterative determination of the equilibrium number densities is initialized
in the Cassie state, and one then increases the concentration 
$c^{L_{A}}_{A}$ starting from its value at the bulk coexistence line step by step,
an abrupt intrusion, i.e., an abrupt transition to the Wenzel state, is observed between
$c^{L_{A}}_{A} \approx 0.951$ and $c^{L_{A}}_{A} \approx 0.952$ (see Fig. \ref{i81030-1}).
Instead, initializing the iteration in the Wenzel state and following the results if 
$c^{L_{A}}_{A}$ is lowered, one finds that the system remains in the Wenzel state
down to $c^{L_{A}}_{A} = 0.950$. The system jumps to the Cassie state once the concentration
of the $A$ particles
in the ambient liquid is reduced a tiny bit further to $c^{L_{A}}_{A} = 0.949$.
This means, that the Wenzel state becomes unstable in the interval 
$0.949 < c^{L_{A}}_{A} < 0.950$,
i.e. there is a small hysteresis.

\begin{figure}
\hspace{-1.8cm}\includegraphics[scale = 0.39]{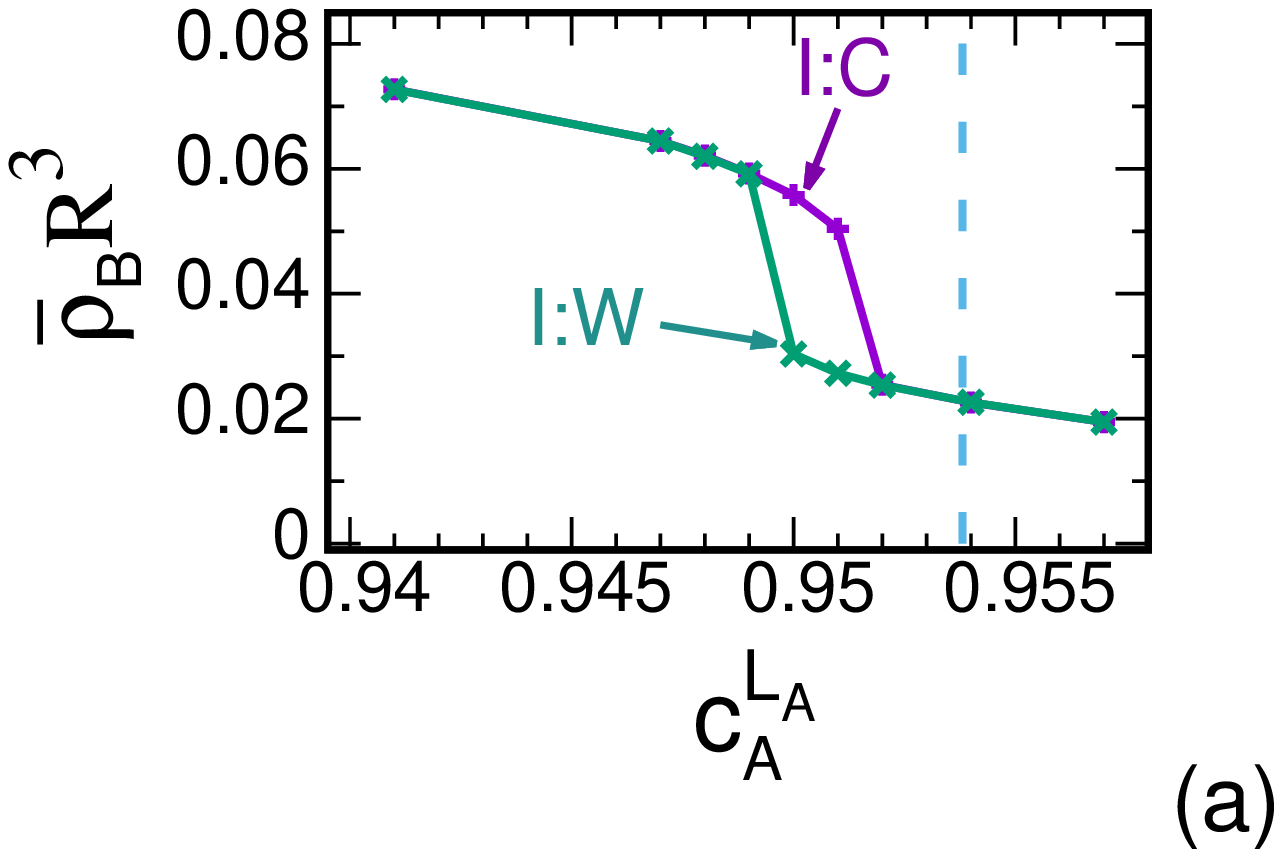}
\hspace{0.4cm}\includegraphics[scale = 0.39]{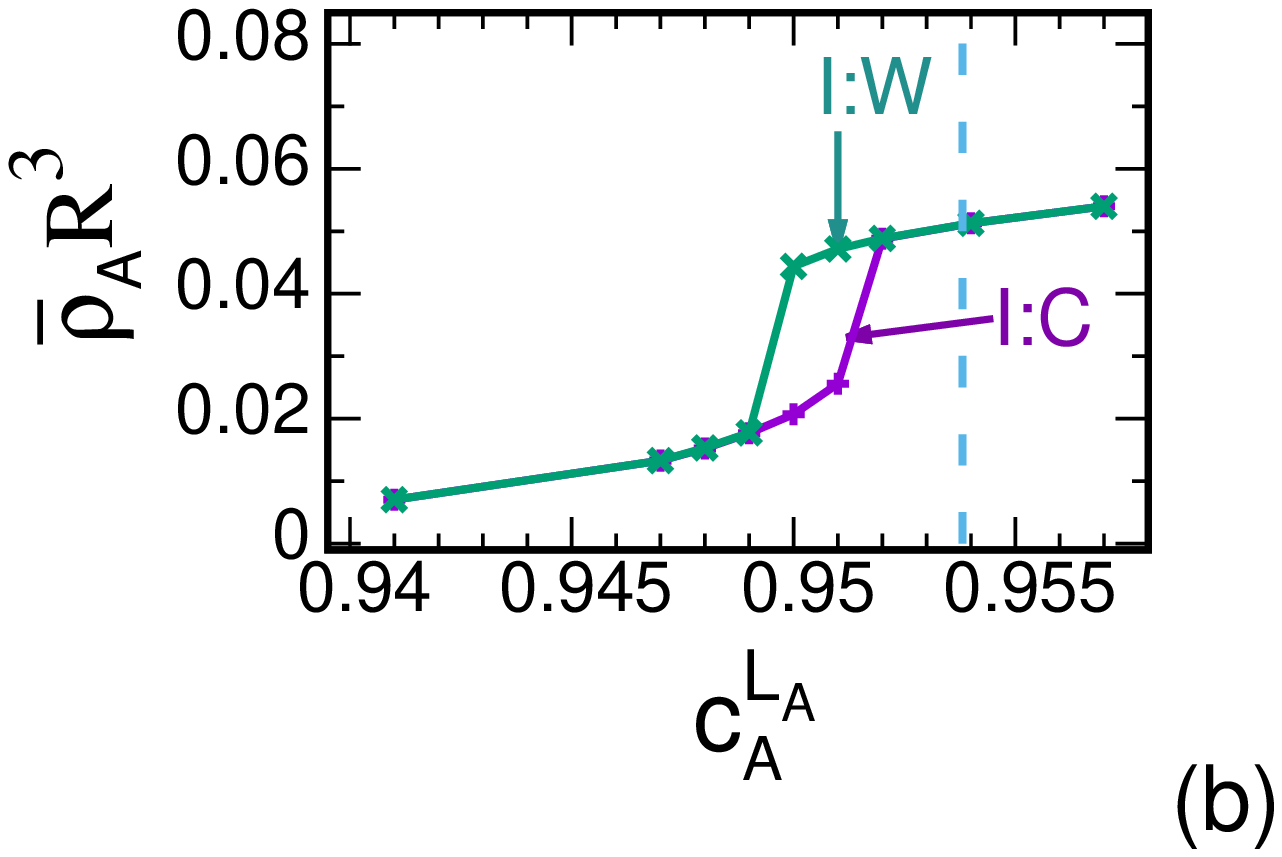}
{\vspace*{0.00cm}\hspace*{0.40cm}\includegraphics[scale = 0.39]{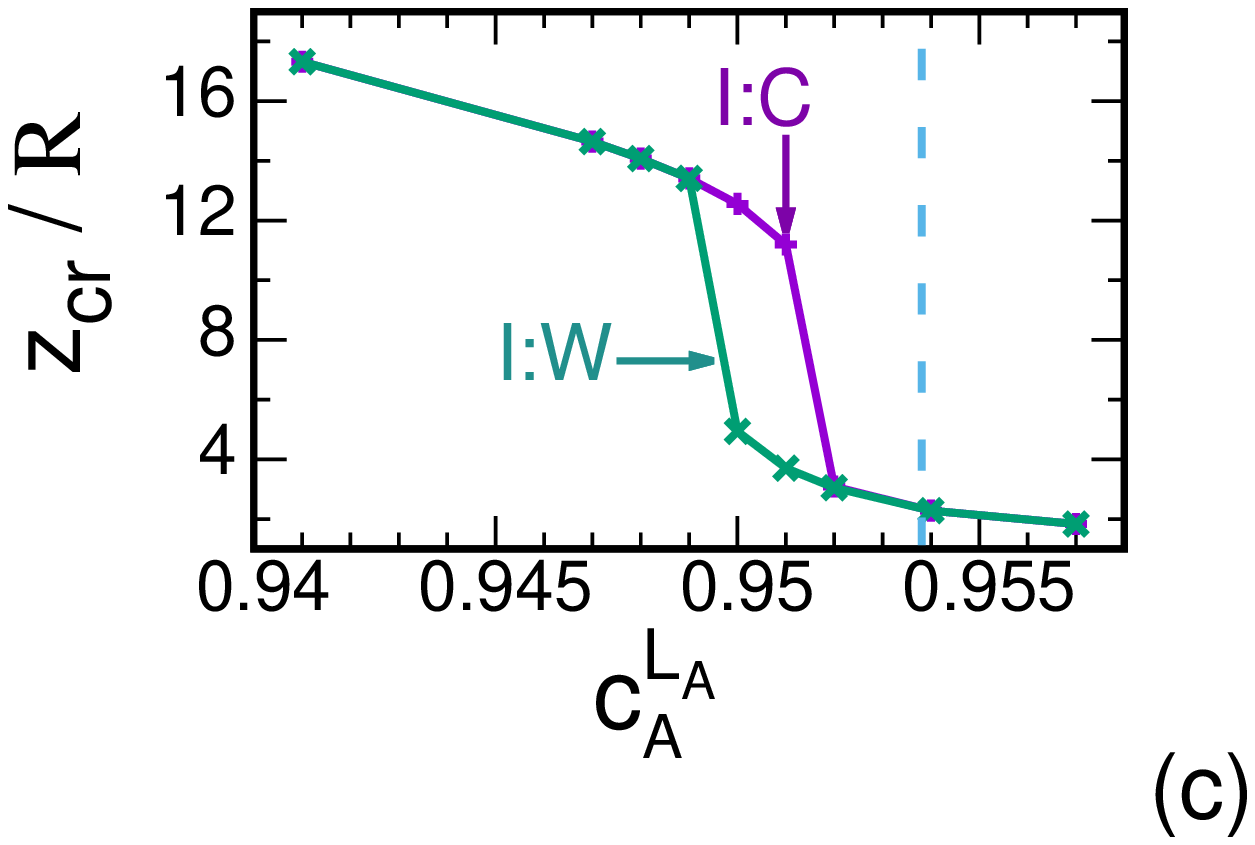}}
{\vspace*{0.0cm}
\caption{\label{fig7}  \baselineskip=2\baselineskip
Averaged number densities of the $B$ particles (panel (a)) and of the $A$ particles (panel (b)) 
inside the pit, and the position $z_{cr}$ of the $L_{A}$--$L_{B}$ interface (panel (c)) 
as the concentration $c^{L_{A}}_A$ of the
$A$ particles in the ambient liquid is varied along path $P_1$ in Fig. \ref{pd}.
The results depend on whether the iterations are initialized in a Cassie type state (I:C)
or in a Wenzel type state (I:W).
The ratio of the wall--$A$ to the wall--$B$ interaction strengths correspond to a contact angle
$\theta_{AB} \approx 30^{\circ}$. 
The width of the pit is $20R = 10\sigma$ and the depth is $16R = 8\sigma$.
The remaining parameters are the same as in all other figures.
The position of the Cassie to Wenzel transition, as expected from 
the macroscopic theory,
is indicated by the dashed line.
                                                         }}
\end{figure}

In Fig. \ref{fig7} we show the average number densities of the $A$ and the 
$B$ particles inside the pit
as well as the position $z_{cr}$ of the interface between the $A$ and the $B$ rich liquids
as a function of $c^{L_{A}}_{A}$ for two different initializations of the iterations,
i.e., in the Cassie state (I:C) or in the Wenzel state (I:W). Within a certain interval
of the concentration $c^{L_{A}}_{A}$, the results do depend on the initial condition.
(It turns out that whether in the computations $c^{L_{A}}_{A}$ is increased or decreased 
is of no consequence; only the initial condition matters.)
The system will go through a full hysteresis cycle, if it is initialized, e.g., in
the Cassie state (i.e., the pit is filled with the lubricant $L_{B}$) and if the $A$ concentration
$c^{L_{A}}_{A}$ in the ambient liquid is increased until the ambient liquid intrudes.
In the above process the curves marked as I:C represent the state of the system.
For decreasing values of $c^{L_{A}}_{A}$, the state of the system is represented by the curves
marked as I:W.
The two sets of curves merge at values of $c^{L_{A}}_{A}$ which are larger than the value at which intrusion
occurs, which is the limit of stability of the Cassie state, and for 
values of $c^{L_{A}}_{A}$ which are smaller than
the value at which the Wenzel state becomes unstable.

\begin{figure}  
\hspace{-1.8cm}\includegraphics[scale = 0.42]{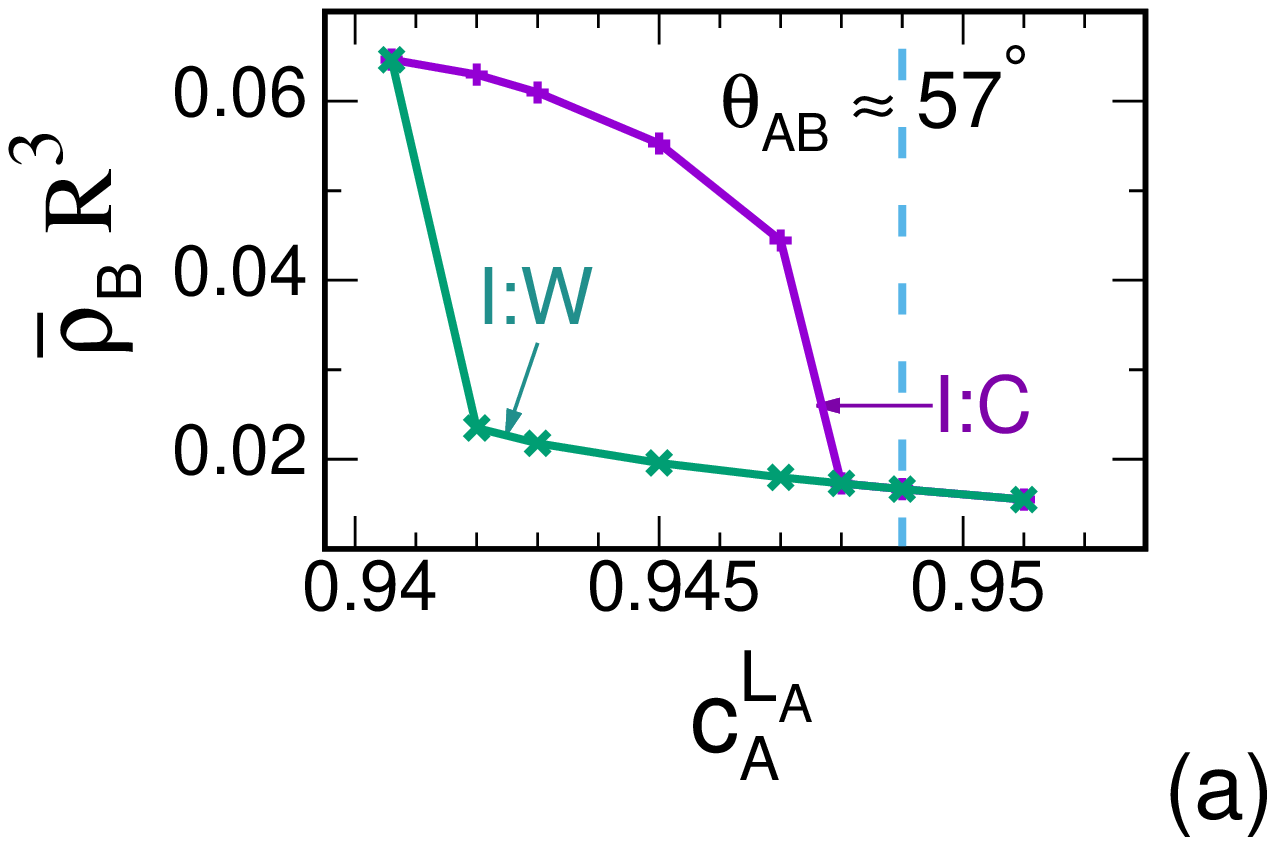}
\hspace{0.4cm}\includegraphics[scale = 0.42]{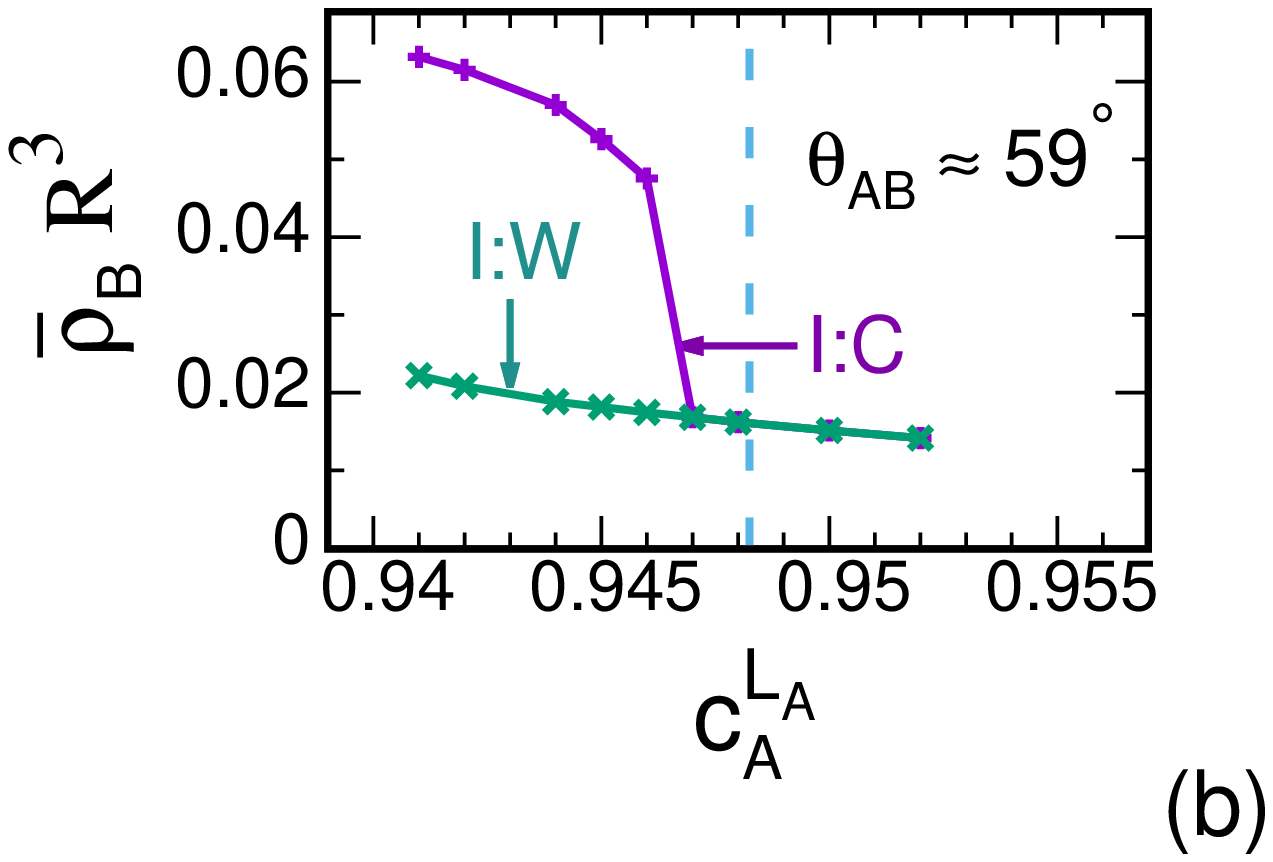}

{\vspace*{0.0cm}
\caption{\label{fig9}  \baselineskip=2\baselineskip
Averaged number densities of the $B$ particles inside the pit as a function of $c^{L_{A}}_{A}$ 
along path $P_1$ in Fig. \ref{pd}.
The results depend on whether the numerical iterations are initialized in a Cassie type state (I:C)
or in a Wenzel type state (I:W).
The ratio of the wall--$A$ to the wall--$B$ interaction strength corresponds to contact angles
$\theta_{AB} \approx 57^{\circ}$ (panel (a)) and $\theta_{AB} \approx 59^{\circ}$ (panel (b)).
The width of the pit is $w = 20R = 10\sigma$ and the depth is $D = 16R = 8\sigma$.
The remaining parameters are the same as in all other figures.
The position of the Cassie to Wenzel transition, as expected from the
macroscopic theory, is indicated by the dashed line.
                                                                                      }}
\end{figure}

In Fig. \ref{fig9} we show hysteresis loops for $\epsilon_{B}/\epsilon_{A} = 2.1$
$(\theta_{AB} \approx 57^\circ)$ and $\epsilon_{B}/\epsilon_{A} = 2.0$
$(\theta_{AB} \approx 59^\circ)$, in terms of the averaged number densities 
of the $B$ particles only.
For $\theta_{AB} \approx 57^\circ$ a wide but still closed hysteresis loop
is observed. However, for $(\theta_{AB} \approx 59^\circ)$ the loop
does not close for values of $c^{L_{A}}_{A}$ above those of bulk coexistence. 
In order to obtain a closed hysteresis loop also for $(\theta_{AB} = 59^\circ)$ one would
have to lower $c^{L_{A}}_{A}$ to values for which $L_{A}$ is, however, only metastable 
in the bulk. 
The widening of the hysteresis loops, which is observed as $\theta_{AB}$ increases,
and the observation of an open hysteresis loop for $\theta_{AB} \approx 59^\circ$,
can be understood in terms of a mechanism related to corner wetting by the lubricant liquid
$L_{B}$. In order to move from the Wenzel to the Cassie state, an 
$L_{A}$--$L_{B}$ interface spanning the pit must be created, which then can be raised
up to the pit entrance. The spanning interface can be created by forming 
'drops' of the lubricant liquid $L_{B}$ in the four corners at the bottom of the pit.
If these corner drops can grow until they merge, the spanning interface is
created. According to macroscopic capillary theory, for a contact angle 
$\theta_{AB} = \arccos(1/\sqrt3) \approx 54.7^\circ$ and at bulk $L_{A}$--$L_{B}$ coexistence, 
the shift of a planar $L_{A}$--$L_{B}$ interface formed at a corner is possible
without changing the free energy.   
For larger contact angles $\theta_{AB}$ 
a growth of the corner drops is not possible without surmounting a free energy
barrier at bulk liquid--liquid coexistence and in the thermodynamic region 
of a stable liquid $L_{A}$. For contact angles $\theta_{AB} < 54.7^\circ$ the barrierless
growth of the corner drops is always possible at bulk liquid--liquid coexistence. 
At concentrations $c^{L_{A}}_{A}$ somewhat above the coexistence value
these drops can grow, without barrier, up to the size for which they merge.
According to the macroscopic theory,
the concentration up to which this mechanism works,  
increases with decreasing contact angle $\theta_{AB}$ and also with a decreasing 
width of the pit.
In our microscopic calculations the transition from a closed hysteresis loop
to an open hysteresis loop is observed in the interval 
$57^\circ < \theta_{AB} < 59^\circ$, i.e., it is shifted to a contact angle 
somewhat above the value of $54.7^\circ$ predicted by the macroscopic theory. 
The mechanism discussed above is very closely related to the one discussed
in detail in Ref. \cite{Alberto2}, for the standard Wenzel to Cassie transition
in a liquid--vapor system.  

The conditions for intrusion of the ambient liquid into the pit filled by the lubricant, 
as determined from the microscopic theory, qualitatively agree with
the macroscopic predictions for capillary coexistence.
At capillary coexistence, the $L_{A}$--$L_{B}$ interface can be moved 'up' and 'down' 
along the capillary (i.e., along the 'vertical' walls of the pit) 
without changing the free energy (neglecting gravity). 
This condition is equivalent to a force balance between the force resulting from the 
pressure difference between the two liquids and the net capillary force
due to the difference between the interfacial tensions $\sigma_{sA}$
and $\sigma_{sB}$.
Within a macroscopic description the same force balance applies
also at the pit entrance through which the liquid intrudes; the liquid--liquid 
interface is already present. The microscopic computations show that the intrusion transition occurs
somewhat closer to bulk coexistence than predicted by the macroscopic theory.
We would like to emphasize again, that although the pressure of the ambient liquid $L_{A}$
decreases as one moves away from coexistence along path $P_1$ (see Fig. \ref{pcp}),
en route, at a certain value of the $L_A$ pressure, the liquid $L_A$
intrudes, because the pressure in the liquid $L_{B}$ 
decreases even more so, as a result of changes in the composition and density of
the liquid $L_{B}$, which are required in order to maintain chemical equilibrium.


\subsubsection{Effect of the pit dimensions}
\begin{figure}  
\hspace{-1.8cm}\includegraphics[scale = 0.42]{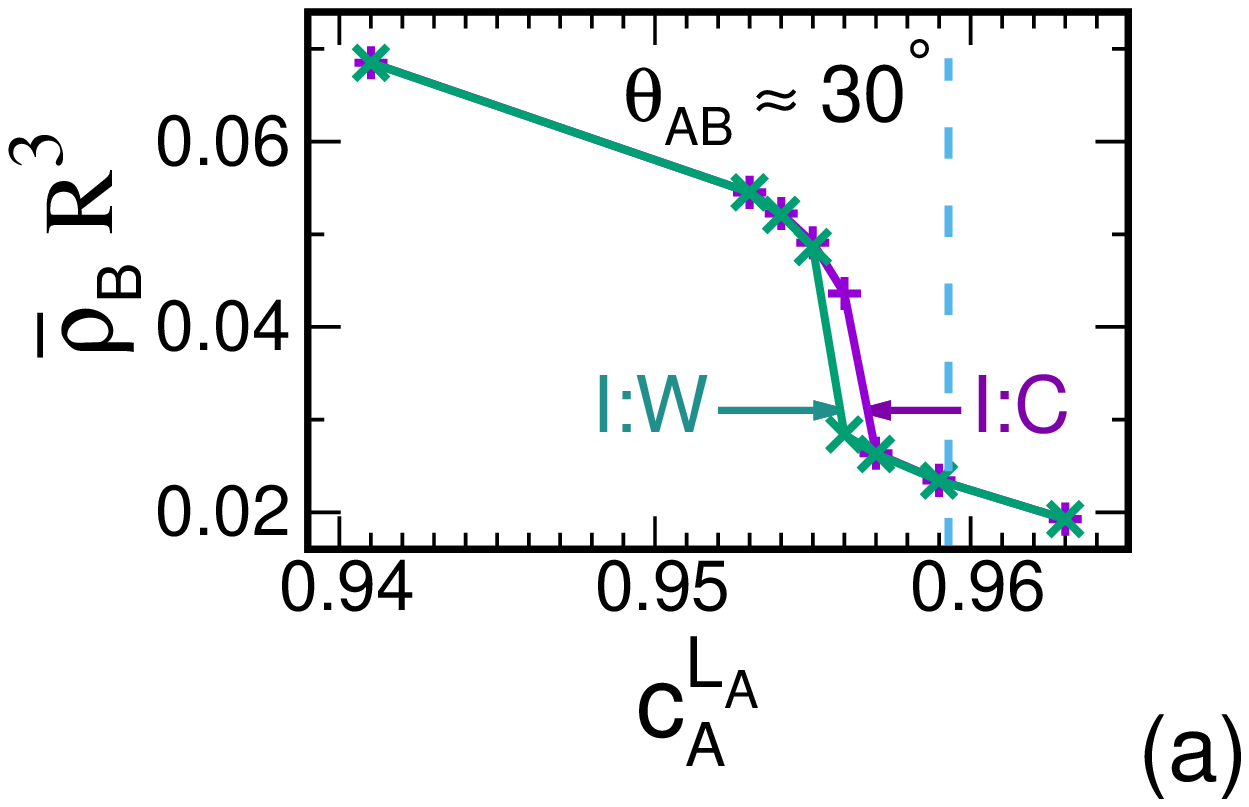}
\hspace{0.4cm}\includegraphics[scale = 0.42]{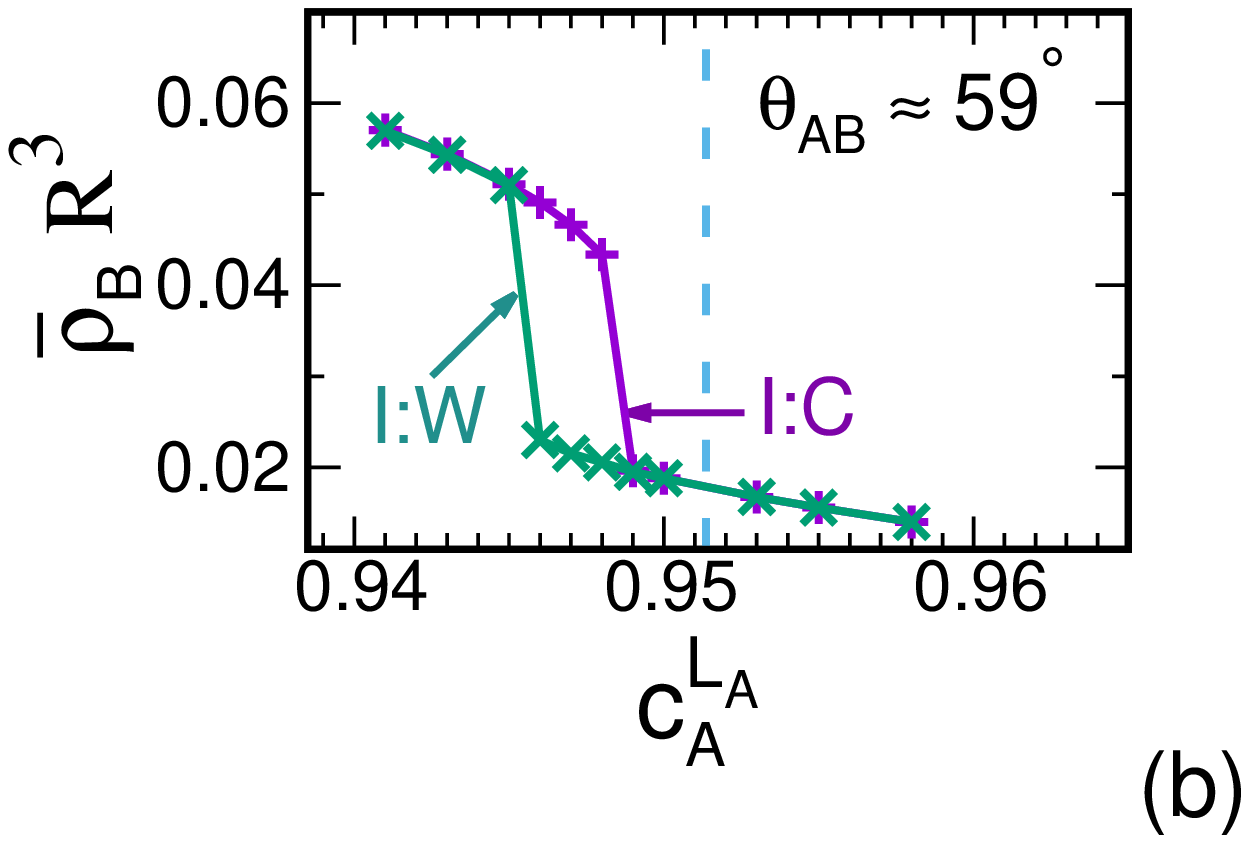}

{\vspace*{0.0cm}
\caption{\label{fig9_w7}   \baselineskip=2\baselineskip
Averaged number densities of the $B$ particles inside the pit as a function of $c^{L_{A}}_{A}$
along path $P_1$ in Fig. \ref{pd}.
The results depend on whether the iterations are initialized in a Cassie type state (I:C)
or in a Wenzel type state (I:W).
The ratio of the wall--$A$ to the wall--$B$ interaction strengths corresponds to the contact angles
$\theta_{AB} \approx 30^{\circ}$ (panel (a)) and $\theta_{AB} \approx 59^{\circ}$ (panel (b)).
The width of the pit is $ w = 14R = 7\sigma$ the depth is $D = 16R = 8\sigma$.
The remaining parameters are the same as in all other figures.
The position of the Cassie to Wenzel transition expected from the macroscopic theory
is indicated by the dashed line.
                                                                                      }}
\end{figure}

In order to study the influence of the pit dimensions on the intrusion of the ambient liquid
into the pit (Cassie to Wenzel transition) and on the reverse process, i.e., the recovery
of the Cassie state from the Wenzel state, we first carried out computations for
pits with a reduced width of $w = 7\sigma$, but with the same depth $D = 8\sigma$
as the one studied above.
In Fig. \ref{fig9_w7} we show hysteresis loops for two selected contact angles
$\theta_{AB} \approx 30^{\circ}$ and $\theta_{AB} \approx 59^{\circ}$ and compare
them with the corresponding ones obtained for the wider pits ($w = 10\sigma$, see Fig. \ref{fig9}),
which have been discussed above.  
As can be seen, the intrusion transition (curves I:C) is shifted to larger concentrations
$c^{L_{A}}_{A}$, i.e., 
farther away from bulk coexistence, as compared with the concentrations
predicted by macroscopic theory (Eq. (\ref{eq:shift})). Like in the case of the wider pits,
the concentration $c^{L_{A}}_{A}$, at which the intrusion transition occurs, is somewhat closer
to the one at bulk coexistence, than the macroscopic theory predicts. The deviation
of the macroscopic prediction from the result of the DFT computations is somewhat
larger for the narrower pits than it is for the wider pits.
Comparing the hysteresis loops obtained for the pits with a width of $w = 7\sigma$
and the ones obtained for the wider pits ($w = 10\sigma$), a striking difference
is found. For $\theta_{AB} \approx 59^{\circ}$ the hysteresis becomes a closed loop for
the narrow pit ($w = 7\sigma$), wheras an 'open' loop is found for the wider pit
($w = 10\sigma$); for $\theta_{AB} \approx 30^{\circ}$ the hysteresis becomes very
slender for the narrow pit ($w = 7\sigma$). 
For the narrow pit ($w = 7\sigma$, $D = 8\sigma$), the hysteresis remains a closed loop
up to the contact angle $\theta_{AB} \approx 70^{\circ}$ and turns into an 'open' loop
within the interval $70^{\circ} < \theta_{AB} < 73^{\circ}$. 
Thus the range of contact angles $\theta_{AB}$ within which the recovery of the
Cassie state (i.e., refilling of the pit with the lubricant) is possible, drastically extends
into the regime of rather large contact angles. According to the macroscopic theory, a 
barrier-free 
transition from the Wenzel state to the Cassie state is not possible for contact angles 
above $\theta_{AB} \approx 54.7^\circ$ in the thermodynamic regime in which the 
liquid $L_{A}$ is stable. 
Apparently, due to various confinement effects,
there is an enhancement of the wettability of the walls by the liquid $L_{B}$ near
the corners, as compared to what is expected on the basis of the contact angle $\theta_{AB}$.
(We note that $\theta_{AB}$ is defined for a macroscopicly extended planar wall.)
The sizes of the corner-droplets, as they merge to form a spanning $L_{A}$--$L_{B}$ interface,
are very small for narrow pits and thus are strongly affected by the confinement effects.
They behave as if they are in contact with a more wettable wall, which favors the destabilization
of the Wenzel state and the transformation into the Cassie state via the corner wetting mechanism.
These considerations explain why the contact angle $\theta_{AB}$, at which the transition
between a closed loop to an 'open' loop hysteresis occurs, shifts to larger values 
upon decreasing the width of the pits. 
This finding is very similar to what is found for the standard Wenzel to Cassie transition
for a liquid--vapor system \cite{Alberto2}. There the contact angle, above which the
Wenzel state may become unstable, shifts to smaller values as compared to the macroscopic
predictions for wedge drying. This shift increases upon decreasing the width of the pit.
   
In addition to the examples discussed in detail above, we have carried out a number of further
DFT studies for various contact angles $\theta_{AB}$ (i.e., various ratios $\epsilon_{B}/\epsilon_{A}$)
for pits with dimensions $w = 7\sigma$ and $D = 8\sigma$.
The key results of these systems are compiled in Table \Rnum{1} and can be compared with the corresponding
ones for wider pits ($w = 10\sigma$ and $D = 8\sigma$) listed in the same table.
%

\begin{figure}  
\hspace{-1.8cm}\includegraphics[scale = 0.42]{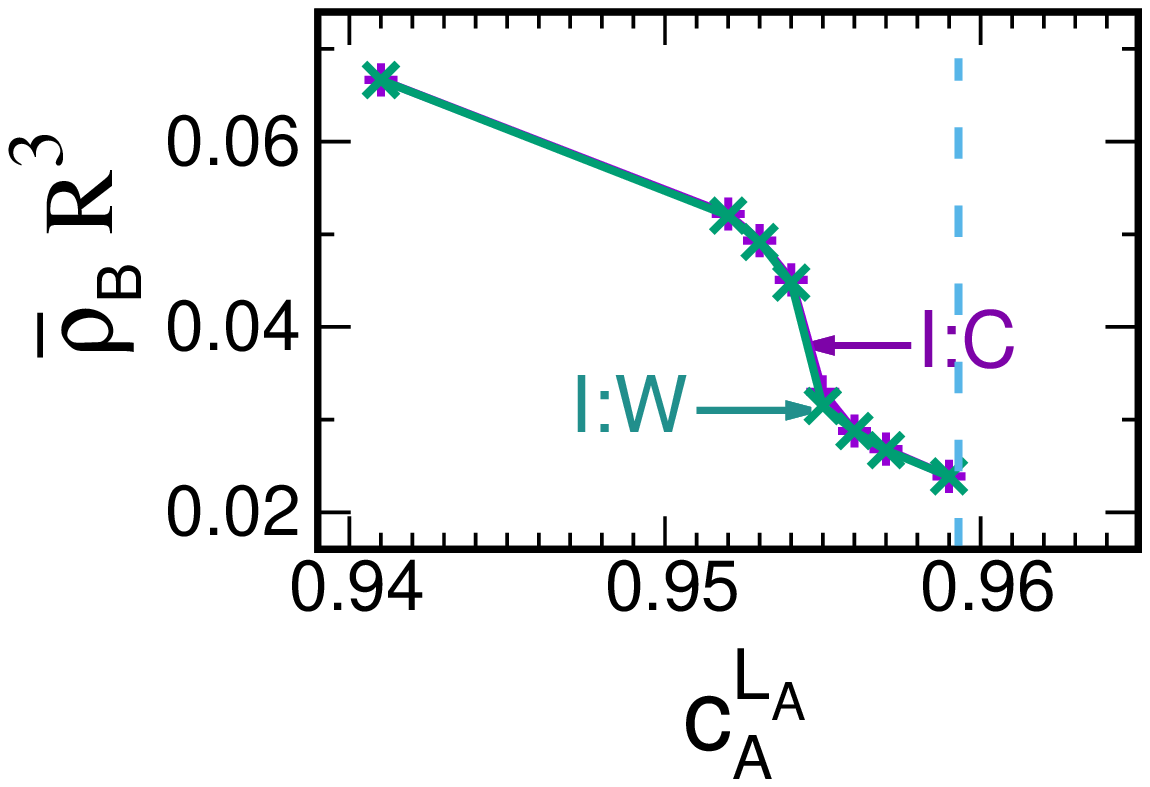}

{\vspace*{0.0cm}
\caption{\label{fig9_w7_D6}   \baselineskip=2\baselineskip
Averaged number density of the $B$ particles inside the pit as a function of $c^{L_{A}}_{A}$
along path $P_1$ in Fig. \ref{pd}.
The ratio of the wall--$A$ to the wall--$B$ interaction strengths corresponds to a contact angle
$\theta_{AB} \approx 30^{\circ}$.
Here, the results do not depend on whether the iterations are initialized in a Cassie type state (I:C)
or in a Wenzel type state (I:W).
The width of the pit is $w = 14R = 7\sigma$ and the depth is $D =12R = 6\sigma$.
The remaining parameters are the same as the ones in all other figures.
The position of the Cassie to Wenzel transition, as expected from the macroscopic theory,
is indicated by the dashed line.
                                                                                      }}
\end{figure}

In a further series of DFT computations, we have reduced the depth from $D = 8\sigma$
to $D = 6\sigma$, keeping the width fixed at $w = 7\sigma$. 
As an example, in Fig. \ref{fig9_w7_D6} we show, for the contact angle 
$\theta_{AB} \approx 30^{\circ}$, 
the averaged number density of the $B$ particles inside  the pit as the concentration $c^{L_{A}}_{A}$ 
in the ambient liquid is varied at constant packing fraction (path $P_1$ in Fig. \ref{pd}).
The hysteresis, still present for the deeper pit, disappears completely for the shallower pit,
and the intrusion transition is slightly shifted to a smaller concentration $c^{L_{A}}_{A}$.
Macroscopic capillarity theory does not and cannot predict a dependence 
of the intrusion transition on the depth of the
pit, but actually there is one. For shallow pits the features near the pit entrance,
which are responsible for the intrusion behavior, are affected by the presence of the bottom of the pit.
Vice versa, the caracteristics of the corners at the bottom of the pit are affected by the
presence of the pit entrance. 
We have carried out a number of further DFT studies for various contact angles $\theta_{AB}$ 
(i.e., various ratios $\epsilon_{B}/\epsilon_{A}$) for these pit dimensions ($w = 7\sigma$ and $D = 6\sigma$).
The corresponding key results are also compiled in Table \Rnum{1}.
%

\begin{figure}[h]           
\hspace{-0.1cm}\includegraphics[scale = 0.30]{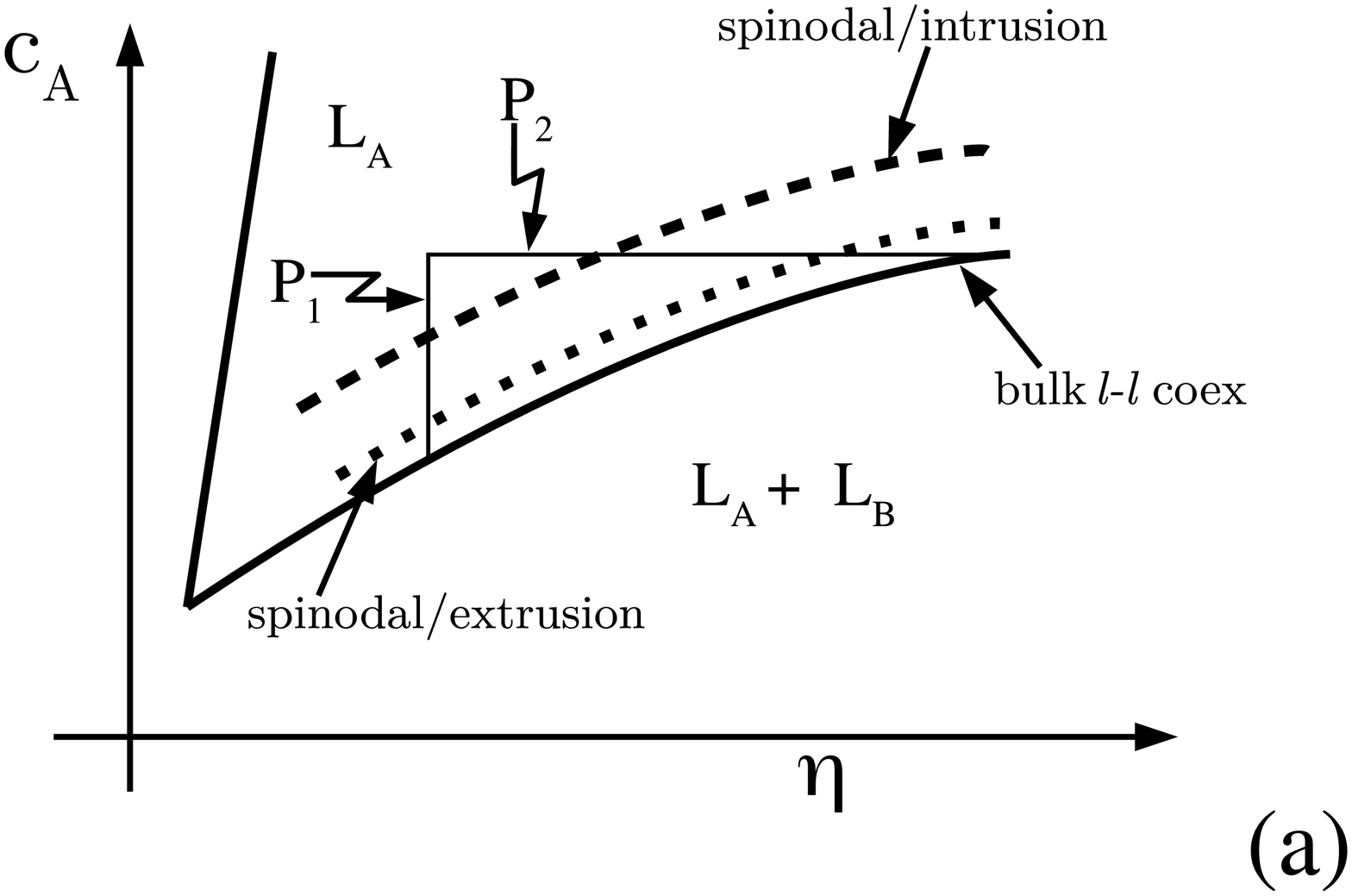}
{\vspace*{-0.30cm}\hspace*{0.00cm}\includegraphics[scale = 0.30]{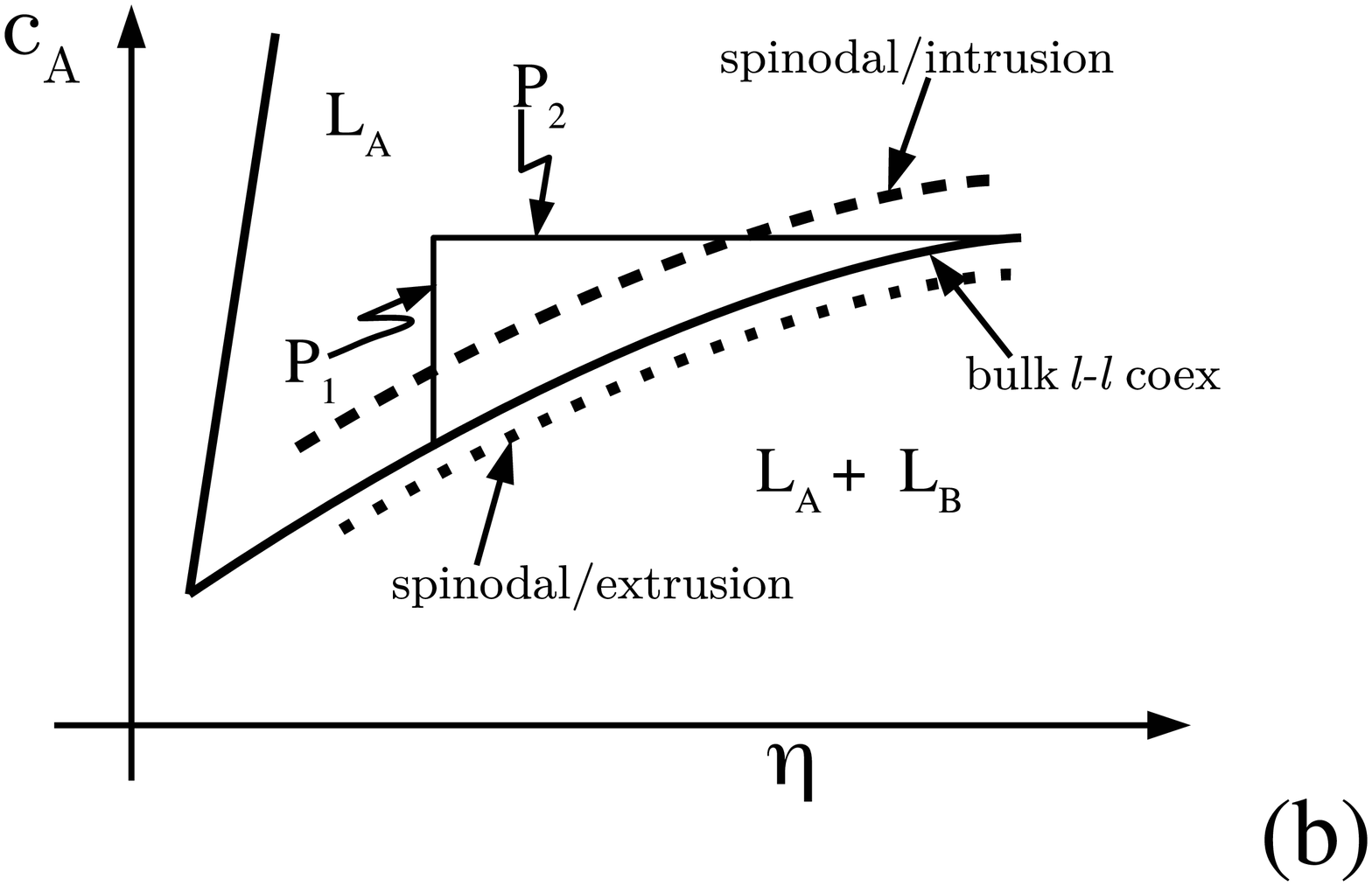}}
{\vspace*{ 0.5cm}
\caption{\label{fig_sk}    \baselineskip=2\baselineskip
Sketch of the relative locations of the spinodal line for the intrusion transition (dashed line), 
the spinodal line for the
reverse transition, i.e., the 'extrusion transition' (dotted curve), and of the bulk 
liquid--liquid coexistence line (full line).
Panel (a) [(b)] holds for contact angles $\theta_{AB}$ below [above] a limiting contact angle, 
which is $\theta_{AB} \approx 54.7^o$ for macroscopic pits. For nanoscopic pits,
this limiting contact angle depends on the width of the pit.
The thin lines correspond to the thermodynamic paths 1 and 2, respectively. 
                                                        }}
\end{figure}

\begin {table}
\caption {  
	Macroscopic and DFT results for intrusion as a function of $c^{L_{A}}_{A}$,
at fixed total packing fraction $\eta^{L_{A}} = 0.4069093$, 
for various contact angles $\theta_{AB}$, and for three sets of geometric parameters.
The macroscopic prediction for the concentration $c^{L_{A}}_{A}$, at which the Cassie to Wenzel transition 
occurs, is identified with the macroscopic prediction of liquid--liquid capillary coexistence.
For complete wetting, i.e., $\theta_{AB} = 0^{\circ}$ a lower bound for 
this concentration is given, which follows from Eq. (\ref{eq:shift}) 
with $\cos\theta_{AB} = 1$. 
(In the bulk, $L_{A}$ and $L_{B}$ coexist at $c^{L_{A}}_{A} \approx 0.941$.)
For the reverse transition (I:W) a detailed macroscopic theory is not available.
In this regime the transitions may be gradual within a wide concentration
interval. In these cases the interval given in the table contains about 70\% of the
changes in the average number densities inside the pit, the rest of the changes
is roughly equally distributed between the regimes in which the concentrations are
either below or above the given interval. In the other cases there is a jump from the
Cassie state at the lower concentration value of the given interval to the Wenzel state
at the next higher concentration, which is varied in steps of 0.001.
} 
\begin{center}
\begin{tabular}{| *{6}{c|}}
\hline
          &                     & macroscopic                           & \multicolumn{3}{c|}{DFT results}   \\ 
\( \epsilon_{B}/\epsilon_{A} \)  & \(\theta_{AB}\) & \(c^{L_{A}}_{A}\) at transition & \multicolumn{2}{c|}{\(c^{L_{A}}_{A}\) interval of transition} & hysteresis \\ 
 \hline
   &  & I:C & I:C & I:W & \\ 
\hline
\multicolumn{6}{|c|}{ $w = 10\sigma, D = 8\sigma$ (Fig. 10) }  \\
 \hline
4.0 &\(0^{\circ}\)        &             0.9561 &     0.954 - 0.967 & 0.954 - 0.967    & no  \\ 
 \hline
3.5 &\(0^{\circ}\)        &             0.9561 &     0.953 - 0.956 & 0.953 - 0.956 & no  \\ 
 \hline
3.4 &\(0^{\circ}\)        &             0.9561 &     0.955 - 0.956 & 0.955 - 0.956 & no \\ 
 \hline
3.2 & \( 12^{\circ}\) &          0.9558 & 0.954 - 0.955  & 0.954 - 0.955 & no    \\ 
 \hline
3.1 & \( 18^{\circ}\) &          0.9553 & 0.953 - 0.954  & 0.952 - 0.953 & yes    \\ 
 \hline
2.8 & \( 30^{\circ}\) &             0.9530 &  0.951 - 0.952 & 0.948 - 0.949 & yes  \\ 
 \hline
2.2 & \( 54^{\circ}\) &             0.9498 &  0.947 - 0.948 & 0.942 - 0.943 &  yes  \\ 
 \hline
2.1 & \( 57^{\circ}\) &             0.9490 &  0.947 - 0.948 & 0.941 - 0.942 & yes  \\ 
 \hline
2.0 & \( 59^{\circ}\) &             0.9480 &  0.946 - 0.947 &  --- & 'open loop'  \\ 
\hline
\multicolumn{6}{|c|}{ $w = 7\sigma, D = 8\sigma$ (Fig. 11) }  \\
\hline
%
4.0 & \(0^{\circ}\)        &             0.9626  &   0.963 - 0.970 &   0.963 - 0.970  & no   \\ 
 \hline
3.4 & \(0^{\circ}\)        &             0.9626  &   0.960 - 0.962 &   0.960 - 0.962  & no   \\ 
 \hline
3.3 & \(0^{\circ}\)        &             0.9626  &   0.959 - 0.960 &   0.959 - 0.960  & no   \\ 
 \hline
3.0 & \(23^{\circ}\) &             0.9608  & 0.957 - 0.958 &   0.957 - 0.958  & no   \\ 
 \hline
2.9 & \(26^{\circ}\) &             0.9601  & 0.956 - 0.957 &   0.956 - 0.957  & no   \\ 
 \hline
2.8 & \(30^{\circ}\) &             0.9593  & 0.956 - 0.957 &   0.955 - 0.956  & yes  \\ 
 \hline
2.0 & \(59^{\circ}\) &             0.9510 &  0.948 - 0.949 &   0.944 - 0.945  & yes  \\ 
 \hline
\multicolumn{6}{|c|}{ $w = 7\sigma, D = 6\sigma$ (Fig. 12) }  \\
\hline
4.0 & \(0^{\circ}\)        &             0.9626 &   0.963 - 0.970 & 0.963 - 0.970 & no   \\ 
 \hline
3.3 & \(0^{\circ}\)        &             0.9626 &   0.958 - 0.960 & 0.958 - 0.960 & no  \\ 
 \hline
3.2 & \(12^{\circ}\) &             0.9557 & 0.958 - 0.960 & 0.958 - 0.960 & no  \\ 
 \hline
2.8 & \(30^{\circ}\) &             0.9593 & 0.954 - 0.955 & 0.954 - 0.955 & no   \\ 
 \hline
2.5 & \(43^{\circ}\) &             0.9560 &  0.951 - 0.952 & 0.951 - 0.952 & no   \\ 
 \hline
2.3 & \(51^{\circ}\) &             0.9540 &  0.949 - 0.950 & 0.948 - 0.949 & yes  \\ 
 \hline
2.0 & \(59^{\circ}\) &             0.9510 & 0.946 - 0.947 & 0.943 - 0.944 & yes  \\ 
 \hline
\end{tabular}
\end{center}
\end {table}
Repeating the same type of computations for a series of packing fractions of the
ambient liquid, for each contact angle $\theta_{AB}$ and 
each pit width $w$ (and depth $D$) one can determine two spinodal lines running 'parallel' to the bulk
liquid--liquid coexistence line (see the sketch in Fig. \ref{fig_sk}). One of these spinodal lines
corresponds to the intrusion transition (i.e., the limit of stability of the Cassie state). 
The second spinodal line describes the stability limit of the Wenzel state. It characterizes the
reverse transition, called 'extrusion transition', i.e., the Wenzel to Cassie transition, at which a pit filled 
with the liquid $L_A$ refills with the liquid $L_B$. For contact angles $\theta_{AB}$ below a certain
threshold value, which depends on
the width of the pit, this second spinodal line might be located between the bulk liquid--liquid coexistence 
line and the spinodal line for intrusion (see Fig. \ref{fig_sk} (a)).  
For small contact angles, in which case no hysteresis is observed,
the line marking this stability limit coincides with the one locating the intrusion transition.
For contact angles above the limiting one the spinodal line for extrusion does no longer lie within 
the domain of the stable bulk liquid $L_{A}$ (see Fig. \ref{fig_sk} (b)), 
which corresponds to the 'open loop' hysteresis 
mentioned above. For concentrations $c^{L_{A}}_{A}$ of the ambient liquid
below the one of capillary liquid--liquid coexistence, but above the one at the extrusion spinodal,
the Wenzel state remains metastable.
(For deep pits the capillary liquid--liquid coexistence is at concentrations $c^{L_{A}}_{A}$ very close to
and slightly below the one of the intrusion spinodal.) 
We refrain from performing such extensive computations. Instead we have 
carried out a few DFT calculations
at constant concentration $c^{L_{A}}_{A}$ of the ambient liquid, changing the packing fraction
(i.e., the pressure) along path $P_2$ in Fig. \ref{pd}. The purpose of these calculations
is twofold. First, we want to corroborate the above picture and to find the spinodal lines
at much higher packing fractions. Second, we want to check whether the picture developed
so far is complete, or whether an unexpected dependence on the path taken in the
phase diagram is observed.

%
\subsection{Cassie--Wenzel and Wenzel--Cassie transition at fixed concentration}

In this subsection, we present results of DFT calculations carried out at 
constant concentration $c^{L_{A}}_{A} = 0.941$ 
in the ambient liquid. The packing fraction in the ambient liquid is increased in steps from $\eta^{L_{A}} = 0.40696$,
along path $P_2$ in Fig. \ref{pd}, towards the one at which liquid--liquid coexistence occurs in the bulk, 
at the chosen concentration. We follow the evolution of the (meta)stable states along this path.
As before we have initialized the iterative determination of the 
number densities in a Cassie type configuration (I:C) as well as in a
Wenzel type configuration (I:W). 
We have studied pits with the same geometric parameters ($w = 10\sigma$, $D = 8\sigma$), 
($w = 7\sigma$, $D = 8\sigma$), and ($w = 7\sigma$, $D = 6\sigma$) as considered in the
previous subsection. We have also carried out computations for various ratios 
$\epsilon_{B}/\epsilon_{A}$ of the interactions between the $B$ and $A$ particles,
respectively, with the wall particles. Instead of specifying this ratio, we
use the contact angle $\theta_{AB}$ at the particular
packing fraction $\eta^{L_{A}} = 0.40696$. This was suitable for the purpose
of the analyses in the previous subsection. Since the contact angle depends on the
packing fraction and because the transitions to be studied occur at high packing fractions,
for a quantitative comparison with the macroscopic theory it might be better to 
determine the contact angle for an appropriate packing fraction. However, these details
are not relevant for our purpose and, by always referring to the same packing fraction 
$\eta^{L_{A}} = 0.40696$, equal contact angles
correspond to equal ratios $\epsilon_{B}/\epsilon_{A}$, which is beneficial from another point 
of view.

Based on the picture developed in the previous section, we expect the following
evolution of configurations upon moving along path $P_2$ in Fig. \ref{pd}. At 
($c^{L_{A}}_{A} = 0.941$, $\eta^{L_{A}} = 0.40696$) one knows that, whether the
iteration scheme is started in the Cassie state (I:C) or in the Wenzel state (I:W),
the Wenzel state as the equilibrium one will be eventually reached. Increasing the
packing fraction upon moving along path $P_2$, the same behavior will be observed
until the spinodal line marking the intrusion transition (i.e., the limit of
stability of the Cassie state) is crossed (see Fig. \ref{fig_sk}). 
Beyond this crossing, the Cassie state remains stable
for the I:C initial condition. The occurrence of the phenomena, which follow from
choosing the I:W initial condition, depends
on the contact angle and on the geometric parameters. For small contact angles the
resulting equilibrium state is the same for both initial conditions, I:C and I:W.
For higher contact angles one finds that the Wenzel state remains (meta)stable
(for the I:W computations) up to a packing fraction at which the second spinodal, 
marking the limit of (meta)stability of the Wenzel state, is crossed (see Fig. \ref{fig_sk} (a)). 
Beyond that, the Cassie state is found to be independent of whether I:C or I:W 
initial conditions are used. 
For even larger contact angles the Wenzel state remains (meta)stable up
to the bulk liquid--liquid coexistence line (see Fig. \ref{fig_sk} (b)). 

\begin{figure}[h]           
\hspace{-0.0cm}\includegraphics[scale = 0.39]{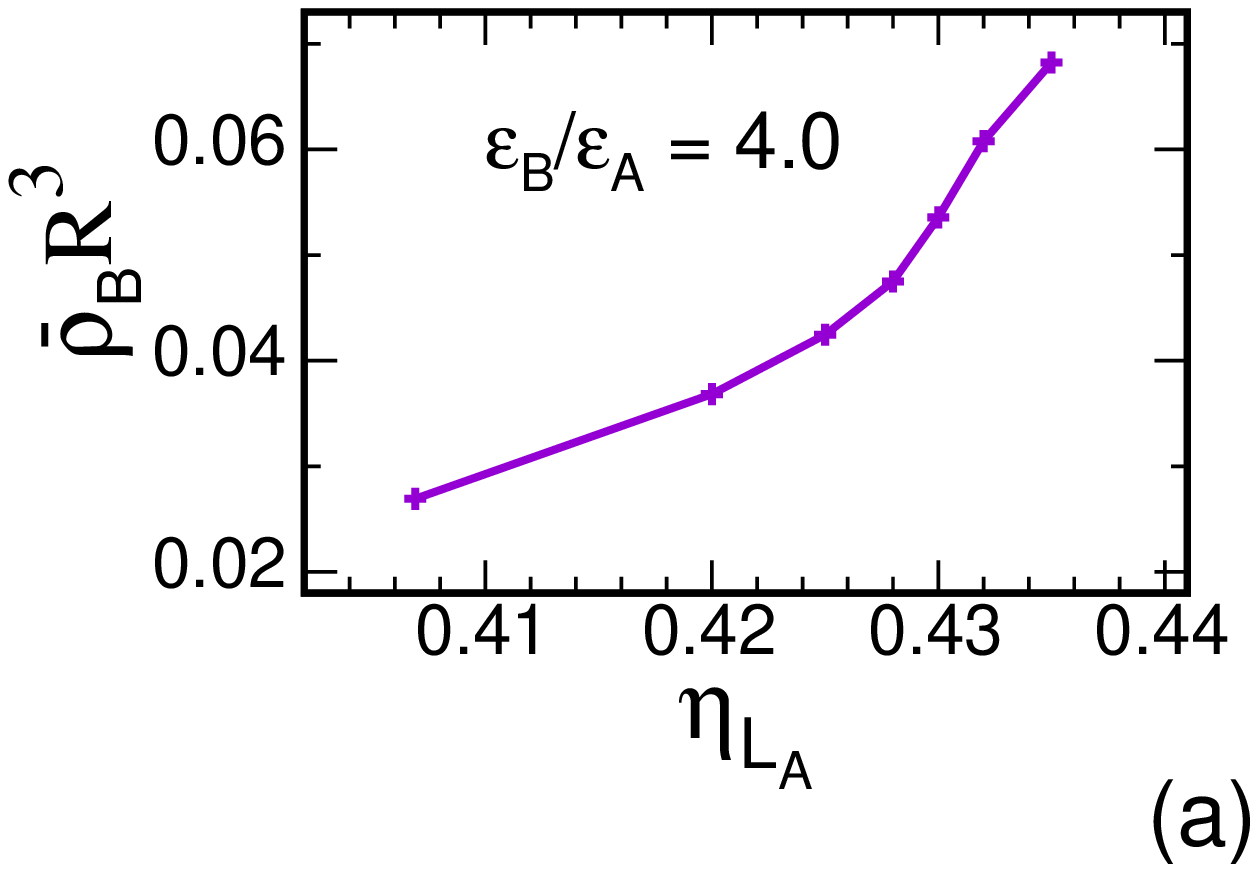}
{\vspace*{-0.30cm}\hspace*{0.00cm}\includegraphics[scale = 0.42]{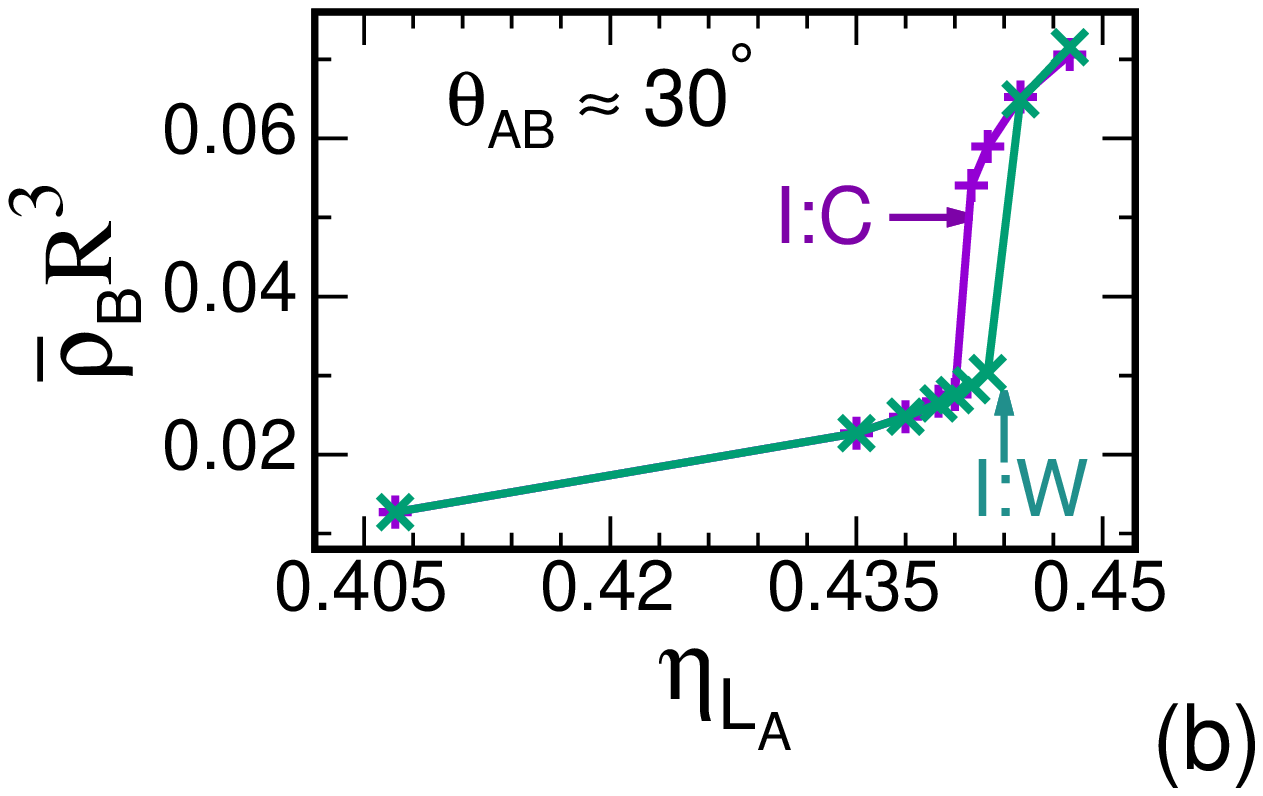}}
{\vspace*{-0.30cm}\hspace*{0.00cm}\includegraphics[scale = 0.42]{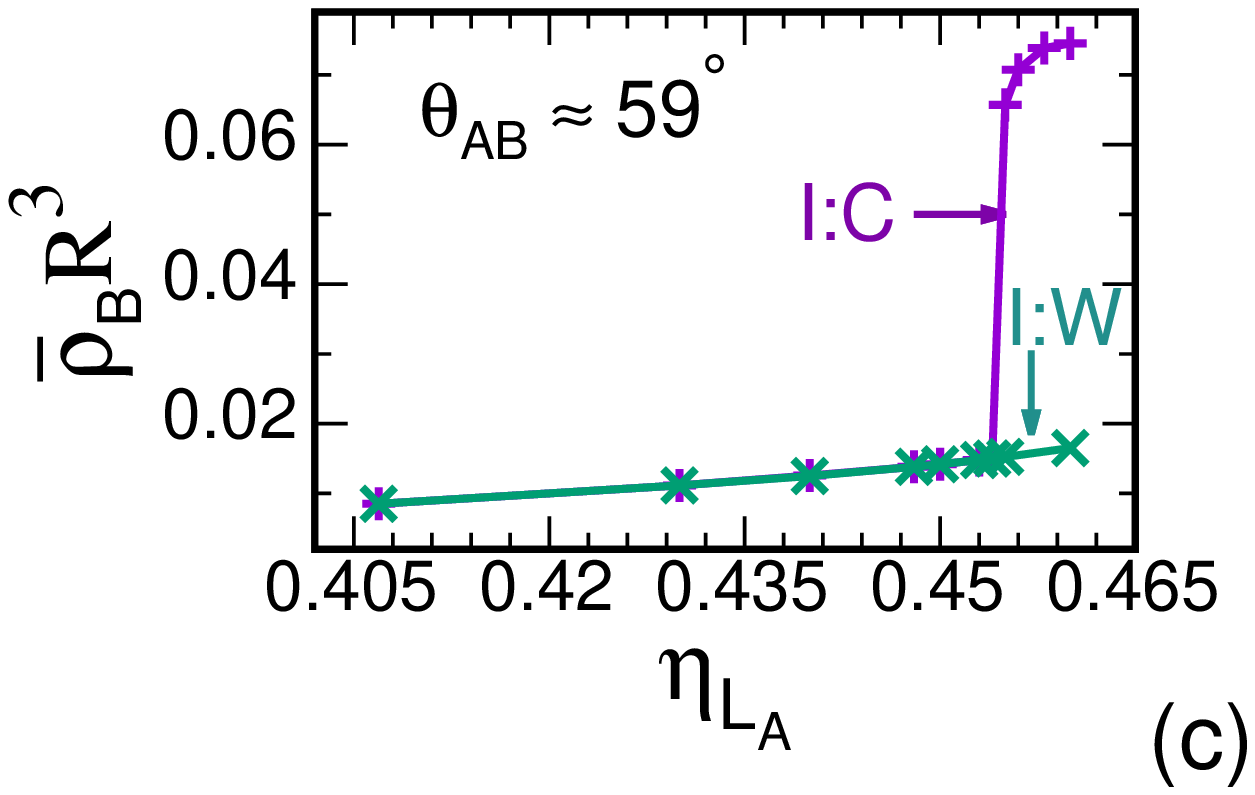}}
{\vspace*{ 0.5cm}
\caption{\label{fig19}    \baselineskip=2\baselineskip
Averaged number densities of $B$ particles inside the pit, obtained from DFT, as a function 
of $\eta^{L_{A}}$ for $w = 10\sigma$, $D = 8\sigma$, and  $c^{L_{A}}_{A} = 0.967$.
The ratio $\epsilon_{B}/\epsilon_{A}$ of the wall--$B$ to the wall--$A$ interaction strengths 
corresponds to complete wetting $\theta_{AB} = 0^{\circ}$ 
($\epsilon_{B}/\epsilon_{A} = 4.0$, panel (a); the curves I:C and I:W
coincide), to a contact angle
$\theta_{AB} \approx 30^{\circ}$ (panel (b)), and to a contact angle $\theta_{AB} \approx 59^{\circ}$ 
(panel (c)).
                                                        }}
\end{figure}

The DFT computations fully confirm the expected behavior. No unexpected features
of the system emerge by following path $P_2$ instead of path $P_1$. For example, 
in Fig. \ref{fig19}
the averaged number densities inside a wide pit ($w = 10\sigma$, $D = 8\sigma$) are 
shown as function of the packing fraction in the ambient liquid 
for a complete wetting situation ($\epsilon_B/\epsilon_A = 4.0$)  
and for the contact angles $\theta_{AB} \approx 30^{\circ}$ and $\theta_{AB} \approx 59^{\circ}$. 
In the first case one finds, without any hysteresis,  
a smooth transition to the Cassie state as the packing fraction is increased. 
In the second case, upon increasing the packing fraction, a transition to the 
Cassie state first appears within the I:C computations. In the I:W computations
the transition occurs at a slightly higher packing fraction. For the contact angle
$\theta_{AB} \approx 59^{\circ}$ it seems as if one encounters again an 'open loop' hysteresis;
currently this is only a hypothesis because we do not have results for total packing fractions up to the bulk 
liquid--liquid coexistence at $\eta^{L_{A}} = 0.46898504$ for the chosen 
concentration $c^{L_{A}}_A = 0.967$.
At these high total packing fractions strongly structured and peaked number densities form
inside the pit. It becomes difficult to reach convergence in the numerical iterations.  
This particular phenomenon also deserves a study of its own.
%

\begin{figure}[h]           
\hspace{-1.5cm}\includegraphics[scale = 0.50]{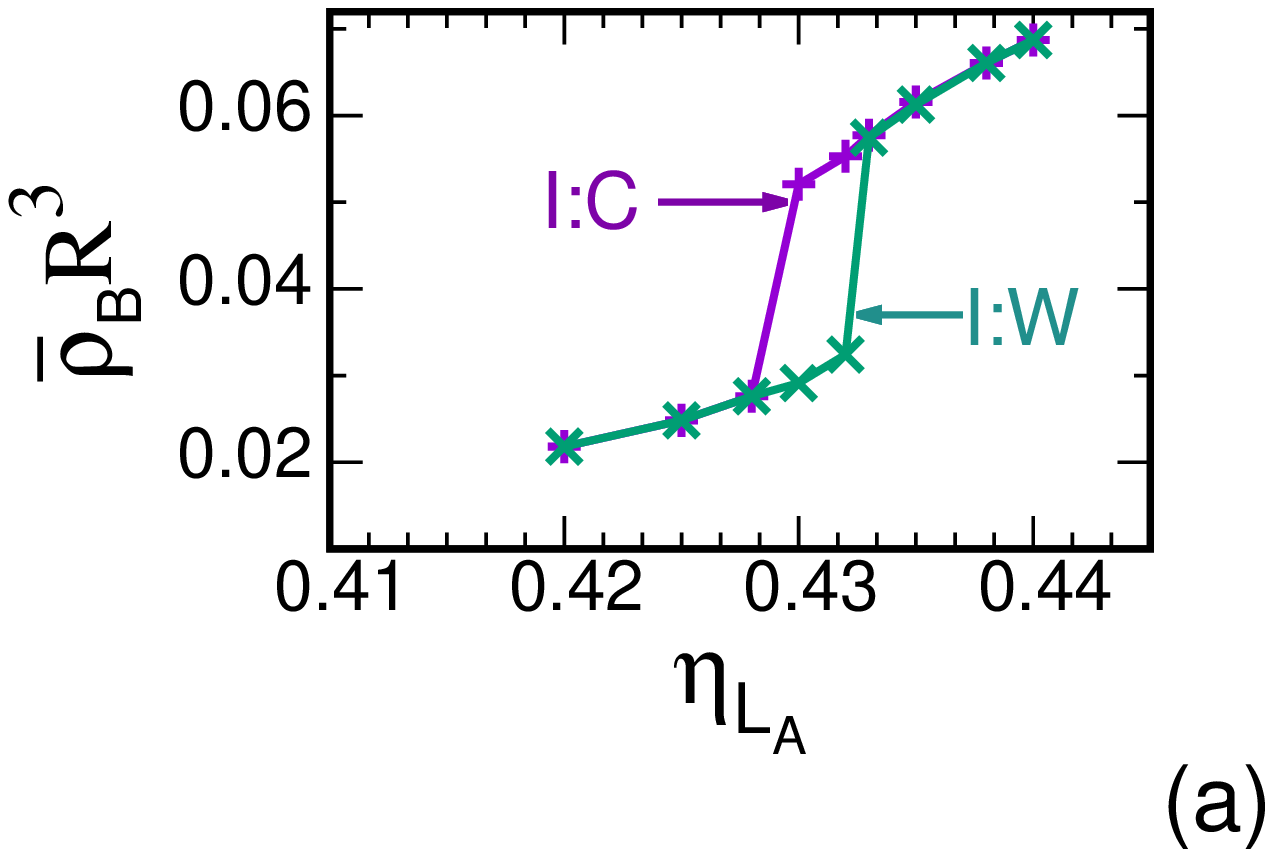}
{\vspace*{-0.00cm}\hspace*{0.20cm}\includegraphics[scale = 0.50]{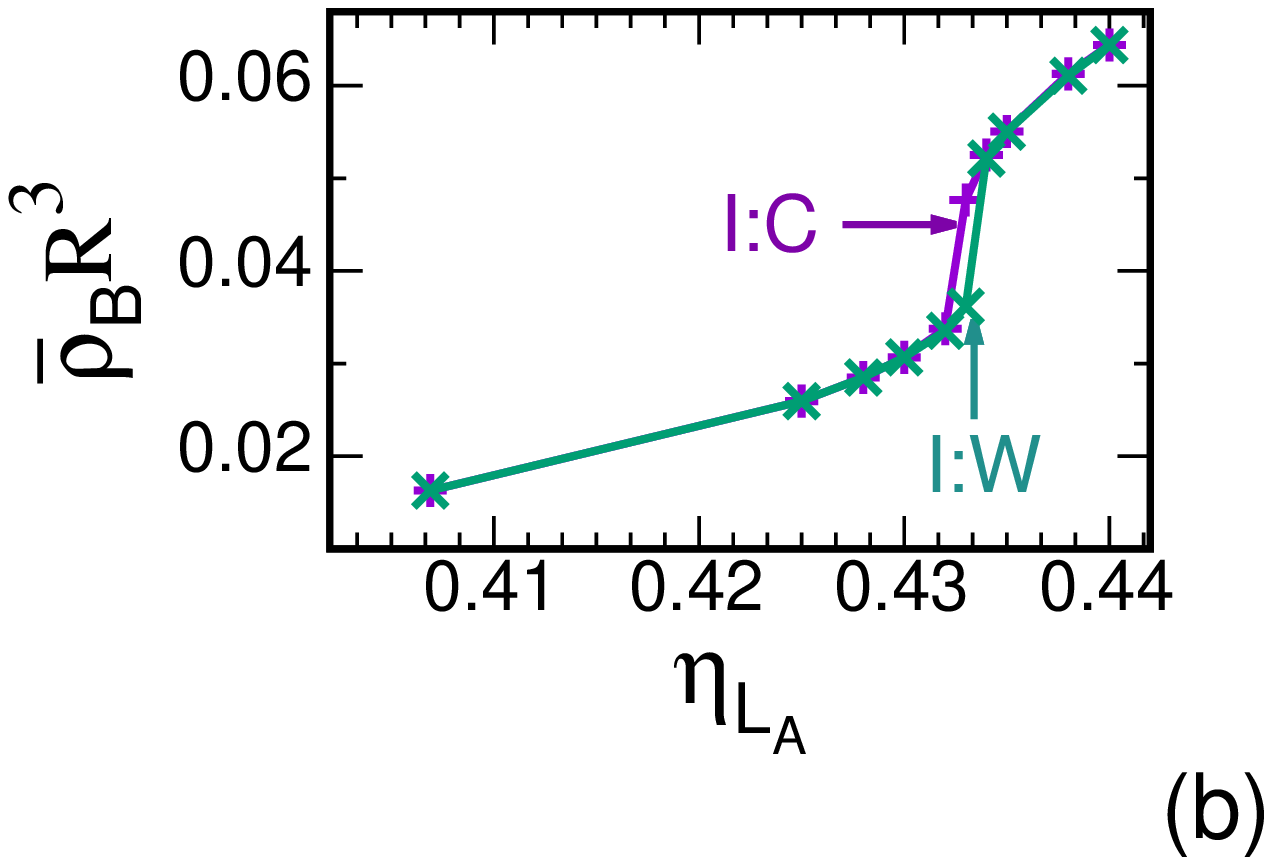}}
{\vspace*{-0.0cm}
\caption{\label{fig19_w7}  \baselineskip=2\baselineskip
Averaged number densities of the $B$ particles inside the pit, calculated by using DFT, 
as a function of $\eta^{L_{A}}$ at $c^{L_{A}}_{A} = 0.967$. The ratio $\epsilon_{B}/\epsilon_{A}$
of the wall--$A$ to the wall--$B$ interaction strengths corresponds
to $\theta_{AB} \approx 30^{\circ}$. Two sets of geometric parameters are considered:
        $w = 7\sigma$ and $D = 8\sigma$ for panel (a), $w = 7\sigma$ and 
        $D = 6\sigma$ for panel (b).
                                          }}
\end{figure}

One also finds, in line with the insights gained in the previous section, that the intrusion
and extrusion transitions
are shifted to lower packing fractions if the width $w$ of the pit is decreased 
(see Fig. \ref{fig19_w7} (a));
the spinodal lines are shifted further away from the bulk coexistence curve as $w$ is decreased.
These shifts can be inferred also from Table \Rnum{2}. In addition, the data 
presented in this table reveal how an open loop hysteresis transforms
into a closed loop hysteresis as $w$ is reduced from $10\sigma$ to $7\sigma$,
with otherwise identical system parameters.
Figure \ref{fig19_w7} (b) demonstrates that shallow pits lead to narrow hysteresis loops.

\begin {table}
\caption {  
	DFT results for the Cassie to Wenzel and the Wenzel to Cassie transitions at fixed concentration
$c^{L_{A}}_{A} = 0.967 $ as a function of the total packing fraction $\eta^{L_{A}}$ 
for various contact angles $\theta_{AB}$ and three sets of geometric parameters $w$ and $D$.
The contact angles given in the table correspond to liquid--liquid coexistence at 
a total packing fraction $\eta^{L_{A}} = 0.4069093$ as in Table \Rnum{1}. 
In the complete wetting regime the transitions might exhibit a gradual behavior 
with about 70\% of the changes within the given interval of
total packing fractions. For the intervals of width 0.001, corresponding to the steps
in which $\eta^{L_{A}}$ is changed, the system is in the Wenzel
state at total packing fractions at and below the lower value of the given interval, and it is in the Cassie state at
and above the upper value of the interval.  
 } 
\begin{center}
\begin{tabular}{| *{5}{c|}}
\hline
 &  & \multicolumn{2}{c|} { } &  \\
\(\epsilon_{B}/\epsilon_{A}\)  & $\theta_{AB}$  & \multicolumn{2}{c|}{{$\eta^{L_{A}}$} interval of transition} & hysteresis  \\ 
                    &                & \multicolumn{2}{c|}{                                       } &             \\ 
\hline
    &	& I:C & I:W &   \\ 
 \hline
\multicolumn{5}{|c|}{ $w = 10\sigma, D = 8\sigma$ (Fig. 14)} \\
\hline
4.0 & \(0^{\circ}\)   &   0.420 - 0.432 & 0.420 - 0.432 & no  \\ 
 \hline
3.6 & \(0^{\circ}\)   &   0.430 - 0.436 & 0.430 - 0.436 & no  \\ 
 \hline
3.5 & \(0^{\circ}\)   &   0.434 - 0.435 & 0.434 - 0.435 & no  \\ 
 \hline
3.4 & \(0^{\circ}\)   &   0.435 - 0.436 & 0.435 - 0.436 & no \\ 
 \hline
3.2 & \( 12^{\circ}\)    &  0.436 - 0.437 & 0.437 - 0.438 & yes \\ 
 \hline
2.8 & \( 30^{\circ}\)   &   0.440 - 0.441 & 0.444 - 0.445 &  yes \\ 
 \hline
2.2 & \( 54^{\circ}\)   &   0.449 - 0.450 & 0.461 - 0.462 & yes \\ 
 \hline
2.0 & \( 59^{\circ}\)   &   0.454 - 0.455 & --- & 'open loop' \\ 
 \hline

\multicolumn{5}{|c|}{ $w = 7\sigma, D = 8\sigma$ (Fig. 15 (a))} \\
\hline
4.0 & \(0^{\circ}\)     &  0.410 - 0.415 & 0.410 - 0.415 & no  \\ 
 \hline
3.7 & \(0^{\circ}\)   &    0.410 - 0.417 &  0.410 - 0.417 & no  \\ 
 \hline
3.4 & \(0^{\circ}\)   &    0.418 - 0.422 & 0.418 - 0.422 & no  \\ 
 \hline
3.3 & \(0^{\circ}\)   &    0.422 - 0.423 & 0.422 -  0.423 & no  \\ 
 \hline
3.1 & \( 18^{\circ}\)   &  0.425 - 0.426 & 0.425 - 0.426 & no  \\ 
 \hline
3.0 & \( 23^{\circ}\)   &  0.426 - 0.427 & 0.427 - 0.428 & yes  \\ 
 \hline
2.8 & \( 30^{\circ}\)   &  0.429 - 0.430 & 0.432 - 0.433 &  yes  \\ 
 \hline
2.0 & \( 59^{\circ}\)   &  0.445 - 0.446 & 0.461 - 0.462 & yes  \\ 
 \hline
\multicolumn{5}{|c|}{ $w = 7\sigma, D = 6\sigma$ (Fig. 15 (b))} \\
\hline
4 .0& \(0^{\circ}\)     &0.410 - 0.420 & 0.410 - 0.420 & no  \\ 
 \hline
3.3 & \(0^{\circ}\)     &0.422 - 0.430 & 0.422 - 0.430 & no  \\ 
 \hline
3.0 & \( 23^{\circ}\)     & 0.425 - 0.431 & 0.425 - 0.431 & no  \\ 
 \hline
2.9 & \( 26^{\circ}\)     & 0.431 - 0.432 & 0.431 - 0.432 & no  \\ 
 \hline
2.8 & \( 30^{\circ}\)     & 0.432 - 0.433 & 0.433 - 0.434 & yes  \\ 
 \hline
2.0 & \( 59^{\circ}\)     & 0.447 - 0.448 & 0.460 - 0.461 & yes  \\ 
 \hline
\end{tabular}
\end{center}
\end {table}
It is interesting to note, that in the present case {\it reducing} the packing
fraction (i.e., reducing the pressure) triggers the Cassie to Wenzel transition. 
In the well known liquid--vapor systems, the Cassie to Wenzel
transition is induced by {\it increasing} the pressure. However, in both cases
the transition occurs upon moving away from bulk coexistence. 
\section{Summary, Conclusions, and Outlook }

By using microscopic density functional theory, we have studied
the intrusion transition of an ambient liquid into cavities 
which are filled with a lubricant liquid. We have considered the case that the lubricant liquid 
adsorbs at the walls of the cavities more strongly than the ambient liquid does.
In order to model the ambient and the lubricant liquids, a binary liquid mixture
composed of $A$ and $B$ particles is investigated which decomposes into 
an $A$ rich liquid, 
(i.e., the ambient liquid), and a $B$ rich liquid, 
(i.e., the lubricant liquid).
The $B$ particles are taken as the ones which are attracted more strongly by the walls. The cavities
are modeled as square pits of width $w$ and depth $D$ ( Fig. \ref{schem}).
The intrusion transition is triggered by changing the composition of the
ambient liquid (see Table I and Figs. \ref{fig5} and \ref{fig7}--\ref{fig9_w7_D6}) 
or its total packing fraction (i.e., the pressure, see Table II and Figs. \ref{fig19} and \ref{fig19_w7}).    
However, these parameters are constrained to that region of the bulk phase diagram where the
ambient liquid is stable.
We have also studied the reverse process in which a pit filled with
the ambient liquid undergoes a transition back to a state in which it
is filled with the lubricant liquid. Also this transition is triggered
by varying the composition or the total packing fraction of the ambient liquid.
Both transitions have been studied for different sizes of the cavities and
for various chemical compositions of the walls (giving rise to distinct substrate
potentials).      
In our studies we have considered stable or metastable equilibria, which
implies that the composition and the total packing fraction of the
lubricant liquid confined in the pits adjust until mechanical and chemical
equilibrium is reached. 

In order to compare the results of the microscopic computations with the predictions of 
the macroscopic capillarity model (see Sect. II E), the relative strength of the interaction between
the $A$ and the $B$ particles, respectively, and the wall is characterized
in terms of the contact angle $\theta_{AB}$ formed by the liquid--liquid interface
between the two coexisting fluid phases and a planar wall, which is 
composed of a freely chosen material.
The intrusion transition is located in that region of the bulk phase diagram where the 
ambient liquid is stable in the bulk. Its thermodynamic location qualitatively follows the
shift (away from the bulk liquid--liquid coexistence line) predicted by
the macroscopic capillarity model with respect to its dependence on the
contact angle $\theta_{AB}$ and the width of the pit. Quantitatively, the
transition found within the microscopic calculations occurs somewhat closer
to the bulk liquid--liquid coexistence than predicted by the macroscopic capillarity model
(see Figs. \ref{fig7} and \ref{fig9} and Table I).
The quantitative discrepancies become larger for narrower cavities
(compare Figs. \ref{fig9} and \ref{fig9_w7}). 
In cases in which the wall is completely wetted by the lubricant liquid
($\theta_{AB} = 0$) the reverse transition follows the same path as for intrusion,
i.e., there is no hysteresis (see Fig. \ref{fig5}). 
In these cases the transition is rather smooth. 
Also for small but nonzero contact angles there is no hysteresis, 
but the transition can be rather sharp. 
For larger contact angles, one has to move closer to the bulk liquid--liquid
coexistence line, as compared to the intrusion transition, in order to trigger
the inverse transition from a pit filled with ambient liquid to one
filled with the lubricant liquid. In these cases the intrusion transition as well as the
reverse transition occur abruptly.
For large contact angles the reverse transition
may not be found within the regime in which the bulk of the ambient liquid 
is stable (see Fig. \ref{fig9} (b)).
The contact angle $\theta_{AB}$, above which the reverse transition is not
found, depends on the width (and to a certain extent also on the depth) of the pit. 
For a small width the reverse transition is observed up to quite high
contact angles (up to $\theta_{AB} \approx 73^\circ$ for a width of $w = 7\sigma$
and a depth of $D = 8\sigma$). 
We can associate the reverse transition with a mechanism, which is related to wetting 
by the lubricant of the corners at the bottom of the pits. A spontaneous
reverse transition occurs once the individual 'droplets' of lubricant forming
at the corners 
are large enough to merge, which creates a liquid--liquid interface spanning
the whole pit. For the geometry studied, according to the macroscopic theory,
this mechanism is only possible if $\theta_{AB} < 54.7^\circ$,
within the regime in which the ambient liquid is stable in the bulk. 
However, nanoscale effects 
shift the contact angle, up to which the reverse transition occurs, 
to much higher values. 
This shift becomes more pronounced for smaller widths of the pit.
For narrower pits the merging lubricant 'droplets', which are localized at the corners
of the bottom of the pits, are small and 
thus they are more strongly affected by nanoscale effects near the corners than the
larger merging droplets in the wider pits.
Nanoscale effects originate, for instance, from the extended range of the
fluid--wall and the fluid--fluid interactions as well as from density oscillations induced
by the walls.

For a nonzero contact angle $\theta_{AB}$, the transitions discussed here typically
occur at a spinodal marking the limit of stability of a metastable state.
In the case of the intrusion transition -- as the thermodynamic conditions for 
the ambient liquid are gradually shifted away from those for liquid--liquid coexistence
in the bulk -- the state, in which the pit is filled 
with lubricant, becomes a metastable state before the intrusion transition occurs.
In order to reach the stable state, in which the pit is indeed filled with the ambient liquid,
a free energy barrier has to be surmounted. 
The liquid--liquid interface has to be pushed down to the bottom of the pit, which
costs free energy.
Only after the liquid--liquid and the lubricant--wall interfaces have disappeared and 
are replaced by an ambient-liquid -- wall interface, the free energy is lowered below the
one of the metastable state.  
In the case of the reverse transition, a pit filled with the ambient liquid, will
remain metastable until the lubricant 'droplets', forming at the corners of the 
pit floor,
can merge spontaneously. Thermally assisted barrier crossing can become relevant
already slightly off the spinodals. Our studies also contribute to  an understanding
of these processes.
  
There are similarities between the Cassie to Wenzel and the Wenzel to Cassie transitions
in fluid--vapor systems and the intrusion and reverse transitions
in a liquid--liquid system studied here. However, there are also differences.
For instance, in fluid--vapor systems intrusion of liquid may be induced
by increasing the pressure in the liquid, whereas in the liquid--liquid system
intrusion of ambient liquid may be induced by decreasing the pressure of the
ambient liquid. Both cases, however, have in common that moving away from the
bulk coexistence line (i.e., liquid--vapor and liquid--liquid coexistence, respectively) 
leads to intrusion. 
    
In the present studies we have tuned the liquid--liquid interaction parameters
such that each of the two coexisting liquids contains a small fraction of the 
respective minority component. We refrained from analyzing the limit of very low mutual
solubility of the two liquids. We expect that the main phenomena remain the same
also in that limit.

The picture provided by the present study offers guidelines for designing 
slippery liquid infused porous surfaces. However, also other applications of porous surfaces 
might be envisaged, which make use of the possibility to switch forth and back between wetting states
in order to switch other associated physical properties of the composite surface. 

It would be very interesting, with respect to both basic research and technical applications, 
to study the dynamics of these transitions.
This can be achieved, inter alia, by applying dynamic density functional 
(DDFT, see, e.g., Refs.  \cite{Archer, Rauscher}),
kinetic Monte Carlo simulations (see, e.g., Refs.  \cite{Striolo, Binder, PDa, PDb, DPR, PDO}), 
as well as dissipative particle dynamics, 
which has been used recently to study the intrusion of a mixed binary fluid into a nanochannel \cite{Otter}.
\end{spacing}  
\begingroup


\newpage
\begin{thebibliography}{99}
\bibliographystyle{plain}
\bibitem{Ref1} J. Young,  F. Chen, Q. Yang, Y. Fang, J. Huo, J. Zhang, and X. Hou,  
        Adv. Mat. Interf. {\bf 4}, 1700552 (2017).
\bibitem{Ref2} T. S. Wong, S. H. Kang, S. K. Y. Tang, E. J. Smythe, B. D. Hatton, 
         A. Grinthal, and J. Aizenberg, Nature {\bf 477}, 443 (2011).            
\bibitem{Ref3}  J. Genzer and K. Efimenko, Biofouling {\bf 22}, 339 (2006).
\bibitem{Ref4} S. Nishimoto and B. Bhushan, RSC Adv. {\bf 3}, 671 (2013).
\bibitem{Ref5} P. Ragesh, V. A. Ganesh, S. V. Nair, and A. S. Nair, J. Mater. Chem. A
              {\bf 2}, 14773 (2014).
\bibitem{Ref6} S. Pan, A. K. Kota, J. M. Mabry, and A. Tuteja, J. Am. Chem. Soc.
              {\bf 135}, 578 (2013).
\bibitem{Ref7} M. J. Kreder, J. Alvarenga, P. Kim, and J. Aizenberg, Nat. Rev. Mater.
               {\bf 1}, 15003 (2016).
\bibitem{Ref8} J. Lv, Y. Song, L. Jiang, and J. Wang, ACS Nano {\bf 8}, 3152 (2014).
\bibitem{Ref9} J. D. Smith, R. Dhiman, S. Anand, E. Reza-Garduno, R. E. Cohen, G. H. McKinley, and
               K. K. Varanasi, Soft Matter {\bf 9}, 1772 (2013).      
\bibitem{Ref10} M. K. Fu, I. Arenas, S. Leonardi, M. Hultmark, J. Fluid Mechanics {\bf 824}, 688 (2017).
\bibitem{Ref11} W. Barthlott and C. Neinhuis, Planta {\bf 202}, 1 (1997).
\bibitem{Ref12} R. N. Wenzel, Ind. Eng. Chem. {\bf 28}, 988 (1936).
\bibitem{Ref13} A. B. D. Cassie and S. Baxter, Trans. Faraday Soc. {\bf 40}, 546 (1944).
\bibitem{Ref14} C. Neinhuis and W. Barthlott, Ann. Bot. {\bf 79}, 667 (1997).
\bibitem{Ref15} D. Qu\'er\'e, Rep. Prog. Phys. {\bf 68}, 2495 (2005).
\bibitem{Ref16} A. Lafuma and D. Qu\'er\'e, Nat. Mater. {\bf 2}, 457 (2003).
\bibitem{Alberto1} A. Giacomello, L. Schimmele, S. Dietrich, and M. Tasinkevych,
                   Soft Matter {\bf 12}, 8927 (2016).
\bibitem{Alberto2} A. Giacomello, L. Schimmele, S. Dietrich, and M. Tasinkevych,
                   Soft Matter {\bf 15}, 7462 (2019).
\bibitem{Ref17} D. Qu\'er\'e, Ann. Rev. Mater. Res. {\bf 38}, 71 (2008).
\bibitem{Ref18} T. P. N. Nguyen, P. Brunet, Y. Coffinier, and R. Boukherroub,
                Langmuir {\bf 26}, 18369 (2010).
\bibitem{Ref19} L. Bocquet and E. A. Lauga, Nat. Mater. {\bf 10}, 334 (2011).
\bibitem{Ref20} R. Poetes, K. Holtzmann, K. Franze, and U. Steiner, Phys.
                Rev. Lett. {\bf 105}, 166104 (2010).
\bibitem{Ref21} T. Verho, C. Bower , P. Andrew , S. Franssila, O. Ikkala, and R. H. A. Ras,
             Adv. Mater. {\bf 23}, 673 (2011).
\bibitem{Ref22} M. Reyssat, J. M. Yeomans, and D. Qu\'er\'e, Europhys. Lett. {\bf 74}, 299 (2006).
\bibitem{Ref23} T. Deng, K. K. Varanasi, M. Hsu, N. Bhate, C. Keimel, J. Stein, and
                M. Blohm, Appl. Phys. Lett. {\bf 94}, 133109 (2009).
\bibitem{Ref24} H. Bellanger, T. Darmanin, E. T. de Givenchy, and F. Guittard, Chem.
                Rev. {\bf 14}, 2694 (2014).
\bibitem{Ref25} Z. L. Chu and S. Seeger, Chem. Soc. Rev. {\bf 43}, 2784 (2014).
\bibitem{Ref26} T. Jiang, Z. G. Guo, and W. M. Liu, J. Mater. Chem. A {\bf 3},
                1181 (2015).

\bibitem{Ref27} A. Lafuma and D. Qu\'er\'e, EPL {\bf 96}, 56001 (2011).
\bibitem{Ref28} H. F. Bohn and W. Federle, Proc. Natl. Acad. Sci. USA {\bf 101}, 14138 (2004).
\bibitem{Ref29} E. K. Epstein, T. K. Wong, R. A. Belisle, E. M. Boggs, and J. Aizenberg,
                 Proc. Natl. Acad. Sci. USA {\bf 109}, 13182 (2012).
 \bibitem{Ref30} C. Howell, T. L. Vu, J. J. Lin, S. Kolle, N. Juthani,
                E. Watson, J. C Weaver, J. Alvarenga, and J. Aizenberg
                ACS Appl. Mater. Interf. {\bf 6}, 13299 (2014).
%
\bibitem{Ref31}  S. Nishioka, M. Tenjimbayashi, K. Manabe, T. Matsubayashi,
                K. Suwabe, K. Tsukada, and S. Shiratori, RSC Adv. {\bf 6}, 47579 (2016).
\bibitem{Ref34} P. Kim, T.-S. Wong, J. Alvarenga, M. J. Kreder, W. E. 
                Adorno-Martinez, and J. Aizenberg,  ACS Nano {\bf 6}, 6569 (2012).
\bibitem{Ref35} P. W. Wilson, W. Lu, H. Xu, P. Kim, M. J. Kreder,
                J. Alvarenga, and J. Aizenberg, Phys. Chem. Chem. Phys.
                {\bf 15}, 581 (2013).
\bibitem{Ref36} K. Rykaczewski, S. Anand, S. B. Subramanyam, and
                 K. K. Varanasi, Langmuir {\bf 29}, 5230 (2013).
\bibitem{Ref37} K. Rykaczewski, A. T. Paxson, M. Staymates, M. L. Walker,
                 X. Sun, S. Anand, S. Srinivasan, G. H.  McKinley,
                 J. Chinn, J. H.J Scott, and K. K. Varanasi, Sci. Rep. {\bf 4},  4158 (2014).
\bibitem{Ref38} K.-C. Park, P. Kim, A. Grinthal, N. He, D. Fox, J. C. Weaver,
                and J.  Aizenberg,  Nature  {\bf 531}, 78 (2016).
\bibitem{Doris1} T. Kajiya, S. Wooh, Y. Lee, K. Char,  D. Vollmer and  H.-J. Butt, 
	Soft Matter {\bf 12}, 9377 (2016).   	
\bibitem{Ref32} G. H. Zhu, C. Zhang, C. Wang, and N. S. Zacharia, Adv. Mat. Interf.
                {\bf 3}, 1600515 (2016).
\bibitem{Ref33} S. Sunny, N. Vogel, C. Howell, T. L. Vu, and J. Aizenberg,
        Adv. Funct. Mater. {\bf 24}, 6658 (2014).
\bibitem{Ref39} G. H. Zhu, S.-H. Cho, H. Zhang, M. Zhao, and N. S. Zacharia,
                Langmuir {\bf 34}, 4722 (2018).
\bibitem{Ref39b} G. H. Zhu, C.  Zhang, C. Wang, and N. S. Zacharia,
                Adv. Mat. Interf. {\bf 3}, 1600515 (2016).
\bibitem{Ref40} W. Ma, Y. Higaki, H. Otsuka, and A. Takahara, Chem. Commun. {\bf 49}, 597 (2013).

\bibitem{Ref41} P. Wang, D. Zhang, S. Sun, T. Li, and Y. Sun,
                ACS Appl. Mater. Interf. {\bf 9}, 972 (2017).
\bibitem{Ref42} S. Yuan, Z. Li, L. Song, H. Shi, S. Luan, and J. Yin,
                ACS Appl. Mater. Interf. {\bf 8}, 21214 (2016).
\bibitem{Ref43} J. Wang, K. Kato, A. P. Blois, and T.-S. Wong,
                ACS Appl. Mater. Interf. {\bf 8}, 8265 (2016).
\bibitem{Ref44} S. Nishioka, M. Tenjimbayashi, K. Manabe, T. Matsubayashi,
                K. Suwabe, K. Tsukada, and S. Shiratori,  RSC Adv. {\bf 6}, 47579 (2016).
\bibitem{Ref45} X. Yao, J. Ju, S. Yang, J. Wang, and L. Jiang, Adv. Mater.
                {\bf 26}, 1895 (2014).
\bibitem{Ref46} E. Jenner and B. D'Urso, Appl. Phys. Lett. {\bf 103}, 251606 (2013).
\bibitem{Ref47} V. Hejazi and M. Nosonovsky, Langmuir {\bf 28}, 2173 (2012).
\bibitem{Ref50} X. Hou, Y. Hu, A. Grinthal, M. Khan, and J. Aizenberg, Nature {\bf 519},
                70 (2015).
\bibitem{Ref51} M. Ulbricht, Nature {\bf 519}, 41 (2015).
\bibitem{Ref48} A. D. Stroock, V. V. Pagay, M. A. Zwieniecki, and N. M. Holbrook,
                Annu. Rev. Fluid Mech. {\bf 46}, 615 (2014).
\bibitem{Doris2} W. S. Y. Wong, K. I. Hegner, V. Donadei, L. Hauer, A. Naga, and D. Vollmer, Nano Lett.
                  {\bf 20}, 8508 (2020).
\bibitem{Ref49} O. Peleg and R. Y. H. Lim,
                Biol. Chem. {\bf 391}, 719 (2010).
\bibitem{mypre2} S. L. Singh, L. Schimmele, and S. Dietrich, Phys. Rev. E. {\bf 101}, 052115 (2020).
\bibitem{RDN}   K. Reijmer, S. Dietrich, and M. Napi\'orkowski, Phys. Rev. E. {\bf 60}, 4027 (1999).
\bibitem{Alex3}   A. Malijevsky and A. O. Parry, Phys. Rev. Lett. {\bf 120}, 135701 (2018). 
\bibitem{Alex}   A. Malijevsky, Phys. Rev. E {\bf 102}, 012804 (2020).
\bibitem{Alex2}   A. Malijevsky and A. O. Parry, J. Phys.: Condens. Matter {\bf 26}, 355003 (2014).
\bibitem{Weeks}  J. D. Weeks, D. Chandler, and H. C. Andersen,
                 J. Chem. Phys. {\bf 54}, 5237 (1971).                                       %
\bibitem{mypre1} S. L. Singh, L. Schimmele, and S. Dietrich, Phys. Rev. E. {\bf 91}, 032405 (2015).
\bibitem{Evans2} R. Evans, Adv. Phys. {\bf 28}, 143 (1979).
\bibitem{gurug} Y. Singh, Phys. Rep. {\bf 207}, 351 (1991).

\bibitem{Roth3} R. Roth, R. Evans, A. Lang, and G. Kahl, J. Phys.: Condens. Matter
                {\bf 14}, 12063 (2002).
\bibitem{Tarazona} P. Tarazona, J.A. Cuesta, and Y. Mart{\`{i}}nez-Rat{\`{o}}n, in {\it Lecture  Notes in Physics}
            (Springer, Heidelberg, 2008) vol. 753, p. 247.
\bibitem{Yasha1}  Y. Rosenfeld, Phys. Rev. Lett. {\bf 63}, 980 (1989).
\bibitem {Yasha2} Y. Rosenfeld, M. Schmidt, H. L{\"{o}}wen, and P. Tarazona, Phys. Rev. E {\bf 55}, 4245 (1997).
\bibitem{Roth4} R. Roth, J. Phys.: Condens. Matter {\bf 22}, 063102 (2010).
\bibitem{D01} S. Dietrich, {in \it Phase Transitions and Critical Phenomena}, edited by C. Domb and J. L.
	         Lebowitz (Academic, London, 1988), Vol. 12, p.1.
\bibitem{Evans3} R. Evans and U. Marini Bettolo Marconi, J. Chem. Phys. {\bf 86}, 7138 (1987).
\bibitem{Archer} A. J. Archer and R. Evans, J. Chem. Phys. {\bf 121}, 4246 (2004). 
\bibitem{Rauscher} M. Rauscher, Dynamic Density Functional Theory (DDFT), in 
	          {\it Encyclopedia of Microfluidics and Nanofluidics}, edited by D. Li (Springer, Boston, MA. (2008)),
		  pp. 693-699.
\bibitem{Striolo} M. Apostolopoulou, R. Day, R. Hull, M. Stamatakis,1, and A. Striolo, 
                        J. Chem. Phys. {\bf 147}, 134703 (2017). 
\bibitem{Binder} K. Binder and D.W. Heermann, Monte Carlo Simulation in Statistical Physics: 
	An Introduction, 6th Edition (Springer, Berlin (2019)). 
\bibitem{PDa} M. N. Popescu and S. Dietrich, in {\it Interface and Transport Dynamics},
                    Lecture Notes in Computational Science and Engineering, edited by
                    H. Emmerich, B. Nestler, and M. Schreckenberg (Springer, Berlin (2003)),
                    Vol. 32, p. 202. 
\bibitem{PDb} M. N. Popescu and S. Dietrich, Phys. Rev. E {\bf 69}, 061602 (2004).
\bibitem{DPR} S. Dietrich, M. N. Popescu, and M. Rauscher, J. Phys.: Condens. Matter {\bf 17},
                    S577 (2005). 
\bibitem{PDO} M. N. Popescu, S. Dietrich, and G. Oshanin, J. Phys.: Condens. Matter {\bf 17},
                    S4189 (2005).
\bibitem{Otter} T. H. Chakrapani and W. K. den Otter, Langmuir {\bf 36}, 12712 (2020).
\end{thebibliography}
\end{document}